\documentclass[longauth]{aaEC}

\usepackage{booktabs}  
\usepackage{graphicx}
\usepackage{natbib}
\usepackage{scalerel}
\usepackage{lastpage}
\usepackage{pdflscape}
\usepackage[table]{xcolor}
\usepackage{xcolor} 

\usepackage{txfonts}
\usepackage[pdfencoding=auto,psdextra]{hyperref}
\hypersetup{
    colorlinks=true,
    linkcolor=blue,
    filecolor=magenta,      
    urlcolor=blue,
    citecolor=blue
}
\makeatletter
\renewcommand*\aa@pageof{, page \thepage{} of \pageref*{LastPage}}
\makeatother

\usepackage[utf8]{inputenc}

\usepackage[switch, modulo]{lineno}

\usepackage{euclid}

\usepackage{siunitx}

\usepackage{glossaries}

\newacronym{eds}{EDS}{Euclid Deep Survey}
\newacronym{ews}{EWS}{Euclid Wide Survey}
\newacronym{edfn}{EDF-N}{Euclid Deep Field North}
\newacronym{edfs}{EDF-S}{Euclid Deep Field South}
\newacronym{edff}{EDF-F}{Euclid Deep Field Fornax}
\newacronym{cdfs}{CDF-S}{Chandra Deep Field South}
\newacronym{lsst}{LSST}{Legacy Survey of Space and Time}
\newacronym{unions}{UNIONS}{The Ultraviolet Near Infrared Optical Northern Survey}
\newacronym{hsc}{HSC}{Hyper-Suprime Cam}
\newacronym{decam}{DECAM}{Dark Energy Camera}
\newacronym{ou}{OU}{Organisational Units}
\newacronym{sgs}{SGS}{Science Ground Segment}
\newacronym{pdz}{PDZ}{redshift probability distribution}
\newacronym{sas}{SAS}{Science Archive System}

\newacronym{esa}{ESA}{European Space Agency}
\newacronym{cmb}{CMB}{cosmic microwave background}
\newacronym{ecgc}{ECGC}{Euclid Catalogue of Galaxy Clusters}
\newacronym{ec}{EC}{Euclid Consortium}
\newacronym{des}{DES}{Dark Energy Survey}
\newacronym{rm}{RM}{\texttt{RedMaPPer}}
\newacronym{icm}{ICM}{intracluster medium}
\newacronym{sz}{SZ}{Sunyaev--Zeldovich}
\newacronym{mcxc}{MCXC-II}{Meta-Catalogue of X-ray Clusters II}
\newacronym{rosat}{ROSAT}{R\"ontgen Satellite}
\newacronym{rass}{RASS}{ROSAT All-Sky Survey}
\newacronym{mcsz}{MCSZ}{Meta-Catalogue of SZ Clusters}
\newacronym{spt}{SPT}{South Pole Telescope}
\newacronym{act}{ACT}{Atacama Cosmology Telescope}
\newacronym{lc2}{LC$^2$}{Literature catalogues of weak lensing clusters of galaxies}
\newacronym{mccd}{MCCD}{Meta-Catalogue of Cluster Dispersions}
\newacronym{sdss}{SDSS}{Sloan Digital Sky Survey}
\newacronym{desi}{DESI}{Dark Energy Spectroscopic Instrument}
\newacronym{snr}{S/N}{signal-to-noise ratio}
\newacronym{egcp}{EGCP}{Euclid Galaxy Cluster Pipeline}
\newacronym{q1}{Q1}{Euclid Quick Release 1}
\newacronym{dr1}{DR1}{Euclid Data Release 1}
\newacronym{photz}{photo-$z$}{photometric redshifts}
\newacronym{madcows2}{MaDCoWS2}{Massive and Distant Clusters of the Wise Survey 2}
\newacronym{ned}{NED}{NASA/IPAC Extragalactic Database}
\newacronym{wh24}{WH24}{Wen and Han 2024}

\usepackage{cleveref}
\crefname{section}{Sect.}{Sects.}
\Crefname{section}{Section}{Sections}
\crefname{figure}{Fig.}{Figs.}
\Crefname{figure}{Figure}{Figures}
\crefname{table}{Table}{Tables}
\Crefname{table}{Table}{Tables}
\crefname{appendix}{Appendix}{Appendices}
\Crefname{appendix}{Appendix}{Appendices}

\newcommand{\AMICO}{{\tt AMICO}\xspace}
\newcommand{\PZWav}{{\tt PZWav}\xspace}
\newcommand{\zp}{z_\mathrm{p}}
\newcommand{\zs}{z_\mathrm{s}}
\newcommand{\zd}{z_{{\rm DET}}}
\newcommand{\redmapper}{{\tt RedMaPPer}\xspace}
\newcommand{\RICHCL}{\texttt{RICH-CL}\xspace}
\newcommand{\OUSPE}{\texttt{OU-SPE}\xspace}
\newcommand{\snr}{S/N}

\newcommand{\hide}[1]{\textcolor{lightgray}{#1}}

\begin{document} 

\newcommand{\orcid}[1]{} 

\title{Euclid Quick Data Release (Q1)} \subtitle{First detections from the \Euclid galaxy cluster workflow}

\renewcommand{\orcid}[1]{} 
\author{Euclid Collaboration: S.~Bhargava\orcid{0000-0003-3851-7219}\thanks{\email{sunayana.bhargava@oca.eu}}\inst{\ref{aff1}}
\and C.~Benoist\inst{\ref{aff1}}
\and A.~H.~Gonzalez\orcid{0000-0002-0933-8601}\inst{\ref{aff2}}
\and M.~Maturi\orcid{0000-0002-3517-2422}\inst{\ref{aff3},\ref{aff4}}
\and J.-B.~Melin\inst{\ref{aff5}}
\and S.~A.~Stanford\orcid{0000-0003-0122-0841}\inst{\ref{aff6}}
\and E.~Munari\orcid{0000-0002-1751-5946}\inst{\ref{aff7},\ref{aff8}}
\and M.~Vannier\inst{\ref{aff1}}
\and C.~Murray\inst{\ref{aff9}}
\and S.~Maurogordato\inst{\ref{aff1}}
\and A.~Biviano\orcid{0000-0002-0857-0732}\inst{\ref{aff7},\ref{aff8}}
\and J.~Macias-Perez\orcid{0000-0002-5385-2763}\inst{\ref{aff10}}
\and J.~G.~Bartlett\orcid{0000-0002-0685-8310}\inst{\ref{aff9}}
\and F.~Pacaud\orcid{0000-0002-6622-4555}\inst{\ref{aff11}}
\and A.~Widmer\orcid{0009-0005-4111-2716}\inst{\ref{aff9}}
\and M.~Meneghetti\orcid{0000-0003-1225-7084}\inst{\ref{aff12},\ref{aff13}}
\and B.~Sartoris\orcid{0000-0003-1337-5269}\inst{\ref{aff14},\ref{aff7}}
\and M.~Aguena\orcid{0000-0001-5679-6747}\inst{\ref{aff7}}
\and G.~Alguero\inst{\ref{aff10}}
\and S.~Andreon\orcid{0000-0002-2041-8784}\inst{\ref{aff15}}
\and S.~Bardelli\orcid{0000-0002-8900-0298}\inst{\ref{aff12}}
\and L.~Baumont\orcid{0000-0002-1518-0150}\inst{\ref{aff16},\ref{aff7},\ref{aff8}}
\and M.~Bolzonella\orcid{0000-0003-3278-4607}\inst{\ref{aff12}}
\and R.~Cabanac\orcid{0000-0001-6679-2600}\inst{\ref{aff17}}
\and A.~Cappi\inst{\ref{aff12},\ref{aff1}}
\and G.~Castignani\orcid{0000-0001-6831-0687}\inst{\ref{aff12}}
\and C.~Combet\orcid{0000-0001-6487-1866}\inst{\ref{aff10}}
\and J.~Comparat\orcid{0000-0001-9200-1497}\inst{\ref{aff18}}
\and S.~Farrens\orcid{0000-0002-9594-9387}\inst{\ref{aff19}}
\and Z.~Ghaffari\orcid{0000-0002-6467-8078}\inst{\ref{aff7},\ref{aff8}}
\and P.~A.~Giles\orcid{0000-0003-4937-8453}\inst{\ref{aff20}}
\and C.~Giocoli\orcid{0000-0002-9590-7961}\inst{\ref{aff12},\ref{aff13}}
\and M.~Girardi\orcid{0000-0003-1861-1865}\inst{\ref{aff16},\ref{aff7}}
\and N.~A.~Hatch\orcid{0000-0001-5600-0534}\inst{\ref{aff21}}
\and L.~Ingoglia\orcid{0000-0002-7587-0997}\inst{\ref{aff22}}
\and A.~Iovino\orcid{0000-0001-6958-0304}\inst{\ref{aff15}}
\and G.~A.~Mamon\orcid{0000-0001-8956-5953}\inst{\ref{aff23},\ref{aff24}}
\and S.~Mei\orcid{0000-0002-2849-559X}\inst{\ref{aff9},\ref{aff25}}
\and L.~Moscardini\orcid{0000-0002-3473-6716}\inst{\ref{aff26},\ref{aff12},\ref{aff13}}
\and S.~Mourre\orcid{0009-0005-9047-0691}\inst{\ref{aff1},\ref{aff27}}
\and J.~Odier\orcid{0000-0002-1650-2246}\inst{\ref{aff10}}
\and G.~W.~Pratt\inst{\ref{aff19}}
\and P.~Tarr\'io\orcid{0000-0002-0915-0131}\inst{\ref{aff28}}
\and G.~Toni\orcid{0009-0005-3133-1157}\inst{\ref{aff26},\ref{aff12},\ref{aff4}}
\and J.~Weller\orcid{0000-0002-8282-2010}\inst{\ref{aff14},\ref{aff18}}
\and E.~Zucca\orcid{0000-0002-5845-8132}\inst{\ref{aff12}}
\and N.~Aghanim\orcid{0000-0002-6688-8992}\inst{\ref{aff29}}
\and B.~Altieri\orcid{0000-0003-3936-0284}\inst{\ref{aff30}}
\and A.~Amara\inst{\ref{aff31}}
\and N.~Auricchio\orcid{0000-0003-4444-8651}\inst{\ref{aff12}}
\and C.~Baccigalupi\orcid{0000-0002-8211-1630}\inst{\ref{aff8},\ref{aff7},\ref{aff32},\ref{aff33}}
\and M.~Baldi\orcid{0000-0003-4145-1943}\inst{\ref{aff34},\ref{aff12},\ref{aff13}}
\and A.~Balestra\orcid{0000-0002-6967-261X}\inst{\ref{aff35}}
\and P.~Battaglia\orcid{0000-0002-7337-5909}\inst{\ref{aff12}}
\and F.~Bellagamba\inst{\ref{aff34},\ref{aff12}}
\and A.~Bonchi\orcid{0000-0002-2667-5482}\inst{\ref{aff36}}
\and E.~Branchini\orcid{0000-0002-0808-6908}\inst{\ref{aff37},\ref{aff38},\ref{aff15}}
\and M.~Brescia\orcid{0000-0001-9506-5680}\inst{\ref{aff39},\ref{aff40}}
\and J.~Brinchmann\orcid{0000-0003-4359-8797}\inst{\ref{aff41},\ref{aff42}}
\and S.~Camera\orcid{0000-0003-3399-3574}\inst{\ref{aff43},\ref{aff44},\ref{aff45}}
\and V.~Capobianco\orcid{0000-0002-3309-7692}\inst{\ref{aff45}}
\and C.~Carbone\orcid{0000-0003-0125-3563}\inst{\ref{aff46}}
\and J.~Carretero\orcid{0000-0002-3130-0204}\inst{\ref{aff47},\ref{aff48}}
\and S.~Casas\orcid{0000-0002-4751-5138}\inst{\ref{aff49}}
\and M.~Castellano\orcid{0000-0001-9875-8263}\inst{\ref{aff50}}
\and S.~Cavuoti\orcid{0000-0002-3787-4196}\inst{\ref{aff40},\ref{aff51}}
\and K.~C.~Chambers\orcid{0000-0001-6965-7789}\inst{\ref{aff52}}
\and A.~Cimatti\inst{\ref{aff53}}
\and C.~Colodro-Conde\inst{\ref{aff54}}
\and G.~Congedo\orcid{0000-0003-2508-0046}\inst{\ref{aff55}}
\and C.~J.~Conselice\orcid{0000-0003-1949-7638}\inst{\ref{aff56}}
\and L.~Conversi\orcid{0000-0002-6710-8476}\inst{\ref{aff57},\ref{aff30}}
\and Y.~Copin\orcid{0000-0002-5317-7518}\inst{\ref{aff58}}
\and F.~Courbin\orcid{0000-0003-0758-6510}\inst{\ref{aff59},\ref{aff60}}
\and H.~M.~Courtois\orcid{0000-0003-0509-1776}\inst{\ref{aff61}}
\and M.~Cropper\orcid{0000-0003-4571-9468}\inst{\ref{aff62}}
\and A.~Da~Silva\orcid{0000-0002-6385-1609}\inst{\ref{aff63},\ref{aff64}}
\and H.~Degaudenzi\orcid{0000-0002-5887-6799}\inst{\ref{aff65}}
\and G.~De~Lucia\orcid{0000-0002-6220-9104}\inst{\ref{aff7}}
\and A.~M.~Di~Giorgio\orcid{0000-0002-4767-2360}\inst{\ref{aff66}}
\and C.~Dolding\orcid{0009-0003-7199-6108}\inst{\ref{aff62}}
\and H.~Dole\orcid{0000-0002-9767-3839}\inst{\ref{aff29}}
\and F.~Dubath\orcid{0000-0002-6533-2810}\inst{\ref{aff65}}
\and C.~A.~J.~Duncan\orcid{0009-0003-3573-0791}\inst{\ref{aff56}}
\and X.~Dupac\inst{\ref{aff30}}
\and S.~Dusini\orcid{0000-0002-1128-0664}\inst{\ref{aff67}}
\and A.~Ealet\orcid{0000-0003-3070-014X}\inst{\ref{aff58}}
\and S.~Escoffier\orcid{0000-0002-2847-7498}\inst{\ref{aff68}}
\and M.~Farina\orcid{0000-0002-3089-7846}\inst{\ref{aff66}}
\and R.~Farinelli\inst{\ref{aff12}}
\and F.~Faustini\orcid{0000-0001-6274-5145}\inst{\ref{aff50},\ref{aff36}}
\and S.~Ferriol\inst{\ref{aff58}}
\and F.~Finelli\orcid{0000-0002-6694-3269}\inst{\ref{aff12},\ref{aff69}}
\and P.~Fosalba\orcid{0000-0002-1510-5214}\inst{\ref{aff70},\ref{aff71}}
\and S.~Fotopoulou\orcid{0000-0002-9686-254X}\inst{\ref{aff72}}
\and M.~Frailis\orcid{0000-0002-7400-2135}\inst{\ref{aff7}}
\and E.~Franceschi\orcid{0000-0002-0585-6591}\inst{\ref{aff12}}
\and M.~Fumana\orcid{0000-0001-6787-5950}\inst{\ref{aff46}}
\and S.~Galeotta\orcid{0000-0002-3748-5115}\inst{\ref{aff7}}
\and K.~George\orcid{0000-0002-1734-8455}\inst{\ref{aff14}}
\and B.~Gillis\orcid{0000-0002-4478-1270}\inst{\ref{aff55}}
\and P.~G\'omez-Alvarez\orcid{0000-0002-8594-5358}\inst{\ref{aff73},\ref{aff30}}
\and J.~Gracia-Carpio\inst{\ref{aff18}}
\and B.~R.~Granett\orcid{0000-0003-2694-9284}\inst{\ref{aff15}}
\and A.~Grazian\orcid{0000-0002-5688-0663}\inst{\ref{aff35}}
\and F.~Grupp\inst{\ref{aff18},\ref{aff14}}
\and L.~Guzzo\orcid{0000-0001-8264-5192}\inst{\ref{aff74},\ref{aff15},\ref{aff75}}
\and S.~Gwyn\orcid{0000-0001-8221-8406}\inst{\ref{aff76}}
\and S.~V.~H.~Haugan\orcid{0000-0001-9648-7260}\inst{\ref{aff77}}
\and J.~Hoar\inst{\ref{aff30}}
\and H.~Hoekstra\orcid{0000-0002-0641-3231}\inst{\ref{aff78}}
\and W.~Holmes\inst{\ref{aff79}}
\and F.~Hormuth\inst{\ref{aff80}}
\and A.~Hornstrup\orcid{0000-0002-3363-0936}\inst{\ref{aff81},\ref{aff82}}
\and P.~Hudelot\inst{\ref{aff23}}
\and S.~Ili\'c\orcid{0000-0003-4285-9086}\inst{\ref{aff83},\ref{aff17}}
\and K.~Jahnke\orcid{0000-0003-3804-2137}\inst{\ref{aff84}}
\and M.~Jhabvala\inst{\ref{aff85}}
\and B.~Joachimi\orcid{0000-0001-7494-1303}\inst{\ref{aff86}}
\and E.~Keih\"anen\orcid{0000-0003-1804-7715}\inst{\ref{aff87}}
\and S.~Kermiche\orcid{0000-0002-0302-5735}\inst{\ref{aff68}}
\and A.~Kiessling\orcid{0000-0002-2590-1273}\inst{\ref{aff79}}
\and M.~Kilbinger\orcid{0000-0001-9513-7138}\inst{\ref{aff19}}
\and B.~Kubik\orcid{0009-0006-5823-4880}\inst{\ref{aff58}}
\and K.~Kuijken\orcid{0000-0002-3827-0175}\inst{\ref{aff78}}
\and M.~K\"ummel\orcid{0000-0003-2791-2117}\inst{\ref{aff14}}
\and M.~Kunz\orcid{0000-0002-3052-7394}\inst{\ref{aff88}}
\and H.~Kurki-Suonio\orcid{0000-0002-4618-3063}\inst{\ref{aff89},\ref{aff90}}
\and Q.~Le~Boulc'h\inst{\ref{aff91}}
\and A.~M.~C.~Le~Brun\orcid{0000-0002-0936-4594}\inst{\ref{aff92}}
\and D.~Le~Mignant\orcid{0000-0002-5339-5515}\inst{\ref{aff93}}
\and P.~Liebing\inst{\ref{aff62}}
\and S.~Ligori\orcid{0000-0003-4172-4606}\inst{\ref{aff45}}
\and P.~B.~Lilje\orcid{0000-0003-4324-7794}\inst{\ref{aff77}}
\and V.~Lindholm\orcid{0000-0003-2317-5471}\inst{\ref{aff89},\ref{aff90}}
\and I.~Lloro\orcid{0000-0001-5966-1434}\inst{\ref{aff94}}
\and G.~Mainetti\orcid{0000-0003-2384-2377}\inst{\ref{aff91}}
\and D.~Maino\inst{\ref{aff74},\ref{aff46},\ref{aff75}}
\and E.~Maiorano\orcid{0000-0003-2593-4355}\inst{\ref{aff12}}
\and O.~Mansutti\orcid{0000-0001-5758-4658}\inst{\ref{aff7}}
\and S.~Marcin\inst{\ref{aff95}}
\and O.~Marggraf\orcid{0000-0001-7242-3852}\inst{\ref{aff11}}
\and M.~Martinelli\orcid{0000-0002-6943-7732}\inst{\ref{aff50},\ref{aff96}}
\and N.~Martinet\orcid{0000-0003-2786-7790}\inst{\ref{aff93}}
\and F.~Marulli\orcid{0000-0002-8850-0303}\inst{\ref{aff26},\ref{aff12},\ref{aff13}}
\and R.~Massey\orcid{0000-0002-6085-3780}\inst{\ref{aff97}}
\and E.~Medinaceli\orcid{0000-0002-4040-7783}\inst{\ref{aff12}}
\and M.~Melchior\inst{\ref{aff98}}
\and Y.~Mellier\inst{\ref{aff24},\ref{aff23}}
\and E.~Merlin\orcid{0000-0001-6870-8900}\inst{\ref{aff50}}
\and G.~Meylan\inst{\ref{aff99}}
\and A.~Mora\orcid{0000-0002-1922-8529}\inst{\ref{aff100}}
\and M.~Moresco\orcid{0000-0002-7616-7136}\inst{\ref{aff26},\ref{aff12}}
\and R.~Nakajima\orcid{0009-0009-1213-7040}\inst{\ref{aff11}}
\and C.~Neissner\orcid{0000-0001-8524-4968}\inst{\ref{aff101},\ref{aff48}}
\and R.~C.~Nichol\orcid{0000-0003-0939-6518}\inst{\ref{aff31}}
\and S.-M.~Niemi\inst{\ref{aff102}}
\and J.~W.~Nightingale\orcid{0000-0002-8987-7401}\inst{\ref{aff103}}
\and C.~Padilla\orcid{0000-0001-7951-0166}\inst{\ref{aff101}}
\and S.~Paltani\orcid{0000-0002-8108-9179}\inst{\ref{aff65}}
\and F.~Pasian\orcid{0000-0002-4869-3227}\inst{\ref{aff7}}
\and K.~Pedersen\inst{\ref{aff104}}
\and W.~J.~Percival\orcid{0000-0002-0644-5727}\inst{\ref{aff105},\ref{aff106},\ref{aff107}}
\and V.~Pettorino\inst{\ref{aff102}}
\and S.~Pires\orcid{0000-0002-0249-2104}\inst{\ref{aff19}}
\and G.~Polenta\orcid{0000-0003-4067-9196}\inst{\ref{aff36}}
\and M.~Poncet\inst{\ref{aff108}}
\and L.~A.~Popa\inst{\ref{aff109}}
\and L.~Pozzetti\orcid{0000-0001-7085-0412}\inst{\ref{aff12}}
\and F.~Raison\orcid{0000-0002-7819-6918}\inst{\ref{aff18}}
\and R.~Rebolo\orcid{0000-0003-3767-7085}\inst{\ref{aff54},\ref{aff110},\ref{aff111}}
\and A.~Renzi\orcid{0000-0001-9856-1970}\inst{\ref{aff112},\ref{aff67}}
\and J.~Rhodes\orcid{0000-0002-4485-8549}\inst{\ref{aff79}}
\and G.~Riccio\inst{\ref{aff40}}
\and E.~Romelli\orcid{0000-0003-3069-9222}\inst{\ref{aff7}}
\and M.~Roncarelli\orcid{0000-0001-9587-7822}\inst{\ref{aff12}}
\and R.~Saglia\orcid{0000-0003-0378-7032}\inst{\ref{aff14},\ref{aff18}}
\and Z.~Sakr\orcid{0000-0002-4823-3757}\inst{\ref{aff3},\ref{aff17},\ref{aff113}}
\and A.~G.~S\'anchez\orcid{0000-0003-1198-831X}\inst{\ref{aff18}}
\and D.~Sapone\orcid{0000-0001-7089-4503}\inst{\ref{aff114}}
\and J.~A.~Schewtschenko\orcid{0000-0002-4913-6393}\inst{\ref{aff55}}
\and P.~Schneider\orcid{0000-0001-8561-2679}\inst{\ref{aff11}}
\and T.~Schrabback\orcid{0000-0002-6987-7834}\inst{\ref{aff115}}
\and A.~Secroun\orcid{0000-0003-0505-3710}\inst{\ref{aff68}}
\and E.~Sefusatti\orcid{0000-0003-0473-1567}\inst{\ref{aff7},\ref{aff8},\ref{aff32}}
\and G.~Seidel\orcid{0000-0003-2907-353X}\inst{\ref{aff84}}
\and S.~Serrano\orcid{0000-0002-0211-2861}\inst{\ref{aff70},\ref{aff116},\ref{aff71}}
\and P.~Simon\inst{\ref{aff11}}
\and C.~Sirignano\orcid{0000-0002-0995-7146}\inst{\ref{aff112},\ref{aff67}}
\and G.~Sirri\orcid{0000-0003-2626-2853}\inst{\ref{aff13}}
\and J.~Skottfelt\orcid{0000-0003-1310-8283}\inst{\ref{aff117}}
\and L.~Stanco\orcid{0000-0002-9706-5104}\inst{\ref{aff67}}
\and J.~Steinwagner\orcid{0000-0001-7443-1047}\inst{\ref{aff18}}
\and P.~Tallada-Cresp\'{i}\orcid{0000-0002-1336-8328}\inst{\ref{aff47},\ref{aff48}}
\and A.~N.~Taylor\inst{\ref{aff55}}
\and I.~Tereno\inst{\ref{aff63},\ref{aff118}}
\and S.~Toft\orcid{0000-0003-3631-7176}\inst{\ref{aff119},\ref{aff120}}
\and R.~Toledo-Moreo\orcid{0000-0002-2997-4859}\inst{\ref{aff121}}
\and F.~Torradeflot\orcid{0000-0003-1160-1517}\inst{\ref{aff48},\ref{aff47}}
\and I.~Tutusaus\orcid{0000-0002-3199-0399}\inst{\ref{aff17}}
\and L.~Valenziano\orcid{0000-0002-1170-0104}\inst{\ref{aff12},\ref{aff69}}
\and J.~Valiviita\orcid{0000-0001-6225-3693}\inst{\ref{aff89},\ref{aff90}}
\and T.~Vassallo\orcid{0000-0001-6512-6358}\inst{\ref{aff14},\ref{aff7}}
\and G.~Verdoes~Kleijn\orcid{0000-0001-5803-2580}\inst{\ref{aff122}}
\and A.~Veropalumbo\orcid{0000-0003-2387-1194}\inst{\ref{aff15},\ref{aff38},\ref{aff37}}
\and Y.~Wang\orcid{0000-0002-4749-2984}\inst{\ref{aff123}}
\and A.~Zacchei\orcid{0000-0003-0396-1192}\inst{\ref{aff7},\ref{aff8}}
\and G.~Zamorani\orcid{0000-0002-2318-301X}\inst{\ref{aff12}}
\and F.~M.~Zerbi\inst{\ref{aff15}}
\and V.~Allevato\orcid{0000-0001-7232-5152}\inst{\ref{aff40}}
\and M.~Ballardini\orcid{0000-0003-4481-3559}\inst{\ref{aff124},\ref{aff125},\ref{aff12}}
\and E.~Bozzo\orcid{0000-0002-8201-1525}\inst{\ref{aff65}}
\and C.~Burigana\orcid{0000-0002-3005-5796}\inst{\ref{aff22},\ref{aff69}}
\and P.~Casenove\orcid{0009-0006-6736-1670}\inst{\ref{aff108}}
\and D.~Di~Ferdinando\inst{\ref{aff13}}
\and J.~A.~Escartin~Vigo\inst{\ref{aff18}}
\and G.~Fabbian\orcid{0000-0002-3255-4695}\inst{\ref{aff126}}
\and L.~Gabarra\orcid{0000-0002-8486-8856}\inst{\ref{aff127}}
\and J.~Mart\'{i}n-Fleitas\orcid{0000-0002-8594-569X}\inst{\ref{aff100}}
\and S.~Matthew\orcid{0000-0001-8448-1697}\inst{\ref{aff55}}
\and N.~Mauri\orcid{0000-0001-8196-1548}\inst{\ref{aff53},\ref{aff13}}
\and R.~B.~Metcalf\orcid{0000-0003-3167-2574}\inst{\ref{aff26},\ref{aff12}}
\and A.~Pezzotta\orcid{0000-0003-0726-2268}\inst{\ref{aff128},\ref{aff18}}
\and M.~P\"ontinen\orcid{0000-0001-5442-2530}\inst{\ref{aff89}}
\and C.~Porciani\orcid{0000-0002-7797-2508}\inst{\ref{aff11}}
\and I.~Risso\orcid{0000-0003-2525-7761}\inst{\ref{aff129}}
\and V.~Scottez\inst{\ref{aff24},\ref{aff130}}
\and M.~Sereno\orcid{0000-0003-0302-0325}\inst{\ref{aff12},\ref{aff13}}
\and M.~Tenti\orcid{0000-0002-4254-5901}\inst{\ref{aff13}}
\and M.~Viel\orcid{0000-0002-2642-5707}\inst{\ref{aff8},\ref{aff7},\ref{aff33},\ref{aff32},\ref{aff131}}
\and M.~Wiesmann\orcid{0009-0000-8199-5860}\inst{\ref{aff77}}
\and Y.~Akrami\orcid{0000-0002-2407-7956}\inst{\ref{aff132},\ref{aff133}}
\and I.~T.~Andika\orcid{0000-0001-6102-9526}\inst{\ref{aff134},\ref{aff135}}
\and S.~Anselmi\orcid{0000-0002-3579-9583}\inst{\ref{aff67},\ref{aff112},\ref{aff136}}
\and M.~Archidiacono\orcid{0000-0003-4952-9012}\inst{\ref{aff74},\ref{aff75}}
\and F.~Atrio-Barandela\orcid{0000-0002-2130-2513}\inst{\ref{aff137}}
\and K.~Benson\inst{\ref{aff62}}
\and P.~Bergamini\orcid{0000-0003-1383-9414}\inst{\ref{aff74},\ref{aff12}}
\and D.~Bertacca\orcid{0000-0002-2490-7139}\inst{\ref{aff112},\ref{aff35},\ref{aff67}}
\and M.~Bethermin\orcid{0000-0002-3915-2015}\inst{\ref{aff138}}
\and A.~Blanchard\orcid{0000-0001-8555-9003}\inst{\ref{aff17}}
\and L.~Blot\orcid{0000-0002-9622-7167}\inst{\ref{aff139},\ref{aff136}}
\and H.~B\"ohringer\orcid{0000-0001-8241-4204}\inst{\ref{aff18},\ref{aff140},\ref{aff141}}
\and S.~Borgani\orcid{0000-0001-6151-6439}\inst{\ref{aff16},\ref{aff8},\ref{aff7},\ref{aff32},\ref{aff131}}
\and M.~L.~Brown\orcid{0000-0002-0370-8077}\inst{\ref{aff56}}
\and S.~Bruton\orcid{0000-0002-6503-5218}\inst{\ref{aff142}}
\and A.~Calabro\orcid{0000-0003-2536-1614}\inst{\ref{aff50}}
\and B.~Camacho~Quevedo\orcid{0000-0002-8789-4232}\inst{\ref{aff70},\ref{aff71}}
\and F.~Caro\inst{\ref{aff50}}
\and C.~S.~Carvalho\inst{\ref{aff118}}
\and T.~Castro\orcid{0000-0002-6292-3228}\inst{\ref{aff7},\ref{aff32},\ref{aff8},\ref{aff131}}
\and Y.~Charles\inst{\ref{aff93}}
\and F.~Cogato\orcid{0000-0003-4632-6113}\inst{\ref{aff26},\ref{aff12}}
\and A.~R.~Cooray\orcid{0000-0002-3892-0190}\inst{\ref{aff143}}
\and M.~Costanzi\orcid{0000-0001-8158-1449}\inst{\ref{aff16},\ref{aff7},\ref{aff8}}
\and O.~Cucciati\orcid{0000-0002-9336-7551}\inst{\ref{aff12}}
\and S.~Davini\orcid{0000-0003-3269-1718}\inst{\ref{aff38}}
\and F.~De~Paolis\orcid{0000-0001-6460-7563}\inst{\ref{aff144},\ref{aff145},\ref{aff146}}
\and G.~Desprez\orcid{0000-0001-8325-1742}\inst{\ref{aff122}}
\and A.~D\'iaz-S\'anchez\orcid{0000-0003-0748-4768}\inst{\ref{aff147}}
\and J.~J.~Diaz\inst{\ref{aff148}}
\and S.~Di~Domizio\orcid{0000-0003-2863-5895}\inst{\ref{aff37},\ref{aff38}}
\and J.~M.~Diego\orcid{0000-0001-9065-3926}\inst{\ref{aff149}}
\and P.~Dimauro\orcid{0000-0001-7399-2854}\inst{\ref{aff50},\ref{aff150}}
\and P.-A.~Duc\orcid{0000-0003-3343-6284}\inst{\ref{aff138}}
\and A.~Enia\orcid{0000-0002-0200-2857}\inst{\ref{aff34},\ref{aff12}}
\and Y.~Fang\inst{\ref{aff14}}
\and A.~G.~Ferrari\orcid{0009-0005-5266-4110}\inst{\ref{aff13}}
\and P.~G.~Ferreira\orcid{0000-0002-3021-2851}\inst{\ref{aff127}}
\and A.~Finoguenov\orcid{0000-0002-4606-5403}\inst{\ref{aff89}}
\and A.~Fontana\orcid{0000-0003-3820-2823}\inst{\ref{aff50}}
\and A.~Franco\orcid{0000-0002-4761-366X}\inst{\ref{aff145},\ref{aff144},\ref{aff146}}
\and K.~Ganga\orcid{0000-0001-8159-8208}\inst{\ref{aff9}}
\and J.~Garc\'ia-Bellido\orcid{0000-0002-9370-8360}\inst{\ref{aff132}}
\and T.~Gasparetto\orcid{0000-0002-7913-4866}\inst{\ref{aff7}}
\and E.~Gaztanaga\orcid{0000-0001-9632-0815}\inst{\ref{aff71},\ref{aff70},\ref{aff151}}
\and F.~Giacomini\orcid{0000-0002-3129-2814}\inst{\ref{aff13}}
\and F.~Gianotti\orcid{0000-0003-4666-119X}\inst{\ref{aff12}}
\and G.~Gozaliasl\orcid{0000-0002-0236-919X}\inst{\ref{aff152},\ref{aff89}}
\and M.~Guidi\orcid{0000-0001-9408-1101}\inst{\ref{aff34},\ref{aff12}}
\and C.~M.~Gutierrez\orcid{0000-0001-7854-783X}\inst{\ref{aff153}}
\and A.~Hall\orcid{0000-0002-3139-8651}\inst{\ref{aff55}}
\and W.~G.~Hartley\inst{\ref{aff65}}
\and C.~Hern\'andez-Monteagudo\orcid{0000-0001-5471-9166}\inst{\ref{aff111},\ref{aff54}}
\and H.~Hildebrandt\orcid{0000-0002-9814-3338}\inst{\ref{aff154}}
\and J.~Hjorth\orcid{0000-0002-4571-2306}\inst{\ref{aff104}}
\and J.~J.~E.~Kajava\orcid{0000-0002-3010-8333}\inst{\ref{aff155},\ref{aff156}}
\and Y.~Kang\orcid{0009-0000-8588-7250}\inst{\ref{aff65}}
\and V.~Kansal\orcid{0000-0002-4008-6078}\inst{\ref{aff157},\ref{aff158}}
\and D.~Karagiannis\orcid{0000-0002-4927-0816}\inst{\ref{aff124},\ref{aff159}}
\and K.~Kiiveri\inst{\ref{aff87}}
\and C.~C.~Kirkpatrick\inst{\ref{aff87}}
\and S.~Kruk\orcid{0000-0001-8010-8879}\inst{\ref{aff30}}
\and J.~Le~Graet\orcid{0000-0001-6523-7971}\inst{\ref{aff68}}
\and L.~Legrand\orcid{0000-0003-0610-5252}\inst{\ref{aff160},\ref{aff161}}
\and M.~Lembo\orcid{0000-0002-5271-5070}\inst{\ref{aff124},\ref{aff125}}
\and F.~Lepori\orcid{0009-0000-5061-7138}\inst{\ref{aff162}}
\and G.~Leroy\orcid{0009-0004-2523-4425}\inst{\ref{aff163},\ref{aff97}}
\and G.~F.~Lesci\orcid{0000-0002-4607-2830}\inst{\ref{aff26},\ref{aff12}}
\and J.~Lesgourgues\orcid{0000-0001-7627-353X}\inst{\ref{aff49}}
\and L.~Leuzzi\orcid{0009-0006-4479-7017}\inst{\ref{aff26},\ref{aff12}}
\and T.~I.~Liaudat\orcid{0000-0002-9104-314X}\inst{\ref{aff164}}
\and S.~J.~Liu\orcid{0000-0001-7680-2139}\inst{\ref{aff66}}
\and A.~Loureiro\orcid{0000-0002-4371-0876}\inst{\ref{aff165},\ref{aff166}}
\and G.~Maggio\orcid{0000-0003-4020-4836}\inst{\ref{aff7}}
\and M.~Magliocchetti\orcid{0000-0001-9158-4838}\inst{\ref{aff66}}
\and E.~A.~Magnier\orcid{0000-0002-7965-2815}\inst{\ref{aff52}}
\and F.~Mannucci\orcid{0000-0002-4803-2381}\inst{\ref{aff167}}
\and R.~Maoli\orcid{0000-0002-6065-3025}\inst{\ref{aff168},\ref{aff50}}
\and C.~J.~A.~P.~Martins\orcid{0000-0002-4886-9261}\inst{\ref{aff169},\ref{aff41}}
\and L.~Maurin\orcid{0000-0002-8406-0857}\inst{\ref{aff29}}
\and M.~Migliaccio\inst{\ref{aff170},\ref{aff171}}
\and M.~Miluzio\inst{\ref{aff30},\ref{aff172}}
\and P.~Monaco\orcid{0000-0003-2083-7564}\inst{\ref{aff16},\ref{aff7},\ref{aff32},\ref{aff8}}
\and A.~Montoro\orcid{0000-0003-4730-8590}\inst{\ref{aff71},\ref{aff70}}
\and C.~Moretti\orcid{0000-0003-3314-8936}\inst{\ref{aff33},\ref{aff131},\ref{aff7},\ref{aff8},\ref{aff32}}
\and G.~Morgante\inst{\ref{aff12}}
\and K.~Naidoo\orcid{0000-0002-9182-1802}\inst{\ref{aff151}}
\and A.~Navarro-Alsina\orcid{0000-0002-3173-2592}\inst{\ref{aff11}}
\and S.~Nesseris\orcid{0000-0002-0567-0324}\inst{\ref{aff132}}
\and F.~Passalacqua\orcid{0000-0002-8606-4093}\inst{\ref{aff112},\ref{aff67}}
\and K.~Paterson\orcid{0000-0001-8340-3486}\inst{\ref{aff84}}
\and L.~Patrizii\inst{\ref{aff13}}
\and A.~Pisani\orcid{0000-0002-6146-4437}\inst{\ref{aff68},\ref{aff173}}
\and D.~Potter\orcid{0000-0002-0757-5195}\inst{\ref{aff162}}
\and S.~Quai\orcid{0000-0002-0449-8163}\inst{\ref{aff26},\ref{aff12}}
\and M.~Radovich\orcid{0000-0002-3585-866X}\inst{\ref{aff35}}
\and P.-F.~Rocci\inst{\ref{aff29}}
\and P.~Rosati\orcid{0000-0002-6813-0632}\inst{\ref{aff124},\ref{aff12}}
\and S.~Sacquegna\orcid{0000-0002-8433-6630}\inst{\ref{aff144},\ref{aff145},\ref{aff146}}
\and M.~Sahl\'en\orcid{0000-0003-0973-4804}\inst{\ref{aff174}}
\and D.~B.~Sanders\orcid{0000-0002-1233-9998}\inst{\ref{aff52}}
\and E.~Sarpa\orcid{0000-0002-1256-655X}\inst{\ref{aff33},\ref{aff131},\ref{aff32}}
\and J.~Schaye\orcid{0000-0002-0668-5560}\inst{\ref{aff78}}
\and A.~Schneider\orcid{0000-0001-7055-8104}\inst{\ref{aff162}}
\and M.~Schultheis\inst{\ref{aff1}}
\and D.~Sciotti\orcid{0009-0008-4519-2620}\inst{\ref{aff50},\ref{aff96}}
\and E.~Sellentin\inst{\ref{aff175},\ref{aff78}}
\and L.~C.~Smith\orcid{0000-0002-3259-2771}\inst{\ref{aff176}}
\and K.~Tanidis\orcid{0000-0001-9843-5130}\inst{\ref{aff127}}
\and G.~Testera\inst{\ref{aff38}}
\and R.~Teyssier\orcid{0000-0001-7689-0933}\inst{\ref{aff173}}
\and S.~Tosi\orcid{0000-0002-7275-9193}\inst{\ref{aff37},\ref{aff38},\ref{aff15}}
\and A.~Troja\orcid{0000-0003-0239-4595}\inst{\ref{aff112},\ref{aff67}}
\and M.~Tucci\inst{\ref{aff65}}
\and C.~Valieri\inst{\ref{aff13}}
\and A.~Venhola\orcid{0000-0001-6071-4564}\inst{\ref{aff177}}
\and D.~Vergani\orcid{0000-0003-0898-2216}\inst{\ref{aff12}}
\and G.~Verza\orcid{0000-0002-1886-8348}\inst{\ref{aff178}}
\and P.~Vielzeuf\orcid{0000-0003-2035-9339}\inst{\ref{aff68}}
\and N.~A.~Walton\orcid{0000-0003-3983-8778}\inst{\ref{aff176}}
\and E.~Soubrie\orcid{0000-0001-9295-1863}\inst{\ref{aff29}}
\and D.~Scott\orcid{0000-0002-6878-9840}\inst{\ref{aff179}}}
										   
\institute{Universit\'e C\^{o}te d'Azur, Observatoire de la C\^{o}te d'Azur, CNRS, Laboratoire Lagrange, Bd de l'Observatoire, CS 34229, 06304 Nice cedex 4, France\label{aff1}
\and
Department of Astronomy, University of Florida, Bryant Space Science Center, Gainesville, FL 32611, USA\label{aff2}
\and
Institut f\"ur Theoretische Physik, University of Heidelberg, Philosophenweg 16, 69120 Heidelberg, Germany\label{aff3}
\and
Zentrum f\"ur Astronomie, Universit\"at Heidelberg, Philosophenweg 12, 69120 Heidelberg, Germany\label{aff4}
\and
Universit\'e Paris-Saclay, CEA, D\'epartement de Physique des Particules, 91191, Gif-sur-Yvette, France\label{aff5}
\and
Department of Physics and Astronomy, University of California, Davis, CA 95616, USA\label{aff6}
\and
INAF-Osservatorio Astronomico di Trieste, Via G. B. Tiepolo 11, 34143 Trieste, Italy\label{aff7}
\and
IFPU, Institute for Fundamental Physics of the Universe, via Beirut 2, 34151 Trieste, Italy\label{aff8}
\and
Universit\'e Paris Cit\'e, CNRS, Astroparticule et Cosmologie, 75013 Paris, France\label{aff9}
\and
Univ. Grenoble Alpes, CNRS, Grenoble INP, LPSC-IN2P3, 53, Avenue des Martyrs, 38000, Grenoble, France\label{aff10}
\and
Universit\"at Bonn, Argelander-Institut f\"ur Astronomie, Auf dem H\"ugel 71, 53121 Bonn, Germany\label{aff11}
\and
INAF-Osservatorio di Astrofisica e Scienza dello Spazio di Bologna, Via Piero Gobetti 93/3, 40129 Bologna, Italy\label{aff12}
\and
INFN-Sezione di Bologna, Viale Berti Pichat 6/2, 40127 Bologna, Italy\label{aff13}
\and
Universit\"ats-Sternwarte M\"unchen, Fakult\"at f\"ur Physik, Ludwig-Maximilians-Universit\"at M\"unchen, Scheinerstrasse 1, 81679 M\"unchen, Germany\label{aff14}
\and
INAF-Osservatorio Astronomico di Brera, Via Brera 28, 20122 Milano, Italy\label{aff15}
\and
Dipartimento di Fisica - Sezione di Astronomia, Universit\`a di Trieste, Via Tiepolo 11, 34131 Trieste, Italy\label{aff16}
\and
Institut de Recherche en Astrophysique et Plan\'etologie (IRAP), Universit\'e de Toulouse, CNRS, UPS, CNES, 14 Av. Edouard Belin, 31400 Toulouse, France\label{aff17}
\and
Max Planck Institute for Extraterrestrial Physics, Giessenbachstr. 1, 85748 Garching, Germany\label{aff18}
\and
Universit\'e Paris-Saclay, Universit\'e Paris Cit\'e, CEA, CNRS, AIM, 91191, Gif-sur-Yvette, France\label{aff19}
\and
Department of Physics \& Astronomy, University of Sussex, Brighton BN1 9QH, UK\label{aff20}
\and
School of Physics and Astronomy, University of Nottingham, University Park, Nottingham NG7 2RD, UK\label{aff21}
\and
INAF, Istituto di Radioastronomia, Via Piero Gobetti 101, 40129 Bologna, Italy\label{aff22}
\and
Institut d'Astrophysique de Paris, UMR 7095, CNRS, and Sorbonne Universit\'e, 98 bis boulevard Arago, 75014 Paris, France\label{aff23}
\and
Institut d'Astrophysique de Paris, 98bis Boulevard Arago, 75014, Paris, France\label{aff24}
\and
CNRS-UCB International Research Laboratory, Centre Pierre Binetruy, IRL2007, CPB-IN2P3, Berkeley, USA\label{aff25}
\and
Dipartimento di Fisica e Astronomia "Augusto Righi" - Alma Mater Studiorum Universit\`a di Bologna, via Piero Gobetti 93/2, 40129 Bologna, Italy\label{aff26}
\and
OCA, P.H.C Boulevard de l'Observatoire CS 34229, 06304 Nice Cedex 4, France\label{aff27}
\and
Observatorio Astron\'omico Nacional, IGN, Calle Alfonso XII 3, E-28014 Madrid, Spain\label{aff28}
\and
Universit\'e Paris-Saclay, CNRS, Institut d'astrophysique spatiale, 91405, Orsay, France\label{aff29}
\and
ESAC/ESA, Camino Bajo del Castillo, s/n., Urb. Villafranca del Castillo, 28692 Villanueva de la Ca\~nada, Madrid, Spain\label{aff30}
\and
School of Mathematics and Physics, University of Surrey, Guildford, Surrey, GU2 7XH, UK\label{aff31}
\and
INFN, Sezione di Trieste, Via Valerio 2, 34127 Trieste TS, Italy\label{aff32}
\and
SISSA, International School for Advanced Studies, Via Bonomea 265, 34136 Trieste TS, Italy\label{aff33}
\and
Dipartimento di Fisica e Astronomia, Universit\`a di Bologna, Via Gobetti 93/2, 40129 Bologna, Italy\label{aff34}
\and
INAF-Osservatorio Astronomico di Padova, Via dell'Osservatorio 5, 35122 Padova, Italy\label{aff35}
\and
Space Science Data Center, Italian Space Agency, via del Politecnico snc, 00133 Roma, Italy\label{aff36}
\and
Dipartimento di Fisica, Universit\`a di Genova, Via Dodecaneso 33, 16146, Genova, Italy\label{aff37}
\and
INFN-Sezione di Genova, Via Dodecaneso 33, 16146, Genova, Italy\label{aff38}
\and
Department of Physics "E. Pancini", University Federico II, Via Cinthia 6, 80126, Napoli, Italy\label{aff39}
\and
INAF-Osservatorio Astronomico di Capodimonte, Via Moiariello 16, 80131 Napoli, Italy\label{aff40}
\and
Instituto de Astrof\'isica e Ci\^encias do Espa\c{c}o, Universidade do Porto, CAUP, Rua das Estrelas, PT4150-762 Porto, Portugal\label{aff41}
\and
Faculdade de Ci\^encias da Universidade do Porto, Rua do Campo de Alegre, 4150-007 Porto, Portugal\label{aff42}
\and
Dipartimento di Fisica, Universit\`a degli Studi di Torino, Via P. Giuria 1, 10125 Torino, Italy\label{aff43}
\and
INFN-Sezione di Torino, Via P. Giuria 1, 10125 Torino, Italy\label{aff44}
\and
INAF-Osservatorio Astrofisico di Torino, Via Osservatorio 20, 10025 Pino Torinese (TO), Italy\label{aff45}
\and
INAF-IASF Milano, Via Alfonso Corti 12, 20133 Milano, Italy\label{aff46}
\and
Centro de Investigaciones Energ\'eticas, Medioambientales y Tecnol\'ogicas (CIEMAT), Avenida Complutense 40, 28040 Madrid, Spain\label{aff47}
\and
Port d'Informaci\'{o} Cient\'{i}fica, Campus UAB, C. Albareda s/n, 08193 Bellaterra (Barcelona), Spain\label{aff48}
\and
Institute for Theoretical Particle Physics and Cosmology (TTK), RWTH Aachen University, 52056 Aachen, Germany\label{aff49}
\and
INAF-Osservatorio Astronomico di Roma, Via Frascati 33, 00078 Monteporzio Catone, Italy\label{aff50}
\and
INFN section of Naples, Via Cinthia 6, 80126, Napoli, Italy\label{aff51}
\and
Institute for Astronomy, University of Hawaii, 2680 Woodlawn Drive, Honolulu, HI 96822, USA\label{aff52}
\and
Dipartimento di Fisica e Astronomia "Augusto Righi" - Alma Mater Studiorum Universit\`a di Bologna, Viale Berti Pichat 6/2, 40127 Bologna, Italy\label{aff53}
\and
Instituto de Astrof\'{\i}sica de Canarias, V\'{\i}a L\'actea, 38205 La Laguna, Tenerife, Spain\label{aff54}
\and
Institute for Astronomy, University of Edinburgh, Royal Observatory, Blackford Hill, Edinburgh EH9 3HJ, UK\label{aff55}
\and
Jodrell Bank Centre for Astrophysics, Department of Physics and Astronomy, University of Manchester, Oxford Road, Manchester M13 9PL, UK\label{aff56}
\and
European Space Agency/ESRIN, Largo Galileo Galilei 1, 00044 Frascati, Roma, Italy\label{aff57}
\and
Universit\'e Claude Bernard Lyon 1, CNRS/IN2P3, IP2I Lyon, UMR 5822, Villeurbanne, F-69100, France\label{aff58}
\and
Institut de Ci\`{e}ncies del Cosmos (ICCUB), Universitat de Barcelona (IEEC-UB), Mart\'{i} i Franqu\`{e}s 1, 08028 Barcelona, Spain\label{aff59}
\and
Instituci\'o Catalana de Recerca i Estudis Avan\c{c}ats (ICREA), Passeig de Llu\'{\i}s Companys 23, 08010 Barcelona, Spain\label{aff60}
\and
UCB Lyon 1, CNRS/IN2P3, IUF, IP2I Lyon, 4 rue Enrico Fermi, 69622 Villeurbanne, France\label{aff61}
\and
Mullard Space Science Laboratory, University College London, Holmbury St Mary, Dorking, Surrey RH5 6NT, UK\label{aff62}
\and
Departamento de F\'isica, Faculdade de Ci\^encias, Universidade de Lisboa, Edif\'icio C8, Campo Grande, PT1749-016 Lisboa, Portugal\label{aff63}
\and
Instituto de Astrof\'isica e Ci\^encias do Espa\c{c}o, Faculdade de Ci\^encias, Universidade de Lisboa, Campo Grande, 1749-016 Lisboa, Portugal\label{aff64}
\and
Department of Astronomy, University of Geneva, ch. d'Ecogia 16, 1290 Versoix, Switzerland\label{aff65}
\and
INAF-Istituto di Astrofisica e Planetologia Spaziali, via del Fosso del Cavaliere, 100, 00100 Roma, Italy\label{aff66}
\and
INFN-Padova, Via Marzolo 8, 35131 Padova, Italy\label{aff67}
\and
Aix-Marseille Universit\'e, CNRS/IN2P3, CPPM, Marseille, France\label{aff68}
\and
INFN-Bologna, Via Irnerio 46, 40126 Bologna, Italy\label{aff69}
\and
Institut d'Estudis Espacials de Catalunya (IEEC),  Edifici RDIT, Campus UPC, 08860 Castelldefels, Barcelona, Spain\label{aff70}
\and
Institute of Space Sciences (ICE, CSIC), Campus UAB, Carrer de Can Magrans, s/n, 08193 Barcelona, Spain\label{aff71}
\and
School of Physics, HH Wills Physics Laboratory, University of Bristol, Tyndall Avenue, Bristol, BS8 1TL, UK\label{aff72}
\and
FRACTAL S.L.N.E., calle Tulip\'an 2, Portal 13 1A, 28231, Las Rozas de Madrid, Spain\label{aff73}
\and
Dipartimento di Fisica "Aldo Pontremoli", Universit\`a degli Studi di Milano, Via Celoria 16, 20133 Milano, Italy\label{aff74}
\and
INFN-Sezione di Milano, Via Celoria 16, 20133 Milano, Italy\label{aff75}
\and
NRC Herzberg, 5071 West Saanich Rd, Victoria, BC V9E 2E7, Canada\label{aff76}
\and
Institute of Theoretical Astrophysics, University of Oslo, P.O. Box 1029 Blindern, 0315 Oslo, Norway\label{aff77}
\and
Leiden Observatory, Leiden University, Einsteinweg 55, 2333 CC Leiden, The Netherlands\label{aff78}
\and
Jet Propulsion Laboratory, California Institute of Technology, 4800 Oak Grove Drive, Pasadena, CA, 91109, USA\label{aff79}
\and
Felix Hormuth Engineering, Goethestr. 17, 69181 Leimen, Germany\label{aff80}
\and
Technical University of Denmark, Elektrovej 327, 2800 Kgs. Lyngby, Denmark\label{aff81}
\and
Cosmic Dawn Center (DAWN), Denmark\label{aff82}
\and
Universit\'e Paris-Saclay, CNRS/IN2P3, IJCLab, 91405 Orsay, France\label{aff83}
\and
Max-Planck-Institut f\"ur Astronomie, K\"onigstuhl 17, 69117 Heidelberg, Germany\label{aff84}
\and
NASA Goddard Space Flight Center, Greenbelt, MD 20771, USA\label{aff85}
\and
Department of Physics and Astronomy, University College London, Gower Street, London WC1E 6BT, UK\label{aff86}
\and
Department of Physics and Helsinki Institute of Physics, Gustaf H\"allstr\"omin katu 2, 00014 University of Helsinki, Finland\label{aff87}
\and
Universit\'e de Gen\`eve, D\'epartement de Physique Th\'eorique and Centre for Astroparticle Physics, 24 quai Ernest-Ansermet, CH-1211 Gen\`eve 4, Switzerland\label{aff88}
\and
Department of Physics, P.O. Box 64, 00014 University of Helsinki, Finland\label{aff89}
\and
Helsinki Institute of Physics, Gustaf H{\"a}llstr{\"o}min katu 2, University of Helsinki, Helsinki, Finland\label{aff90}
\and
Centre de Calcul de l'IN2P3/CNRS, 21 avenue Pierre de Coubertin 69627 Villeurbanne Cedex, France\label{aff91}
\and
Laboratoire d'etude de l'Univers et des phenomenes eXtremes, Observatoire de Paris, Universit\'e PSL, Sorbonne Universit\'e, CNRS, 92190 Meudon, France\label{aff92}
\and
Aix-Marseille Universit\'e, CNRS, CNES, LAM, Marseille, France\label{aff93}
\and
SKA Observatory, Jodrell Bank, Lower Withington, Macclesfield, Cheshire SK11 9FT, UK\label{aff94}
\and
University of Applied Sciences and Arts of Northwestern Switzerland, School of Computer Science, 5210 Windisch, Switzerland\label{aff95}
\and
INFN-Sezione di Roma, Piazzale Aldo Moro, 2 - c/o Dipartimento di Fisica, Edificio G. Marconi, 00185 Roma, Italy\label{aff96}
\and
Department of Physics, Institute for Computational Cosmology, Durham University, South Road, Durham, DH1 3LE, UK\label{aff97}
\and
University of Applied Sciences and Arts of Northwestern Switzerland, School of Engineering, 5210 Windisch, Switzerland\label{aff98}
\and
Institute of Physics, Laboratory of Astrophysics, Ecole Polytechnique F\'ed\'erale de Lausanne (EPFL), Observatoire de Sauverny, 1290 Versoix, Switzerland\label{aff99}
\and
Aurora Technology for European Space Agency (ESA), Camino bajo del Castillo, s/n, Urbanizacion Villafranca del Castillo, Villanueva de la Ca\~nada, 28692 Madrid, Spain\label{aff100}
\and
Institut de F\'{i}sica d'Altes Energies (IFAE), The Barcelona Institute of Science and Technology, Campus UAB, 08193 Bellaterra (Barcelona), Spain\label{aff101}
\and
European Space Agency/ESTEC, Keplerlaan 1, 2201 AZ Noordwijk, The Netherlands\label{aff102}
\and
School of Mathematics, Statistics and Physics, Newcastle University, Herschel Building, Newcastle-upon-Tyne, NE1 7RU, UK\label{aff103}
\and
DARK, Niels Bohr Institute, University of Copenhagen, Jagtvej 155, 2200 Copenhagen, Denmark\label{aff104}
\and
Waterloo Centre for Astrophysics, University of Waterloo, Waterloo, Ontario N2L 3G1, Canada\label{aff105}
\and
Department of Physics and Astronomy, University of Waterloo, Waterloo, Ontario N2L 3G1, Canada\label{aff106}
\and
Perimeter Institute for Theoretical Physics, Waterloo, Ontario N2L 2Y5, Canada\label{aff107}
\and
Centre National d'Etudes Spatiales -- Centre spatial de Toulouse, 18 avenue Edouard Belin, 31401 Toulouse Cedex 9, France\label{aff108}
\and
Institute of Space Science, Str. Atomistilor, nr. 409 M\u{a}gurele, Ilfov, 077125, Romania\label{aff109}
\and
Consejo Superior de Investigaciones Cientificas, Calle Serrano 117, 28006 Madrid, Spain\label{aff110}
\and
Universidad de La Laguna, Departamento de Astrof\'{\i}sica, 38206 La Laguna, Tenerife, Spain\label{aff111}
\and
Dipartimento di Fisica e Astronomia "G. Galilei", Universit\`a di Padova, Via Marzolo 8, 35131 Padova, Italy\label{aff112}
\and
Universit\'e St Joseph; Faculty of Sciences, Beirut, Lebanon\label{aff113}
\and
Departamento de F\'isica, FCFM, Universidad de Chile, Blanco Encalada 2008, Santiago, Chile\label{aff114}
\and
Universit\"at Innsbruck, Institut f\"ur Astro- und Teilchenphysik, Technikerstr. 25/8, 6020 Innsbruck, Austria\label{aff115}
\and
Satlantis, University Science Park, Sede Bld 48940, Leioa-Bilbao, Spain\label{aff116}
\and
Centre for Electronic Imaging, Open University, Walton Hall, Milton Keynes, MK7~6AA, UK\label{aff117}
\and
Instituto de Astrof\'isica e Ci\^encias do Espa\c{c}o, Faculdade de Ci\^encias, Universidade de Lisboa, Tapada da Ajuda, 1349-018 Lisboa, Portugal\label{aff118}
\and
Cosmic Dawn Center (DAWN)\label{aff119}
\and
Niels Bohr Institute, University of Copenhagen, Jagtvej 128, 2200 Copenhagen, Denmark\label{aff120}
\and
Universidad Polit\'ecnica de Cartagena, Departamento de Electr\'onica y Tecnolog\'ia de Computadoras,  Plaza del Hospital 1, 30202 Cartagena, Spain\label{aff121}
\and
Kapteyn Astronomical Institute, University of Groningen, PO Box 800, 9700 AV Groningen, The Netherlands\label{aff122}
\and
Infrared Processing and Analysis Center, California Institute of Technology, Pasadena, CA 91125, USA\label{aff123}
\and
Dipartimento di Fisica e Scienze della Terra, Universit\`a degli Studi di Ferrara, Via Giuseppe Saragat 1, 44122 Ferrara, Italy\label{aff124}
\and
Istituto Nazionale di Fisica Nucleare, Sezione di Ferrara, Via Giuseppe Saragat 1, 44122 Ferrara, Italy\label{aff125}
\and
School of Physics and Astronomy, Cardiff University, The Parade, Cardiff, CF24 3AA, UK\label{aff126}
\and
Department of Physics, Oxford University, Keble Road, Oxford OX1 3RH, UK\label{aff127}
\and
INAF - Osservatorio Astronomico di Brera, via Emilio Bianchi 46, 23807 Merate, Italy\label{aff128}
\and
INAF-Osservatorio Astronomico di Brera, Via Brera 28, 20122 Milano, Italy, and INFN-Sezione di Genova, Via Dodecaneso 33, 16146, Genova, Italy\label{aff129}
\and
ICL, Junia, Universit\'e Catholique de Lille, LITL, 59000 Lille, France\label{aff130}
\and
ICSC - Centro Nazionale di Ricerca in High Performance Computing, Big Data e Quantum Computing, Via Magnanelli 2, Bologna, Italy\label{aff131}
\and
Instituto de F\'isica Te\'orica UAM-CSIC, Campus de Cantoblanco, 28049 Madrid, Spain\label{aff132}
\and
CERCA/ISO, Department of Physics, Case Western Reserve University, 10900 Euclid Avenue, Cleveland, OH 44106, USA\label{aff133}
\and
Technical University of Munich, TUM School of Natural Sciences, Physics Department, James-Franck-Str.~1, 85748 Garching, Germany\label{aff134}
\and
Max-Planck-Institut f\"ur Astrophysik, Karl-Schwarzschild-Str.~1, 85748 Garching, Germany\label{aff135}
\and
Laboratoire Univers et Th\'eorie, Observatoire de Paris, Universit\'e PSL, Universit\'e Paris Cit\'e, CNRS, 92190 Meudon, France\label{aff136}
\and
Departamento de F{\'\i}sica Fundamental. Universidad de Salamanca. Plaza de la Merced s/n. 37008 Salamanca, Spain\label{aff137}
\and
Universit\'e de Strasbourg, CNRS, Observatoire astronomique de Strasbourg, UMR 7550, 67000 Strasbourg, France\label{aff138}
\and
Center for Data-Driven Discovery, Kavli IPMU (WPI), UTIAS, The University of Tokyo, Kashiwa, Chiba 277-8583, Japan\label{aff139}
\and
Ludwig-Maximilians-University, Schellingstrasse 4, 80799 Munich, Germany\label{aff140}
\and
Max-Planck-Institut f\"ur Physik, Boltzmannstr. 8, 85748 Garching, Germany\label{aff141}
\and
California Institute of Technology, 1200 E California Blvd, Pasadena, CA 91125, USA\label{aff142}
\and
Department of Physics \& Astronomy, University of California Irvine, Irvine CA 92697, USA\label{aff143}
\and
Department of Mathematics and Physics E. De Giorgi, University of Salento, Via per Arnesano, CP-I93, 73100, Lecce, Italy\label{aff144}
\and
INFN, Sezione di Lecce, Via per Arnesano, CP-193, 73100, Lecce, Italy\label{aff145}
\and
INAF-Sezione di Lecce, c/o Dipartimento Matematica e Fisica, Via per Arnesano, 73100, Lecce, Italy\label{aff146}
\and
Departamento F\'isica Aplicada, Universidad Polit\'ecnica de Cartagena, Campus Muralla del Mar, 30202 Cartagena, Murcia, Spain\label{aff147}
\and
Instituto de Astrof\'isica de Canarias (IAC); Departamento de Astrof\'isica, Universidad de La Laguna (ULL), 38200, La Laguna, Tenerife, Spain\label{aff148}
\and
Instituto de F\'isica de Cantabria, Edificio Juan Jord\'a, Avenida de los Castros, 39005 Santander, Spain\label{aff149}
\and
Observatorio Nacional, Rua General Jose Cristino, 77-Bairro Imperial de Sao Cristovao, Rio de Janeiro, 20921-400, Brazil\label{aff150}
\and
Institute of Cosmology and Gravitation, University of Portsmouth, Portsmouth PO1 3FX, UK\label{aff151}
\and
Department of Computer Science, Aalto University, PO Box 15400, Espoo, FI-00 076, Finland\label{aff152}
\and
Instituto de Astrof\'\i sica de Canarias, c/ Via Lactea s/n, La Laguna 38200, Spain. Departamento de Astrof\'\i sica de la Universidad de La Laguna, Avda. Francisco Sanchez, La Laguna, 38200, Spain\label{aff153}
\and
Ruhr University Bochum, Faculty of Physics and Astronomy, Astronomical Institute (AIRUB), German Centre for Cosmological Lensing (GCCL), 44780 Bochum, Germany\label{aff154}
\and
Department of Physics and Astronomy, Vesilinnantie 5, 20014 University of Turku, Finland\label{aff155}
\and
Serco for European Space Agency (ESA), Camino bajo del Castillo, s/n, Urbanizacion Villafranca del Castillo, Villanueva de la Ca\~nada, 28692 Madrid, Spain\label{aff156}
\and
ARC Centre of Excellence for Dark Matter Particle Physics, Melbourne, Australia\label{aff157}
\and
Centre for Astrophysics \& Supercomputing, Swinburne University of Technology,  Hawthorn, Victoria 3122, Australia\label{aff158}
\and
Department of Physics and Astronomy, University of the Western Cape, Bellville, Cape Town, 7535, South Africa\label{aff159}
\and
DAMTP, Centre for Mathematical Sciences, Wilberforce Road, Cambridge CB3 0WA, UK\label{aff160}
\and
Kavli Institute for Cosmology Cambridge, Madingley Road, Cambridge, CB3 0HA, UK\label{aff161}
\and
Department of Astrophysics, University of Zurich, Winterthurerstrasse 190, 8057 Zurich, Switzerland\label{aff162}
\and
Department of Physics, Centre for Extragalactic Astronomy, Durham University, South Road, Durham, DH1 3LE, UK\label{aff163}
\and
IRFU, CEA, Universit\'e Paris-Saclay 91191 Gif-sur-Yvette Cedex, France\label{aff164}
\and
Oskar Klein Centre for Cosmoparticle Physics, Department of Physics, Stockholm University, Stockholm, SE-106 91, Sweden\label{aff165}
\and
Astrophysics Group, Blackett Laboratory, Imperial College London, London SW7 2AZ, UK\label{aff166}
\and
INAF-Osservatorio Astrofisico di Arcetri, Largo E. Fermi 5, 50125, Firenze, Italy\label{aff167}
\and
Dipartimento di Fisica, Sapienza Universit\`a di Roma, Piazzale Aldo Moro 2, 00185 Roma, Italy\label{aff168}
\and
Centro de Astrof\'{\i}sica da Universidade do Porto, Rua das Estrelas, 4150-762 Porto, Portugal\label{aff169}
\and
Dipartimento di Fisica, Universit\`a di Roma Tor Vergata, Via della Ricerca Scientifica 1, Roma, Italy\label{aff170}
\and
INFN, Sezione di Roma 2, Via della Ricerca Scientifica 1, Roma, Italy\label{aff171}
\and
HE Space for European Space Agency (ESA), Camino bajo del Castillo, s/n, Urbanizacion Villafranca del Castillo, Villanueva de la Ca\~nada, 28692 Madrid, Spain\label{aff172}
\and
Department of Astrophysical Sciences, Peyton Hall, Princeton University, Princeton, NJ 08544, USA\label{aff173}
\and
Theoretical astrophysics, Department of Physics and Astronomy, Uppsala University, Box 515, 751 20 Uppsala, Sweden\label{aff174}
\and
Mathematical Institute, University of Leiden, Einsteinweg 55, 2333 CA Leiden, The Netherlands\label{aff175}
\and
Institute of Astronomy, University of Cambridge, Madingley Road, Cambridge CB3 0HA, UK\label{aff176}
\and
Space physics and astronomy research unit, University of Oulu, Pentti Kaiteran katu 1, FI-90014 Oulu, Finland\label{aff177}
\and
Center for Computational Astrophysics, Flatiron Institute, 162 5th Avenue, 10010, New York, NY, USA\label{aff178}
\and
Department of Physics and Astronomy, University of British Columbia, Vancouver, BC V6T 1Z1, Canada\label{aff179}}

%
%
\abstract{The first survey data release by the \Euclid mission covers approximately $63\,\mathrm{deg^2}$ in the Euclid Deep Fields to the same depth as the Euclid Wide Survey. This paper showcases, for the first time, the performance of cluster finders on \Euclid data and presents examples of validated clusters in the Quick Release 1 (Q1) imaging data. We identify clusters using two algorithms (\AMICO and \PZWav) implemented in the \Euclid cluster-detection pipeline. We explore the internal consistency of detections from the two codes, and cross-match detections with known clusters from other surveys using external multi-wavelength and spectroscopic data sets. This enables assessment of the \Euclid photometric redshift accuracy and also of systematics such as mis-centring between the optical cluster centre and centres based on  X-ray and/or Sunyaev--Zeldovich observations. We report 426 joint \PZWav and \AMICO-detected clusters with high signal-to-noise ratios over the full Q1 area in the redshift range $0.2 \leq z \leq 1.5$. The chosen redshift and signal-to-noise thresholds are motivated by the photometric quality of the early \Euclid data. We provide richness estimates for each of the \Euclid-detected clusters and show its correlation with various external cluster mass proxies. Due to the limited area and evolving data quality, the sample is not intended to serve as a reference for cosmological applications, but to verify and validate the cluster workflow. Out of the full sample, 77 systems are potentially new to the literature. Overall, the Q1 cluster catalogue demonstrates a successful validation of the workflow ahead of the Euclid Data Release 1, based on the consistency of internal and external properties of \Euclid-detected clusters.}

%
%
%
   \titlerunning{\Euclid\/: First detections from the galaxy cluster workflow}
   \authorrunning{Euclid Collaboration:
S. Bhargava et al.}
   
   \maketitle


\section{Introduction}
The European Space Agency (ESA) \Euclid mission is designed to map the geometry of the Universe in unprecedented detail using optical and near-infrared wide-field imaging and spectroscopy; the completed survey will cover \num{14000}\,deg${^2}$ of the extragalactic sky, observing over 1.5 billion galaxies out to a redshift $z\approx 2$ \citep{EuclidSkyOverview}. Furthermore, \Euclid is expected to detect and characterise approximately $10^{5}$ galaxy clusters and groups \citep{Sartoris2016,Adam-EP3}, driving a tremendous leap in cosmological constraints from joint cluster abundance, clustering, and weak lensing mass estimates \cite[e.g.,][]{Maturi2019, lesci20,DESCosmo2020, Costanzi2020, Romanello2023}. 

This paper presents a validation of the \Euclid galaxy cluster workflow on the \gls{q1} data set \citep{Q1cite}, constituting the first survey data released by the mission \citep{Q1-TP001}. The data consists of three non-contiguous fields that cover approximately 63\,deg$^2$ -- one in the northern and two in the southern hemisphere. We detect clusters and characterise candidates with high \gls{snr} across the three fields, favouring a conservative approach during this early phase of \Euclid data.

Numerous simulation-based evaluations have been conducted on the accuracy of recovering \Euclid cluster masses \citep[e.g.][]{EP-Ingoglia,Giocoli-EP30}, as well as multi-wavelength cluster observables \citep{EP-Ragagnin}. Moreover, extensive tests of the cluster workflow have been performed on the Flagship simulation \citep{EuclidSkyFlagship} to better understand the \Euclid selection function for clusters ({\color{blue}Cabanac et al., in preparation}). This paper explores the impact of galaxy selection and quality cuts on the cluster workflow's ability to robustly detect and characterise galaxy clusters on real \Euclid data for the first time. 

We examine the distribution in \gls{snr}, optical richness and redshift for our cluster detections. In addition, cluster redshift estimates are compared to external photometric and spectroscopic data. We utilised the extensive multi-wavelength coverage at near-infrared, optical, millimetre, and X-ray wavelengths of the Q1 fields in the literature to identify counterparts of the \Euclid detections. We evaluate centring offset distributions and study the properties of missed objects. 

Despite the limited area of the Q1 data release, \Euclid's high galaxy density allows us to perform a detailed characterisation of the first clusters detected by the \Euclid mission. This provides a secure training ground for the \gls{dr1}, which will cover approximately \num{1900}\,deg$^2$ and form the basis of the first cosmological analysis.

The paper outline is as follows. In \cref{sec:input} we describe the input galaxy selection performed on the \Euclid source catalogues and associated masks. In \cref{sec:methodology} we describe the two cluster detection algorithms implemented in the mission \gls{sgs}, and we detail the construction of the combined cluster catalogue. \cref{sec:external} presents cross-matches to external cluster catalogues. In \cref{sec:specz} we describe spectroscopic validation of the Q1 cluster catalogue using ground-based spectroscopic surveys. We explore the scientific return of this first validation of \Euclid clusters and the outlook for the future in \cref{sec:conclusions}. Unless otherwise stated, we adopt a flat $\Lambda$CDM model following \cite{Planck2018}, with $H_0=67.4\,\si{\kmsMpc}$ and $\Omega_{\rm m}=0.315$.

\section{Input data}\label{sec:input}

\subsection{The three Euclid Q1 patches} 
The \gls{q1} release consists of three non-contiguous fields that, upon completion of the survey, will constitute the \gls{eds}, as described in \citet{EuclidSkyOverview}. Although the purpose of these fields is to gain approximately two magnitudes in depth compared to the \gls{ews}, the Q1 fields have been imaged to the nominal \gls{ews} depth for this first data release.

The Q1 survey area comprises the \gls{edfn}, the \gls{edfs}, and the \gls{edff}, covering a total area of 63\,deg{$^2$}. Based on the galaxy catalogues used in this analysis, the total area for each field is reported in \cref{fig:counts}. The \gls{edfn} is an approximately 23\,deg{$^2$} circular field located at the northern ecliptic pole. The \gls{edff} is a 12\,deg{$^2$} circular field including the \gls{cdfs}, which has numerous ground- and space-based ancillary observations (as described in \cref{sec:external}). Lastly, the \gls{edfs} is a 28\,deg$^2$ field with an extended shape that encompasses two adjacent deep-drilling fields of the Vera C. Rubin \gls{lsst}. The exact spatial distribution of the fields is illustrated in Figure 1 of the Q1 overview paper, \cite{Q1-TP001}.

The effective area, average photometric redshift scatter and bias corresponding to these fields are computed based on the selection criteria used for cluster detection (see \cref{sec:galsel}) and the external photometric bands available for each of the three fields. The latter are: The Ultraviolet Near Infrared Optical Northern Survey\footnote{\url{https://www.skysurvey.cc}} \citep[UNIONS,][]{UNIONS2025}, \gls{decam}\footnote{\url{https://noirlab.edu/public/programs/ctio/victor-blanco-4m-telescope/decam/}}, described in \cite{DES2021}, and \gls{hsc}\footnote{\url{https://subarutelescope.org/en/subaru2/instrument/hsc/}}, described in \cite{HSC2022}.

\subsection{The galaxy selection}\label{sec:galsel}
Galaxy selection in the Q1 fields is primarily based on the information provided by the OU-MER and OU-LE3 \gls{ou} of the \Euclid \gls{sgs}. We refer the reader to \citet{Q1-TP001} for a detailed description the input products. Detailed descriptions of the VIS and NISP instruments can be found in \cite{EuclidSkyVIS} and \cite{EuclidSkyNISP}, respectively. The sources used in our analysis satisfy the following criteria.
\begin{itemize}
    \item They are detected in the VIS image (\texttt{VIS\_DET $= 1$}). This choice is motivated by the presence of persistence in the NISP detector that is not yet fully solved. This effect currently prevents us from using sources detected in the NISP images only, and therefore limits our cluster search to redshifts lower than the nominal redshift $z=2$. This restriction will be lifted in future releases.   
    \item They are covered by images in all bands used in the considered sky patch (\texttt{FLUX\_$*$\_2FWHM\_APER $\neq$ NaN}).
    \item They are not saturated (\texttt{Flag\_$* \neq 3$}).
    \item They are not classified as stars at the 99\% confidence level (\texttt{POINT\_LIKE\_PROB  $\leq 0.95$}).
    \item They are not within stellar polygons designed to mask regions around bright stars and their diffraction spikes in the VIS and stacked NIR images (\texttt{DET\_QUALITY\_FLAG[bits 7 and 8] $= 0$}).
    \item They are not located in \texttt{HEALPix}\footnote{\url{http://healpix.sourceforge.net}} \citep{healpix,Zonca2019} pixels of the inter-band (Euclid VIS+NIR and external bands) COVERAGE map where the pixel coverage is below 80\%.
    \item They are limited to an \HE magnitude of 24\footnote{corresponding to ${\tt FLUX\_H\_UNIF} > 0.912$ and obtained from $\HE = -2.5 \log_{10} {\tt FLUX\_H\_UNIF}{\mathrm[\mu Jy]} -23.9$ } to ensure galaxy completeness and homogeneous data quality. At these magnitudes the average signal to noise expected for \HE-band photometry is $\sim 5$ \citep{Q1-SP031}.

\end{itemize}

A more detailed description of the individual flags used in the galaxy selection can be found in \cite{Q1-TP004}. In addition to these criteria, three sets of additional masks are applied: larger circular masks around bright stars to reject areas that are spoiled by image artefacts in the ground-based images; elliptical masks to cover bright galaxies in the foreground; and masks around extended stellar systems. The last two sets of masks are derived from the masking performed by the DR10 Legacy Survey.\footnote{\url{https://www.legacysurvey.org/dr10/external}}

\begin{figure}
    \centering
    \includegraphics[width=0.95\linewidth]{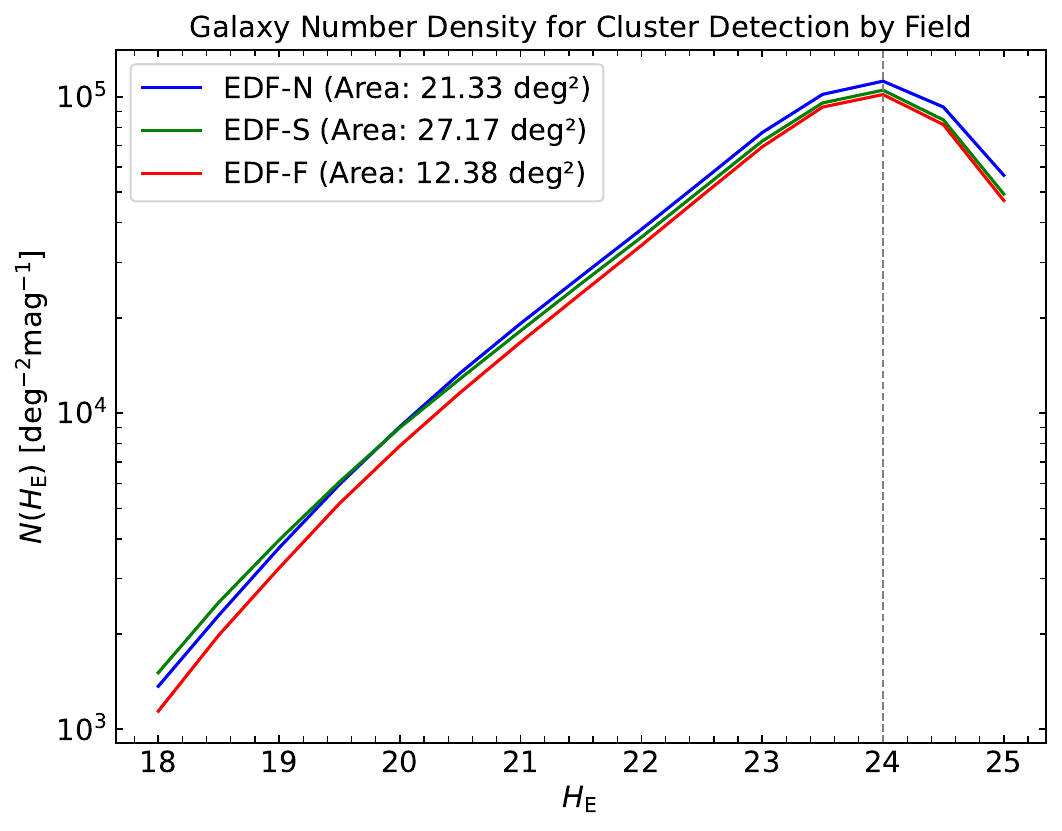}
    \caption{Galaxy number densities for the three \Euclid Q1 patches after applying the filtering described in \cref{sec:galsel}. The number densities are binned as a function of magnitude in the \Euclid NIR-\HE band, as well as by the effective area. The dashed vertical line indicates a magnitude cut of \HE < 24 in the galaxy selection.}
    \label{fig:counts}
\end{figure}

The normalised galaxy number counts as a function of \HE magnitude for the three Q1 patches are presented in \cref{fig:counts}. The galaxy selection leads to consistent number counts that reach a maximum at a \HE magnitude of 24, justifying our choice in limiting magnitude.

\subsection{Photometric redshifts}
\label{subsec:photozs}
High-quality \gls{photz} are an essential ingredient for both cluster detection and richness estimation. The \gls{photz} are derived with the template-fitting code {\tt Phosphoros}, whose application to the \Euclid data is detailed in \cite{Q1-TP005}. Here, we evaluate the \gls{photz} quality for galaxies used in our cluster studies. In particular, we assess the consistency of the computed \gls{pdz} by comparison to a set of \num{26964} publicly available galaxy spectroscopic redshifts in the \gls{edff} and \num{38950} in the \gls{edfn}. The number of available redshifts for the \gls{edfs} is considerably lower (of order 100s). Details of the external spectroscopic data set matched to Q1 can be found in section~5.2 and table~5 of \citet{Q1-TP005}.

\begin{figure*}
    \centering
    \includegraphics[width=0.45\linewidth]{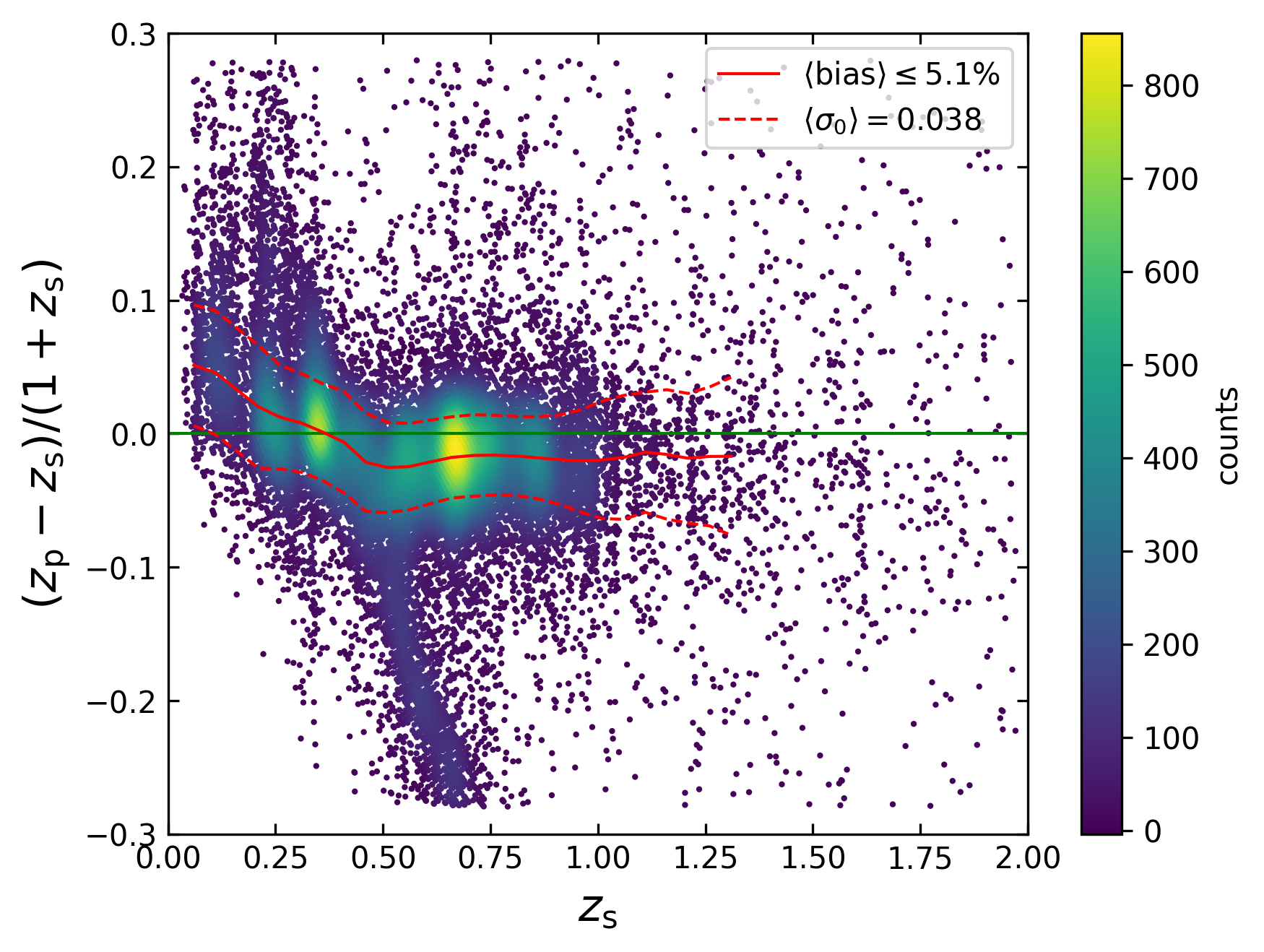}
     \includegraphics[width=0.45\linewidth]{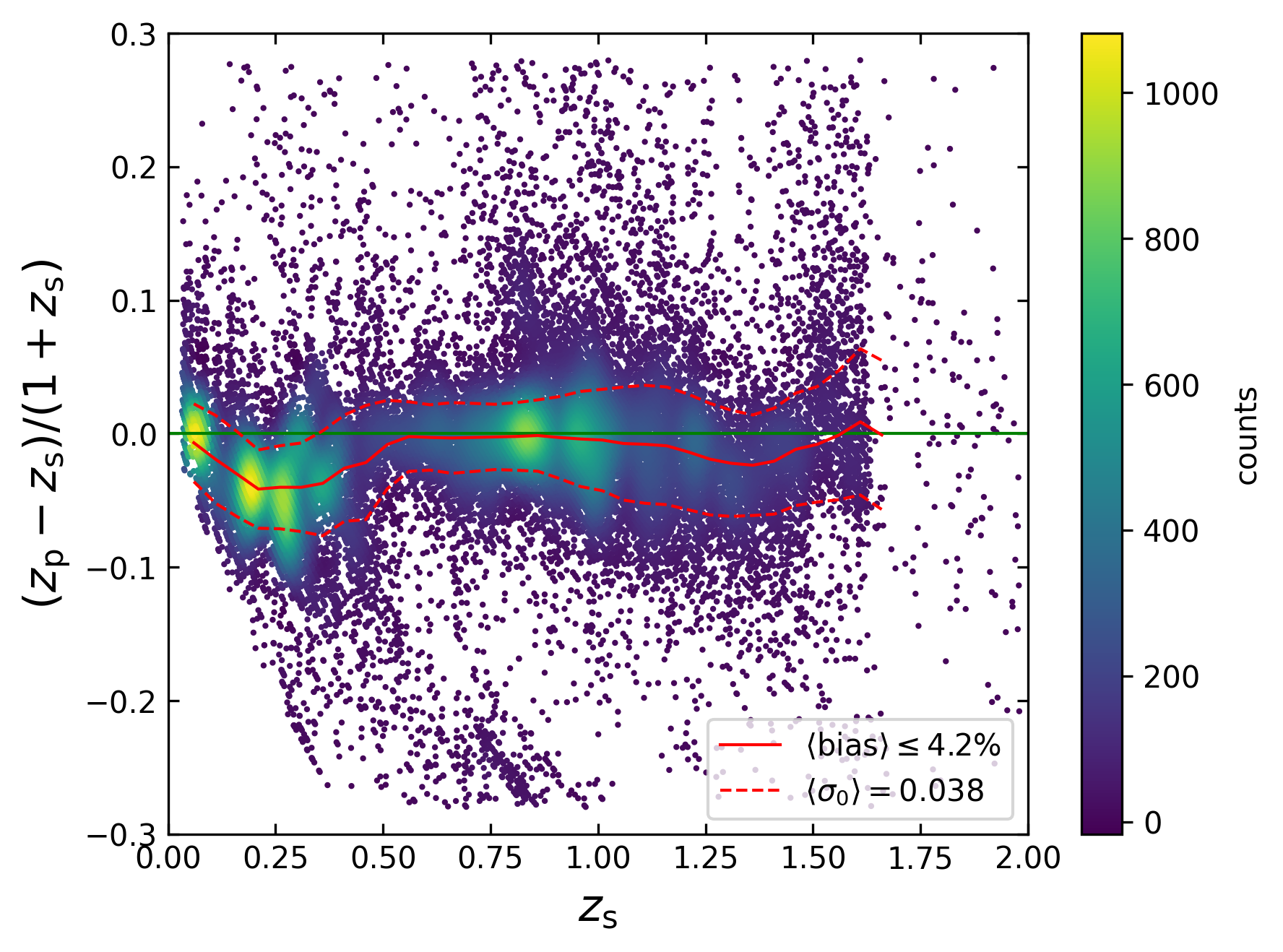}
   \caption{Residuals of the median photometric redshift relative to the spectroscopic redshift for $\num{25416}$ galaxies in the \gls{edff} field (left) and $\num{36564}$ galaxies in the \gls{edfn} (right). The spectroscopic information comes from a compilation of publicly available external spectroscopic surveys, starting from a redshift of $z_{\rm s} = 0.05$ in the two figures. The red lines indicate the associated bias and scatter, estimated in redshift bins containing at least 200 galaxies.}
    \label{fig:zpzs_stats}
\end{figure*}

In \cref{fig:zpzs_stats}, we compare the median values of the \gls{pdz} to the associated spectroscopic redshift. Statistical errors are computed in the redshift range from 0.05 to 1.5 where most of the spectroscopic data is concentrated. Defining the normalised offset $X=(\zp-\zs)/(1+\zs)$, where $\zp$ is the \gls{pdz} median value and $\zs$ is the spectroscopic value, a relatively stable mean scatter of $\sigma_0 \approx 0.04$ and a residual bias of $\pm 2.5\%$ ($\pm 4.1\%$) can be observed for the \gls{edff} (\gls{edfn}). We do not make this comparison for the \gls{edfs} due to the aforementioned much lower number of objects with external spectroscopic counterparts; however, we expect a similar behaviour to \gls{edff} due to the same external band coverage. 

\begin{figure}
    \centering
    \includegraphics[width=0.95\linewidth]{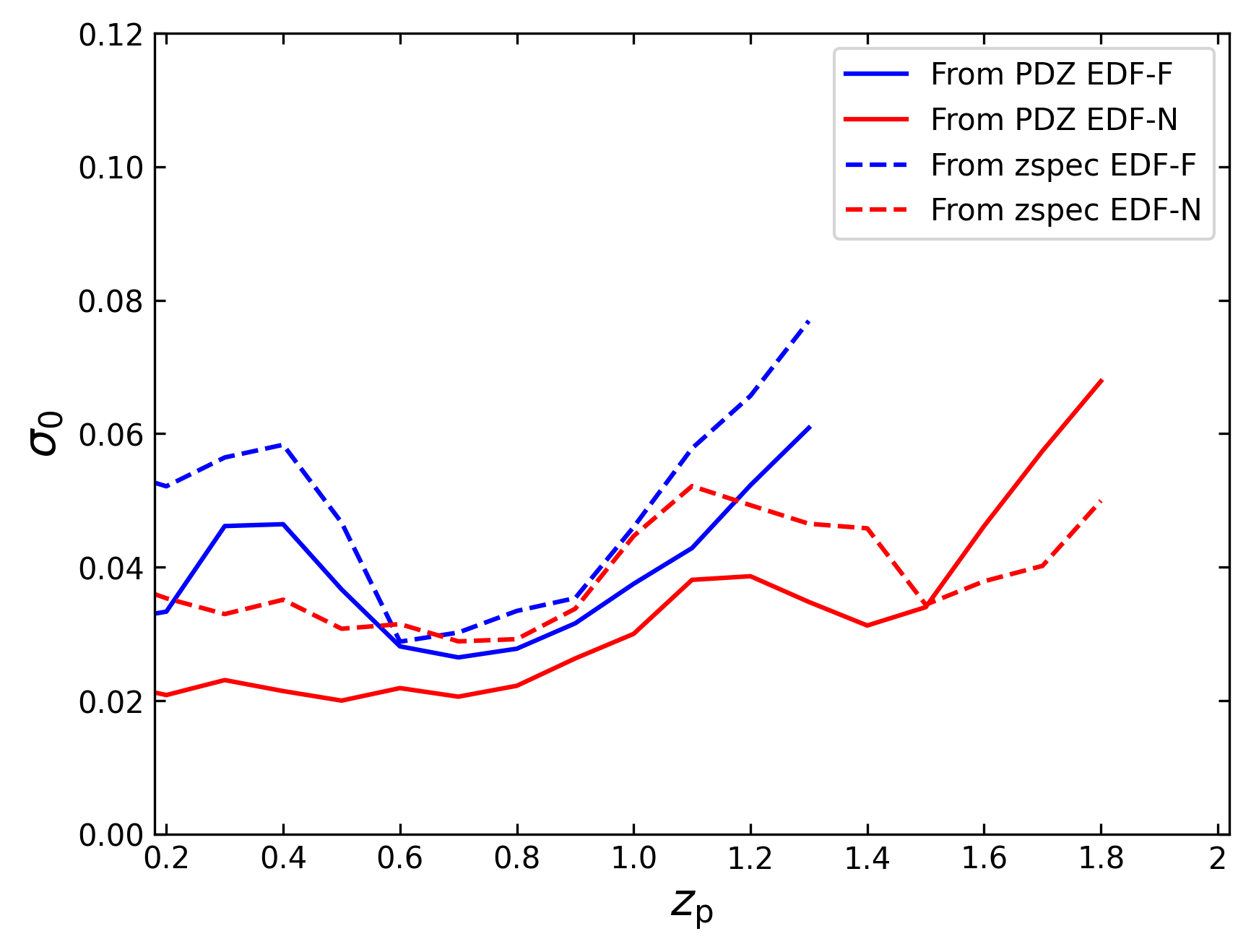}
    \caption{Average scatter of the photometric redshifts in bins of photometric redshift, derived from the comparison to spectroscopic redshifts (dashed lines) or from the mean width of 15th-to-85th percentiles around the median of the individual \gls{pdz} (solid lines) for the same subset of galaxies. The lines stop when fewer than 200 galaxies are available in the considered redshift bin.}
    \label{fig:pdz_zb}
\end{figure}

In this study, \gls{pdz}s are also critical to associate errors to the photometric redshift measurements, which, in turn, are used to weight galaxies in the cluster detection algorithms presented in \cref{sec:detection}. To evaluate how well the shape of the \gls{pdz}s reflect the `true' photometric redshift errors, we compute the average width of the 15th-to-85th percentiles around the median and compare it to the average offset of the \gls{pdz} median value with its associated spectroscopic redshift, where available. This test is performed in photometric redshift bins from $\zp=$ 0.2 to 2 which corresponds to the \Euclid nominal range for cluster searches. The result is shown in \cref{fig:pdz_zb}. On average, the evolution of the \gls{pdz} width with the median photometric redshift is in good agreement with the $\zp, \zs$ scatter. However, the \gls{pdz} widths are systematically narrower than expected across the entire redshift range by about 65\% for the \gls{edfn} and 15\% for the \gls{edff} field. The underestimation of \gls{photz} uncertainties, particularly in extreme cases with an effective $\sigma_z<0.01$, can lead to undesired spurious detections if not addressed adequately. To mitigate this issue, we tested two different approaches in the detection algorithms, discussed in \cref{sec:detection}. 

As explained in \cite{Q1-TP005}, efforts are ongoing to address the need for photometric bias corrections and the treatment of underestimation of photo-z errors. For a more detailed discussion of the photometric validation, the reader is referred to Sect. 5.2 in the aforementioned paper.

\section{Methodology}\label{sec:methodology}

\subsection{Cluster detection}\label{sec:detection}
The two cluster detection algorithms used in the \Euclid galaxy cluster workflow, \AMICO (see \cref{subsec:amico}) and \PZWav (see \cref{subsec:pzwav}), were selected through a series of cluster-finding challenges \citep{Adam-EP3}. These two algorithms employ complementary approaches to cluster detection. \AMICO is a linear optimal matched-filter code that uses prior knowledge of the properties of the cluster galaxy population for cluster identification; \PZWav favours a more model-independent approach, making minimal assumptions about cluster properties by using only positions and photometric redshift data. 

Within the Q1 framework, the algorithms analyse sky regions known as cluster tiles, consisting of an ensemble of sky tile catalogues and maskes produced by the \gls{sgs} \gls{ou}-MER pipeline. The MER tiles themselves are defined geometrically and have a size of approximately $\ang{0.25;;}\times\ang{0.25;;}$ \citep{Q1-TP004}. Each MER tile has an associated source catalogue and coverage mask that incorporates the features detailed in \cref{sec:galsel}. After running on a cluster tile (one for each of the three fields), the two cluster algorithms produce individual catalogues with a list containing the angular positions, redshift, and \gls{snr} of all detected sources.

\subsubsection{\AMICO}\label{subsec:amico}
We briefly describe the main features of \AMICO (Adaptive Matched Identifier of Clustered Objects), with more details available in \citet{Bellagamba2018} and \cite{Maturi2019}. \AMICO works with catalogues of galaxy angular positions, magnitudes, and full photometric redshift probabilities, avoiding any explicit selection based on the cluster red sequence. The data are modelled as the sum of the desired cluster contribution and a noise component representing contamination from field galaxies. The cluster filter model includes a Schechter luminosity function \citep{Schechter76} and a projected NFW \citep*{Navarro+96} radial profile, while the field galaxy distribution at a certain redshift is approximated from the full galaxy sample, weighted by their redshift probabilities. The filter is optimal in the sense that it minimises variance to estimate the signal amplitude, $A$, calculated on a three-dimensional grid of positions and redshifts (with a chosen resolution of 0.01 in both angular position, in degrees, and redshift). Clusters are identified as the peaks with the highest \snr, followed by iterative `cleaning' to remove the contributions of detected clusters, thus reducing blending for refining subsequent detections. This process continues to a minimum \snr, directly linked to the amplitude. For this data set, a regularisation term has been introduced to address galaxies whose photometric redshifts exhibit excessively small uncertainties, indicating potential issues with their redshift estimates. Such galaxies can produce an excess of signal in the \AMICO amplitude maps, resulting in potential spurious detections. To avoid this issue, the \gls{pdz} of galaxies with photometric redshift uncertainties smaller than $\sigma_z=0.01(1+z_\mathrm{med})$, is modelled as a Gaussian with $\sigma=\sigma_z$ and mode located at $z_\mathrm{med}$. Here, $z_\mathrm{med}$ is the median value of the input \gls{pdz}. We observed an excess in low redshift detections (below $z \leq 0.6$) prior to the regularisation in \AMICO, subsequently obtaining comparable numbers to \PZWav. Going forwards for DR1, we plan to implement the same criteria, but with a sigma value modelled as a polynomial function fitted to the data---produced from a new release of the photometric redshift algorithm---rather than a scalar value. A similar procedure will be applied to \PZWav.

\subsubsection{\PZWav}\label{subsec:pzwav}
The \PZWav detection algorithm applies a difference-of-Gaussians smoothing kernel that is designed to detect over-densities on cluster scales, while minimising the impact of larger scale structures on the detection. The code takes as input the positions and full photometric redshift probability distribution functions, \gls{pdz}, for each galaxy and generates a data cube in position and redshift containing this information. We refer the reader to \citet{Adam-EP3}, \citet{Werner2023}, and \citet{Thongkham2024} for more detail on the \PZWav algorithm. Of particular relevance for the current work, the standard deviations for inner and outer Gaussians in the kernel are set at 300\,kpc and 1.2\,Mpc. This choice of scales has been found to effectively remove large-scale structure contributions and isolate clusters in both previous surveys \citep{Thongkham2024} and tests on the Flagship simulations ({\color{blue}Cabanac et al. in preparation}). Purity and completeness evaluations of the \PZWav algorithm are also performed in these two papers. Moreover, previous cluster detection with \PZWav has found no significant sensitivity to the dynamical state of the cluster. The chosen redshift bin size in the search is set to 0.06. The wider redshift binning compared to \AMICO, by construction, regularises the contribution of galaxies with an excessively narrow \gls{pdz}, which would otherwise degrade the results.

\begin{figure}
    \centering
    \includegraphics[width=0.8\linewidth]{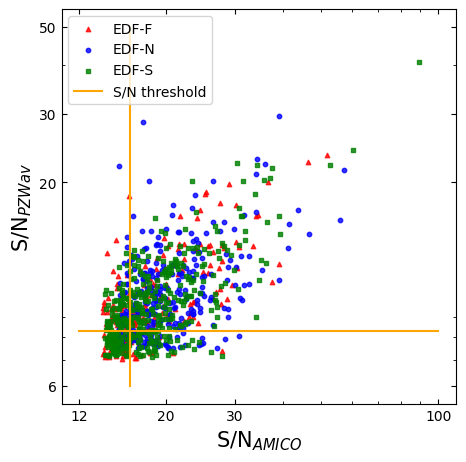}
    \caption{Comparison of the \gls{snr} of the matched \AMICO and \PZWav detections. The yellow lines indicate the adopted \gls{snr} thresholds that ensure 15 detections per square degree by both algorithms in the redshift range $0.2 \leq z \leq 1.5$. The precise \gls{snr} thresholds are stated in \cref{subsec:combinedcat}.}
    \label{fig:snr_snr}
\end{figure}

\begin{figure*}
    \centering
    \includegraphics[width=0.95\linewidth]
    {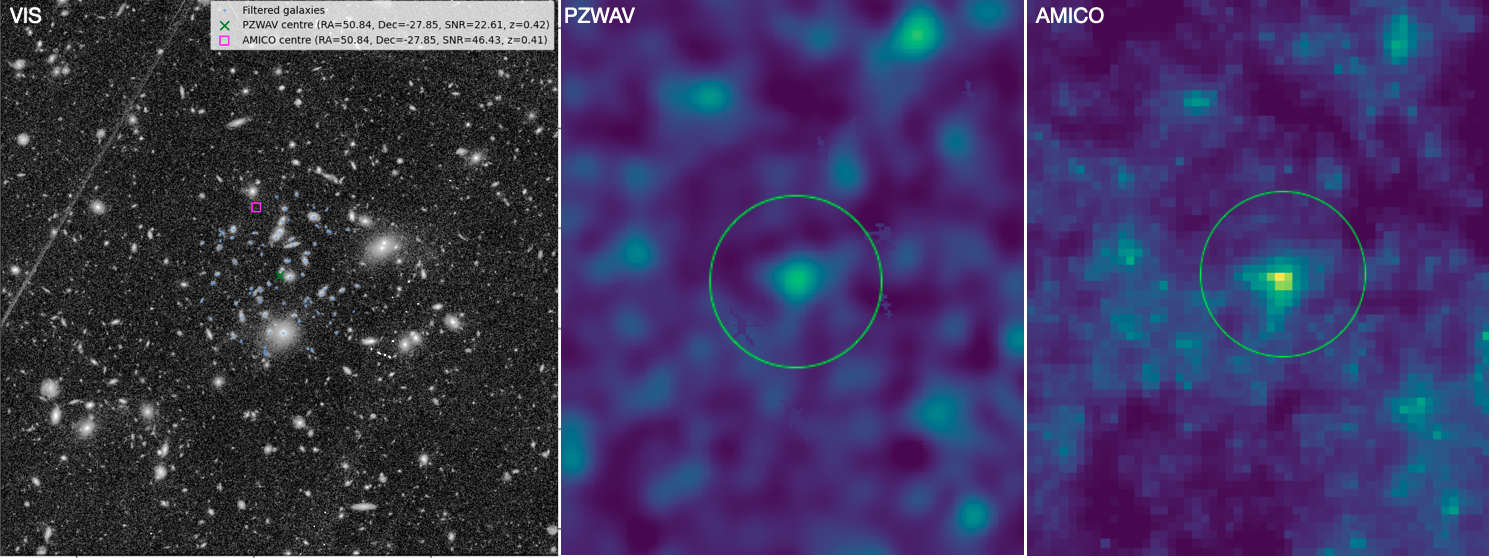}
    \caption{Left: VIS zoom-in of $\ang{;2;}\times\ang{;2;}$  at the location of a \PZWav and \AMICO-detected cluster, at redshift $\zp = 0.42$, located in the \gls{edff}. The cluster has a \Euclid richness of $\lambda = 56.1$. The middle and right panels show \PZWav and \AMICO amplitude maps, respectively centred at the cluster position. The radius of the green circle is 6 arcminutes.} 
    \label{fig:amicopzwavdensitymap}
\end{figure*}

\subsection{Building the combined catalogue}\label{subsec:combinedcat}
The cluster detection algorithms were applied to each of the three Q1 fields. The clusters were selected in the redshift range $0.2 \leq \zp \leq 1.5$, where the lower redshift bound corresponds to the nominal \Euclid cluster catalogue requirement. We note that the upper redshift limit is lower than the one planned for subsequent data releases ($\zp \leq 2.0$), because photometry and photometric redshift estimation still have to be refined for these newly released data; the upper redshift limit for this sample is set by the poorly constrained behaviour of the \Euclid photometric redshifts at $\zp \geq 1.5$, as shown in \cref{fig:zpzs_stats,fig:pdz_zb}. In addition, as described in \cref{sec:galsel}, the provisional requirement that all objects in the galaxy catalogues be detected in the VIS instrument necessarily restricts us to lower redshifts than what is anticipated for future data releases. While the principal catalogue is limited to $z \leq 1.5$, we nevertheless explore objects at $z\geq 1.5$ in the joint catalogue in \cref{subsec:beyondz}. This extension is undertaken with the knowledge of limitations in redshift accuracy, allowing us to assess the reliability of the cluster detections and highlight the most promising candidates in this regime.

For this initial Q1 analysis, we ensured a consistent number density of detections between the two algorithms by adjusting the minimum \gls{snr} accordingly. Specifically, we used thresholds of \snr=$16.1$ ($16.9$) for \AMICO and \snr=$8.3$ ($8.2$) for \PZWav, to guarantee $15$ detections per square degree for each algorithm in the \gls{edff} and \gls{edfs} (\gls{edfn}). The different \gls{snr} thresholds applied to the southern and northern fields is due to the different photometric redshift quality, as discussed in \cref{subsec:photozs}. The absolute difference in values adopted for \AMICO and \PZWav is due to different definitions of the \gls{snr} by each finder. The \gls{snr} thresholds for each of the three fields were determined by visually inspecting a large number of detections to ensure an initial cluster sample that excludes clearly spurious detections. In the following analysis, we prioritise a joint sample of clusters in common between \PZWav and \AMICO to mitigate known issues in the input galaxy photometric redshifts. Focusing on the joint, high \gls{snr} detections for \gls{q1} allows us to perform an initial internal validation of the cluster workflow, enabling us to identify any key issues that may impact the catalogue and subsequent cluster science applications.

To match the cluster detections between \AMICO and \PZWav, a physical distance of 1\,Mpc was chosen, from which an angular search aperture was computed based on the mean cluster redshift of each pair. We imposed an additional constraint on any redshift difference of $\delta z = 0.1(1+z)$ to minimise the number of erroneous matches to clusters in projection. If multiple matches were found, the highest \gls{snr} target was chosen. Fig.~\ref{fig:snr_snr} shows the comparison of the \gls{snr} of the matched systems. The joint catalogue corresponds to the systems in the upper right part of the figure, which consists in 6.4 (\gls{edfn}), 7.0 (\gls{edfs}) and 7.9 (\gls{edff}) common detections out of the adopted nominal 15 detections per square degree for each algorithm. More of these detections are matched but were found at \gls{snr} lower than our thresholds (upper-left and lower-right quadrants of Fig.~\ref{fig:snr_snr}).
The 35 highest \gls{snr} detections in the resulting joint catalogue are presented in Table~\ref{table:catalog}. For each entry, we preserved the positions, redshifts, and \gls{snr}s derived by each detection algorithm, whereas the presented richness is computed at the position of the \PZWav centres. Finally, the occurrence of fragmentation associated with the two detection algorithms is shown to be low over the overall mass range for clusters, with a more detailed assessment performed in \cite{Adam-EP3}.

\begin{figure*}
    \centering \includegraphics[width=0.95\linewidth]{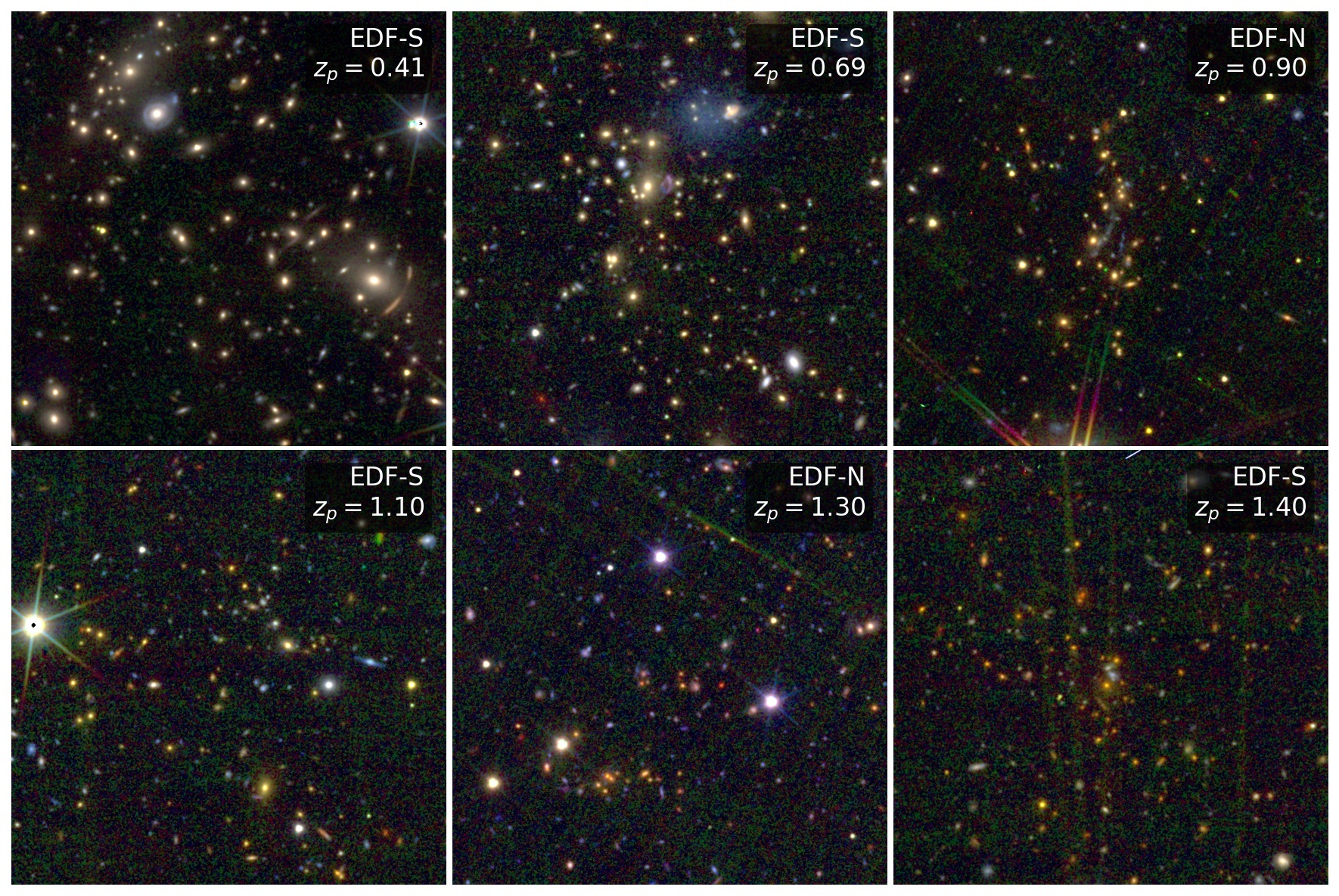}
    \caption{VIS \IE and NISP \YE- and \HE-band combined colour postage stamps of $\ang{;2;}\times\ang{;2;}$ for five of the highest \gls{snr} \PZWav- and \AMICO-detected clusters in six selected redshift intervals ranging from low to high redshift.}
    \label{fig:euclidclustergallery}
\end{figure*}

In \cref{fig:amicopzwavdensitymap}, we show an example of a candidate cluster within the \Euclid VIS image \citep{Q1-TP002} that is detected in both the \PZWav and \AMICO density maps. 
We note that the study of individual detections from the two cluster finders is an important aspect in understanding internal differences in the two algorithms. While the scope of this first paper is to assemble a joint list of reliable cluster detections in the \gls{q1} data set, follow-up analyses of the individual \PZWav and \AMICO catalogues will be pursued prior to \gls{dr1}, including the statistics of the matched and unmatched detections. \cref{fig:euclidclustergallery} shows the visualisation of the \PZWav- and \AMICO-detected clusters with the highest \gls{snr} in five redshift intervals from combined VIS and NISP colour postage stamps.

\begin{figure*}
    \centering
    \includegraphics[width=0.4\linewidth]{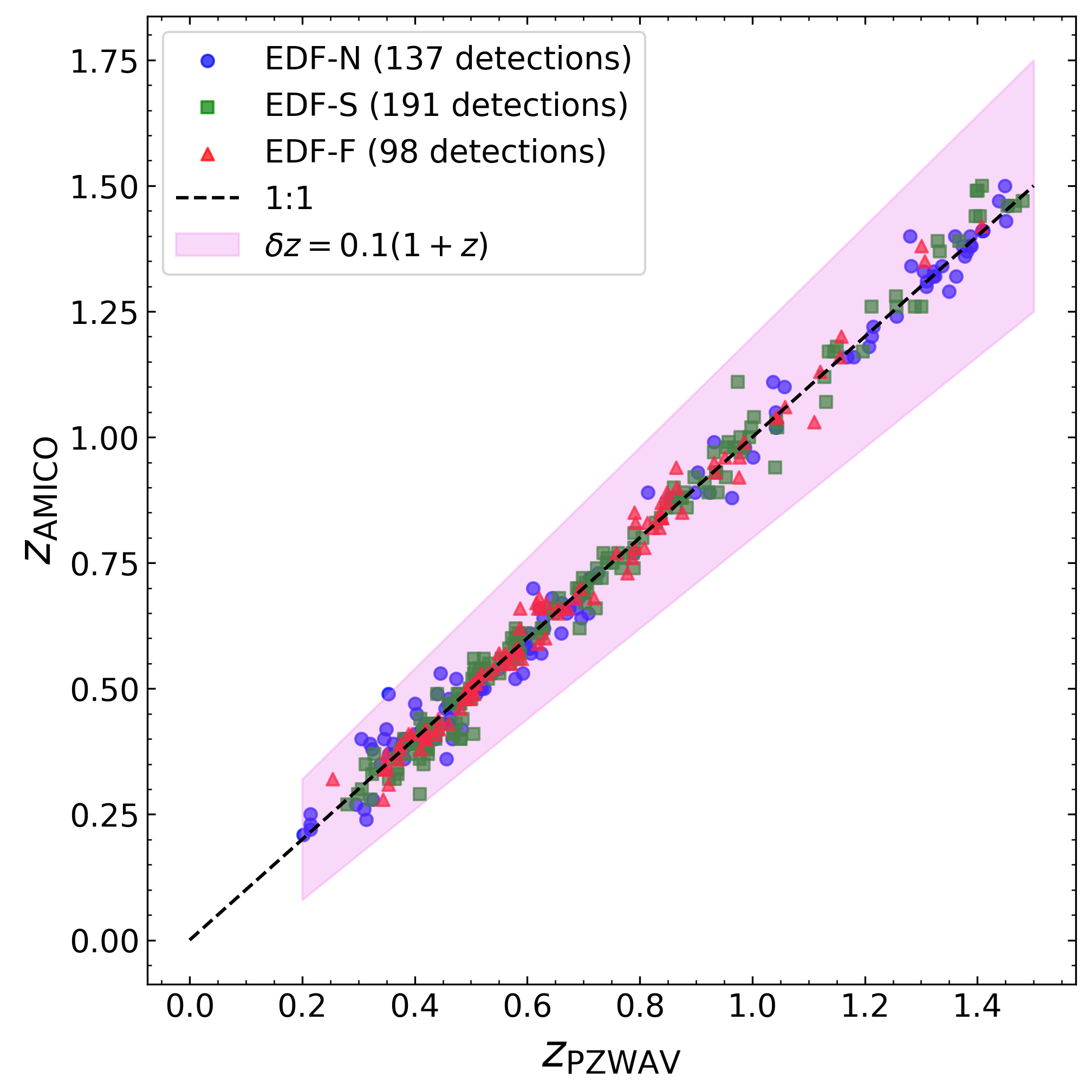}
    \includegraphics[width=0.4\linewidth]{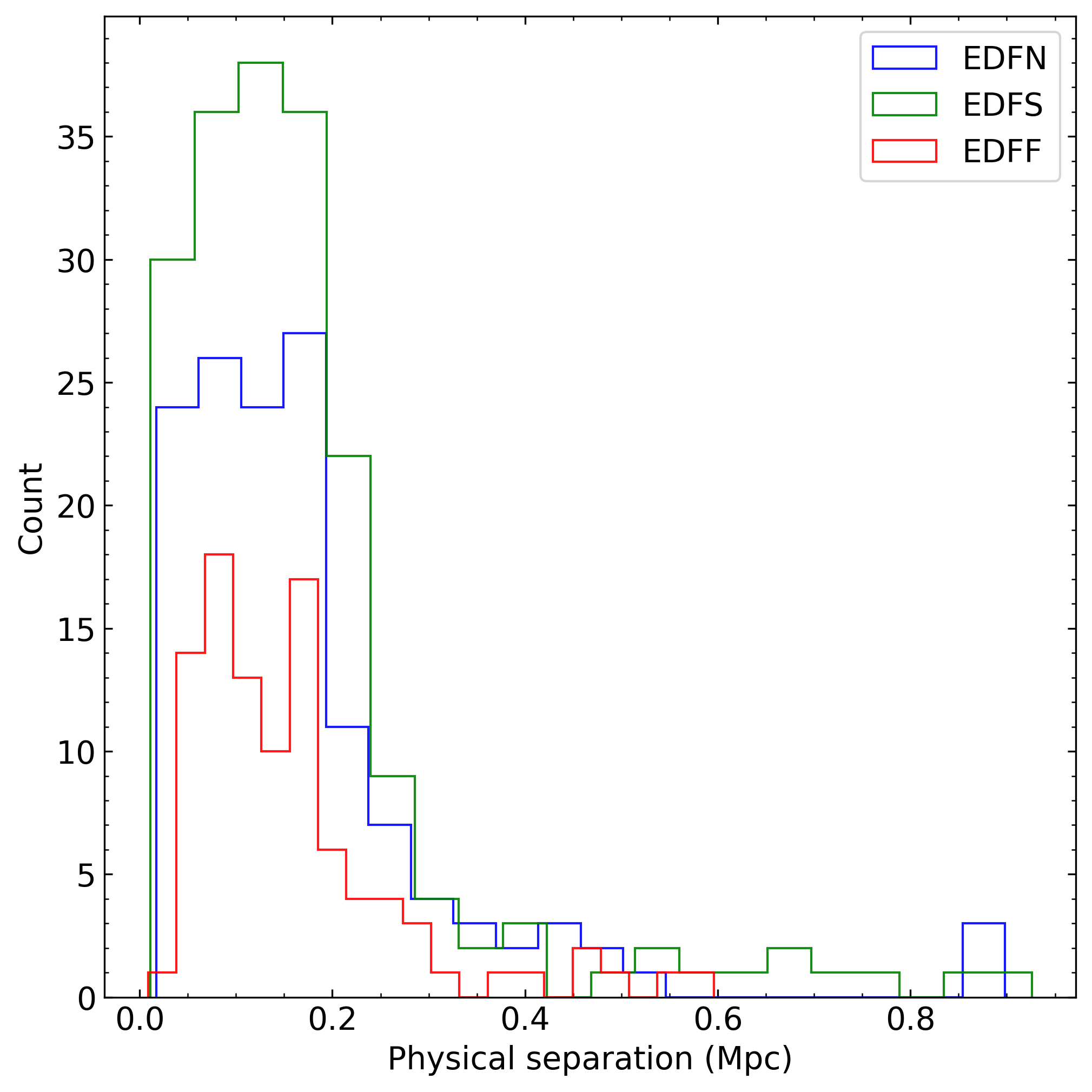}
    \caption{\emph{Left}:
    Redshift correlation between the photometric redshift estimate from the cluster finders \AMICO and \PZWav for all cluster detections in the \gls{q1} region. \emph{Right:} Euclid-based centring offset distribution between \AMICO and \PZWav centres for the catalogues in the three Q1 fields.}
    \label{fig:combinedcat}
\end{figure*}

The left panel of \cref{fig:combinedcat} compares the redshifts estimated by the two algorithms for the matched catalogue. The shaded pink region illustrates the $\delta z=0.1(1+z)$ matching interval, and we see that the redshift difference between matched objects is well within the interval, reinforcing the validity of the matching procedure. The right panel of \cref{fig:combinedcat} shows the internal consistency in the centring performance of the two algorithms. 

\begin{figure}
    \centering
    \includegraphics[width=0.85\linewidth]{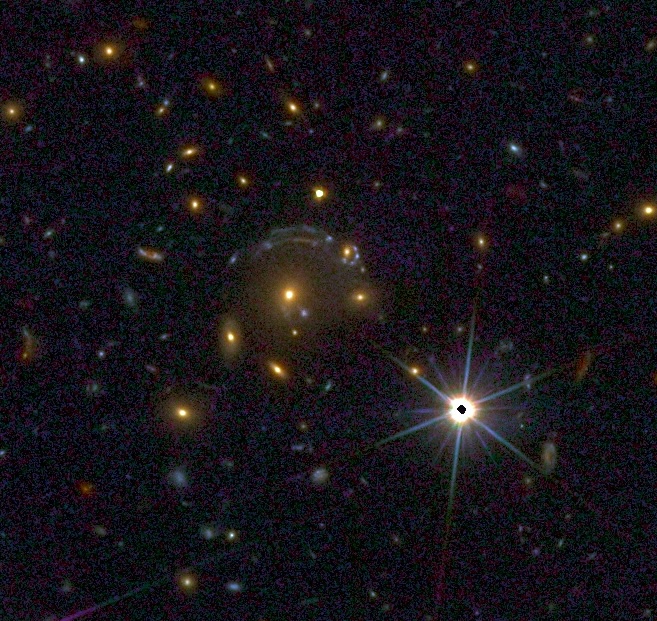}
    \caption{Zoomed-in image of a strong lens in a \PZWav and \AMICO-detected cluster at RA=57\fdg33, Dec=--48\fdg96, with a richness $\lambda = 58.7$ and redshift $\zp = 0.58$ in the \gls{edfs}. This cluster is also reported in the catalogue of giant arcs found in the \Euclid Q1 data release \citep{Q1-SP057}.}
    \label{fig:stronglens}

\end{figure}

Finally, we highlight an example of a strong lens detected in one of the cluster candidates in the \gls{edfs} in \cref{fig:stronglens}, which is located in close proximity to both the \PZWav and \AMICO detection centres. This lens was initially reported in the catalogue of strongly lensed galaxy systems from the Dark Energy Survey Science Verification and Year 1 Observations in \cite{DESLenses2017}. We note that, thanks to \Euclid's superior resolution, in addition to the central lens there is a clearly visible galaxy-galaxy strong lensing event, to the right of the giant arc, involving a cluster member galaxy and likely the same background system which produces the dominant arc. While it is the cluster halo that is responsible for producing strong lensing features on larger scales, cluster galaxies can themselves act as lenses inside the larger lens. Such lensing features are described in greater detail in \cite{Meneghetti2020}.

Cluster richness is estimated using the \gls{sgs} LE3 function \RICHCL, which independently employs photometric redshift and red-sequence-based richness assignment derived from the approaches of \cite{CastignaniBenoist2016} and \cite{Andreon2016}. 
Following the prescription of \cite{CastignaniBenoist2016}, cluster membership probabilities are computed as a function of galaxy cluster-centric distance, \HE-band magnitude, and photometric redshift.  Cluster richness is then calculated as the sum of membership probabilities for members brighter than $\HE^{*}(z_{\rm cluster}) + 1.5$, where $\HE^*$ is the characteristic magnitude marking the knee of the luminosity function. 
In this study, $\HE^*(z)$ is estimated from the passive evolution of a burst galaxy with a formation redshift $z_\mathrm{form}=5$, taken from the PEGASE2 library \citep[\texttt{burst\_sc86\_zo.sed},][]{Fio97}. It is calibrated using the value of ${K}^{\star}$($z=0.25$) derived by 
\citet{2006ApJ...650L..99L} from an observed cluster sample. 

\begin{figure}
    \centering
    \includegraphics[width=0.95\linewidth]{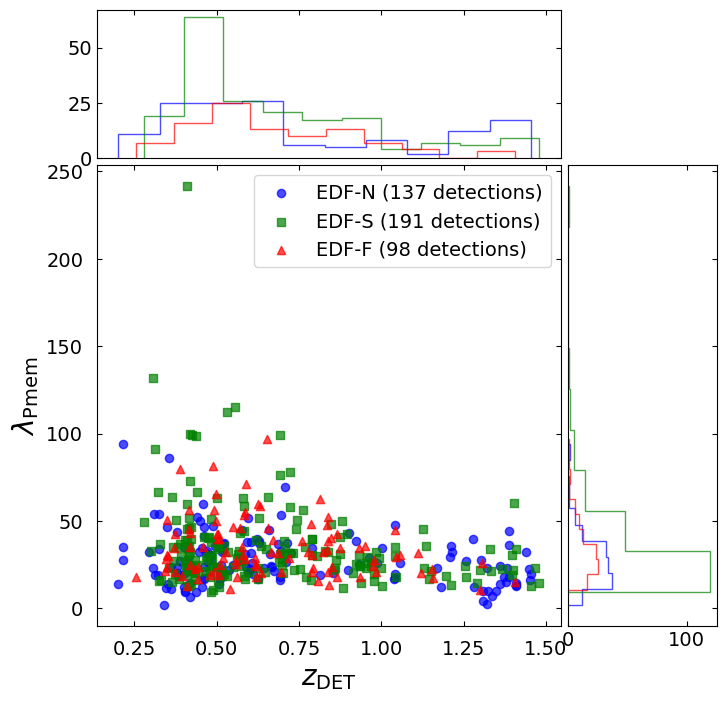}
    \caption{Combined richness and redshift distributions of the cluster detections across the three fields.}
    \label{fig:richnessredshift}
\end{figure}

We computed the richnesses, $\lambda_\mathrm{Pmem}$, using only the \gls{photz} assignments for this study. We calculate the richness based on the \PZWav cluster centre and redshift, but would obtain similar results by using the \AMICO centre and redshift, based on the level of positional and redshift agreement shown in the two panels of \cref{fig:combinedcat}. The combined richness and redshift distributions are presented in \cref{fig:richnessredshift}. 

\begin{figure*}
    \begin{centering}
    \includegraphics[width=0.4\linewidth]{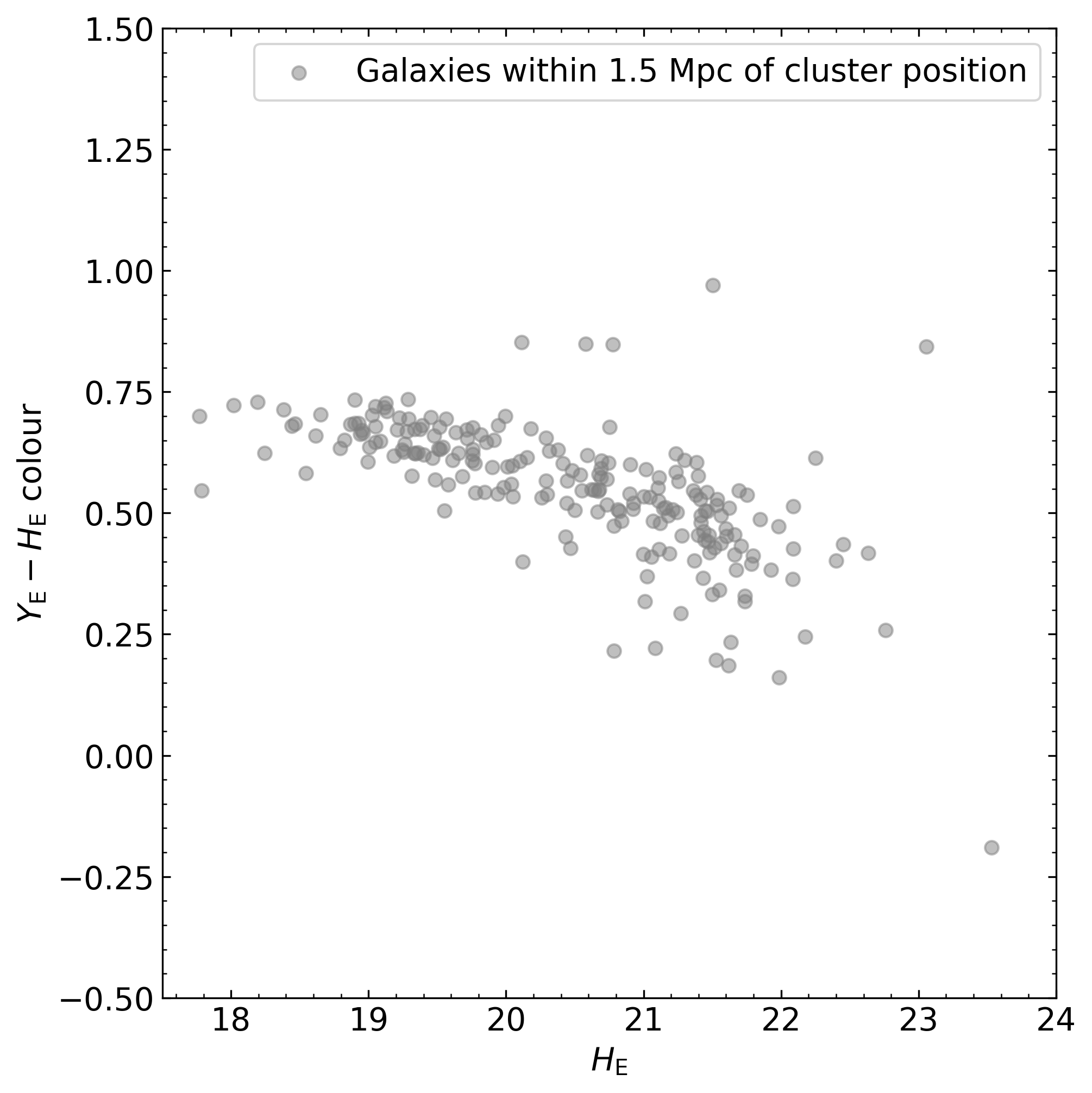}
    \includegraphics[width=0.4\linewidth]{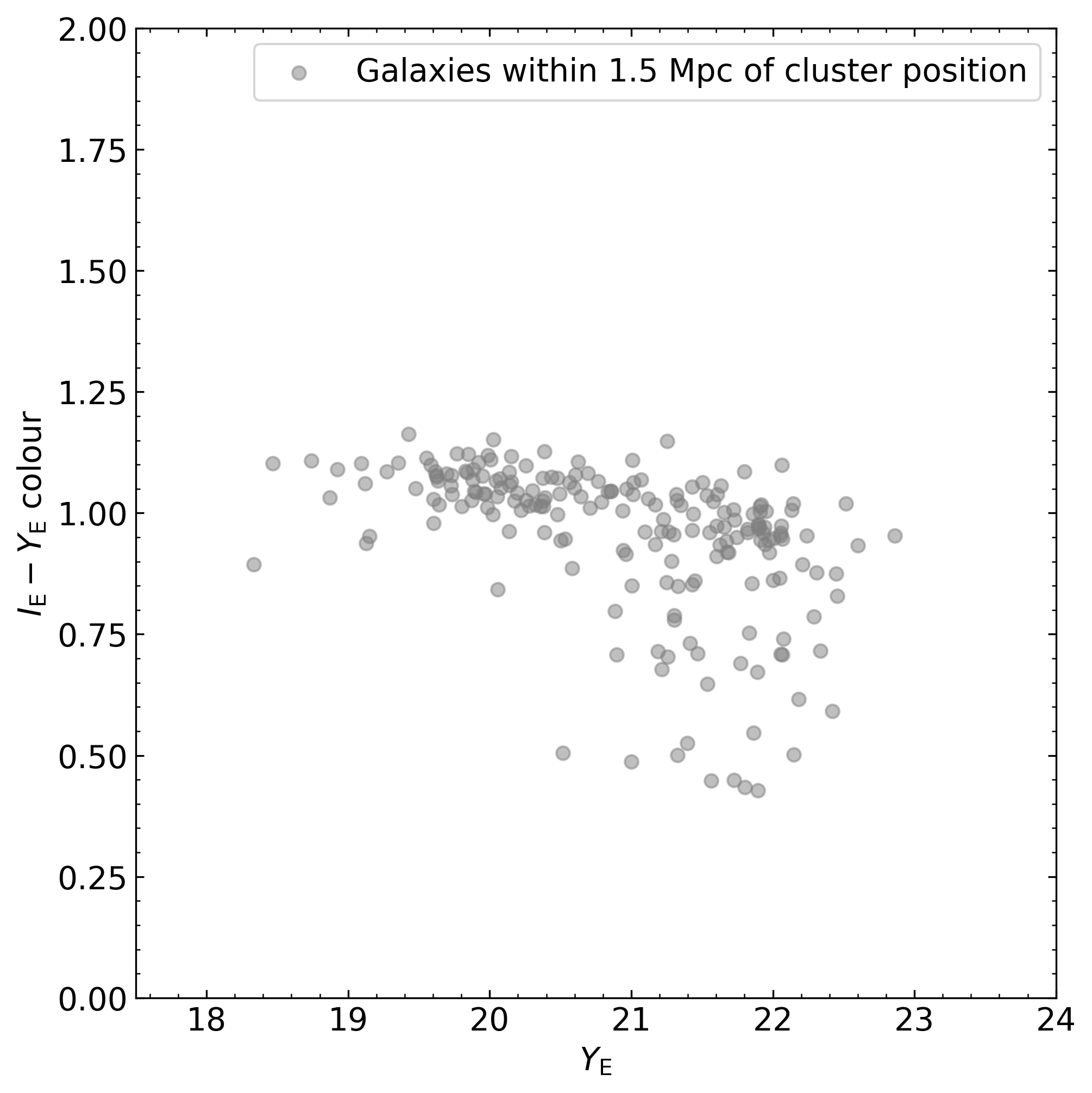}
    \caption{\Euclid-only photometry of the galaxy population (grey) located within 1.5\,Mpc of the \gls{edfs} cluster position at $\zp = 0.41$ as a function of $\YE-\HE$ colour and \HE-band magnitude, and $\IE-\YE$ colour and \YE-band magnitude (left and right panels, respectively). The cluster shown here, EUCL-Q1-CL J041113.88-481928.2, is the richest in the Q1 catalogue ($\lambda_{\rm {Pmem}} = 241.5$) , also observed by SPT (SPT-CL J0411-4819).} 
    \label{fig:colourmagdiagram}
    \end{centering}
\end{figure*}

We illustrate \Euclid's photometric capabilities on the richest cluster detected in the data set, SPT-CLJ0411$-$4819 in the \gls{edfs}, by showing the  $\YE-\HE$ versus $\HE$ and $\IE-\YE$ versus $\YE$ distributions for all galaxies within a distance of \ang{;2;} from the cluster position. We observe a clear red sequence in both diagrams, as shown in \cref{fig:colourmagdiagram}, left and right panels, respectively.  

\section{\label{sec:external}Comparison with external cluster catalogues}

\subsection{Searching for \Euclid counterparts in external data}\label{subsec:euclidcounterparts}

We cross-correlated the joint \gls{q1} cluster catalogue with external meta-catalogues\footnote{A meta-catalogue is a combination of different source catalogues with cross-identification of clusters and homogenisation of key quantities, notably redshift and a mass proxy.} and individual catalogues. We used meta-catalogues and catalogues primarily selected in X-ray -- \gls{mcxc} described in \citealt{Sadibekova2024}, eROSITA~\citep{Bulbul2024,Kluge2024a} -- with the \gls{sz} effect -- \gls{mcsz} described in {\color{blue}Tarrio et al. (in preparation)} -- and in the optical -- \gls{des} Y1 \redmapper~\citep{Rykoff2016,DESCosmo2020}, and the Abell catalogue~\citep{Abell1958, Abell1989}. We also used the joint X-ray and SZ selected catalogue ComPRASS~\citep{Tarrio2019}, the LC$^2$~\citep{Sereno2015}, and the \gls{mccd} described in {\color{blue}Euclid Collaboration: Melin et al. (in preparation)} for clusters with high-quality velocity dispersion measurements.

We applied the cross-correlation method based on a two-way matching. In brief, for each cluster in the external catalogue (or meta-catalogue), we searched for the closest detection (angular distance on the sky) in the \gls{q1} cluster catalogue. Symmetrically, for each detection in the Q1 catalogue, we searched for the closest cluster in the external catalogue. We then only consider \Euclid detection-external cluster pairs that are identical in the two directions, hence the name `two-way matching'. 

We then apply additional criteria on the distance $d$ and redshift difference $\zp-z_{\rm external}$ between the detection and the cluster. For the \gls{mcxc} and the \gls{mcsz}, we imposed $d<\ang{;10;}$, $d/\theta_{500}<2$, and $|\zp-z_{\rm external}|<3 \sigma_{\zp}$, where $\theta_{500}$ is the characteristic angular radius of the external cluster\footnote{$\theta_{500}=R_{500}/D_{\rm A}$, where $R_{500}$ is the radius within which the average cluster mass $M_{500}$ is 500 times the critical density of the Universe at the cluster's redshift and $D_{\rm A}$ is the angular diameter distance to the cluster. $M_{500}$ is obtained from the X-ray luminosity for the \gls{mcxc} and from the SZ flux or the SZ \snr~ for the \gls{mcsz}.}  and $\sigma_{\zp}$ is the quoted error on the photometric redshift of the \Euclid detection. The method is presented in detail in {\color{blue}Euclid Collaboration: Melin et al. (in preparation)}.

In total, we found 4/21/85 counterparts in the external catalogues for the 137/98/191 \Euclid\ detections in the \gls{edfn}, \gls{edff}, and \gls{edfs} fields, respectively. In the \gls{edff} and the \gls{edfs}, the matches are mainly from \redmapper and eROSITA. The \gls{edfn} field is not covered by these two catalogues, which explains the significantly lower number of counterparts that are mainly from the \gls{mcxc}. The total number of detections and matches from the external surveys that fall within the \gls{q1} footprint are listed in \cref{tab:external_datasets}. A detailed breakdown of the individual matches to data sets per Q1 field can be found in Appendix \ref{app:external}.

\begin{table*}
\centering
\caption{Comparison of external data sets with total objects in \gls{q1} footprint and total matches.}
\begin{tabular}{ccccc}
\hline
\hline
\noalign{\vskip 2pt}
\gls{q1} field & No. of \Euclid cluster detections & No. of clusters found in external data sets & No. of unmatched \Euclid detections \\
\hline 
\noalign{\vskip 2pt}
\gls{edfn} & 137 & 80 (4) & 57 \\
\gls{edff} & 98 & 93 (21) & 5 \\
\gls{edfs} & 191 & 176 (85) & 15 \\
\hline
\end{tabular}
\tablefoot{The number of \Euclid detections, matched and unmatched systems correspond to those within the redshift range $0.2 \leq \zp \leq 1.5$. The total number of matches includes the results from the individual catalogues and meta-catalogues, supplementary MaDCoWS2 and WH24 catalogues, and a NED database search (see \cref{subsec:neweuclid}), while the numbers in brackets denote the number of matches using the aforementioned individual catalogues and meta-catalogues only.}
\label{tab:external_datasets}
\end{table*}

\subsection{External photometric redshift comparison}
\begin{figure}
    \centering
    \includegraphics[width=0.95\linewidth]{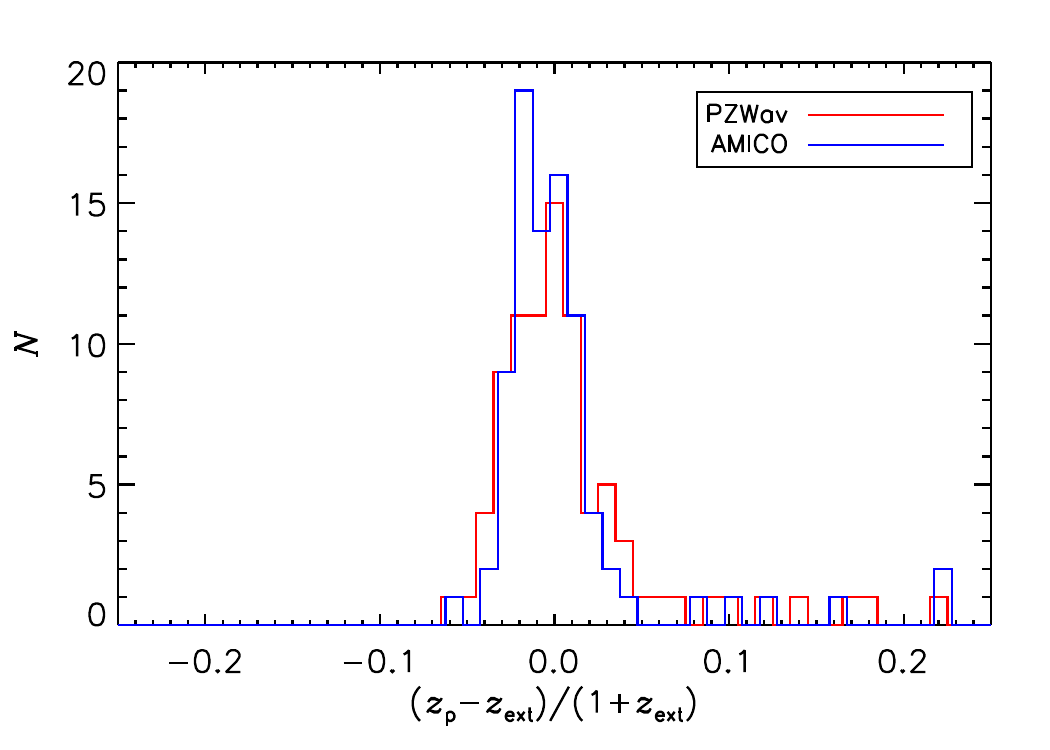}
    \caption{Histogram of the difference  between \Euclid redshift $\zp$ and external redshift $z_{\rm ext}$ (divided by $1+z_{\rm ext}$) for \PZWav and \AMICO in the \gls{edfs} for the 47 cluster detections matched in various multiwavelength catalogues.}
    \label{fig:redshift_external_comparison}
\end{figure}

We concentrate on the \gls{edfs}, for which we found the largest number of counterparts, comparing the \Euclid photometric redshifts with the external cluster redshifts. Only six external redshifts are spectroscopic, so the majority are photometric redshifts. \Cref{fig:redshift_external_comparison} demonstrates that the cluster photometric redshifts have a scatter of around $0.05(1+z)$ relative to the redshifts (mostly photometric) provided for these clusters by other surveys. At $z<0.3$ the redshifts also appear to be systematically higher than those from the other surveys. If we restrict ourselves to clusters with $z>0.3$, the scatter reduces to around $0.03(1+z)$. We note that a limitation of the current pipeline is that the photometric redshift uncertainties produced by both \PZWav and \AMICO are overestimated. Comparing to the scatter relative to the other surveys suggests an overestimation of the \Euclid cluster photometric redshift uncertainties in \gls{q1} of the order of a factor of 1.7 (2.9 for \AMICO). The origin of this discrepancy, likely related to the current photo-$z$ \gls{pdz}, is still under investigation.

\subsection{Centring offsets between \Euclid and external data sets}
\begin{figure*}
    \centering
    \includegraphics[width=0.495\linewidth]{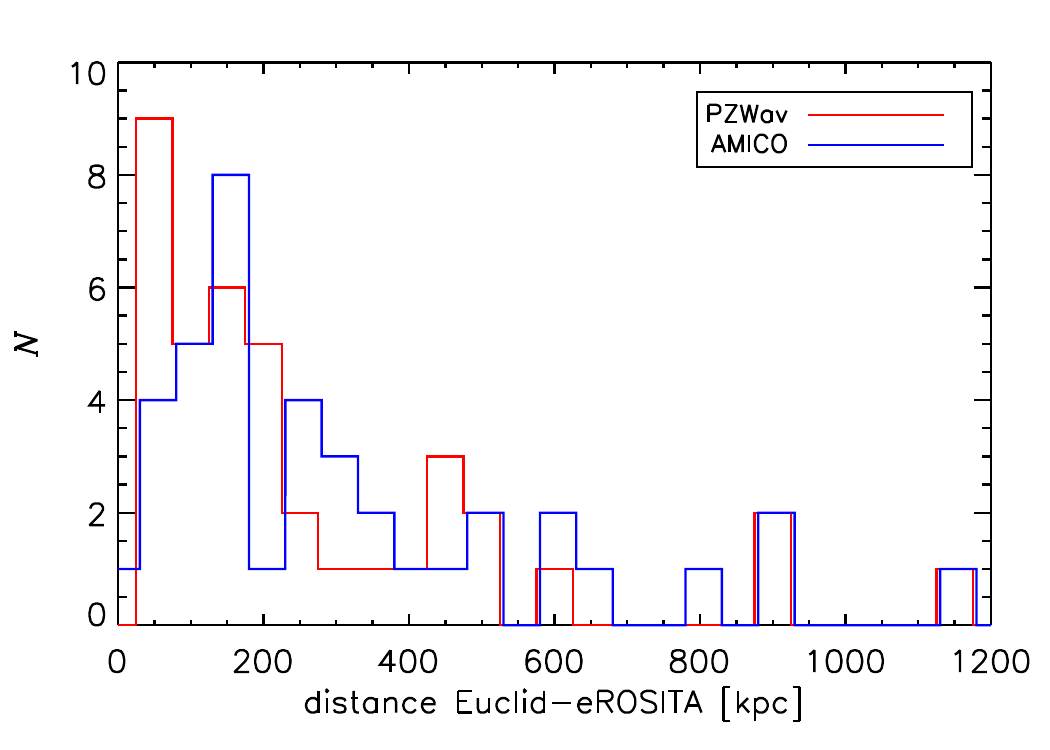}
    \includegraphics[width=0.495\linewidth]{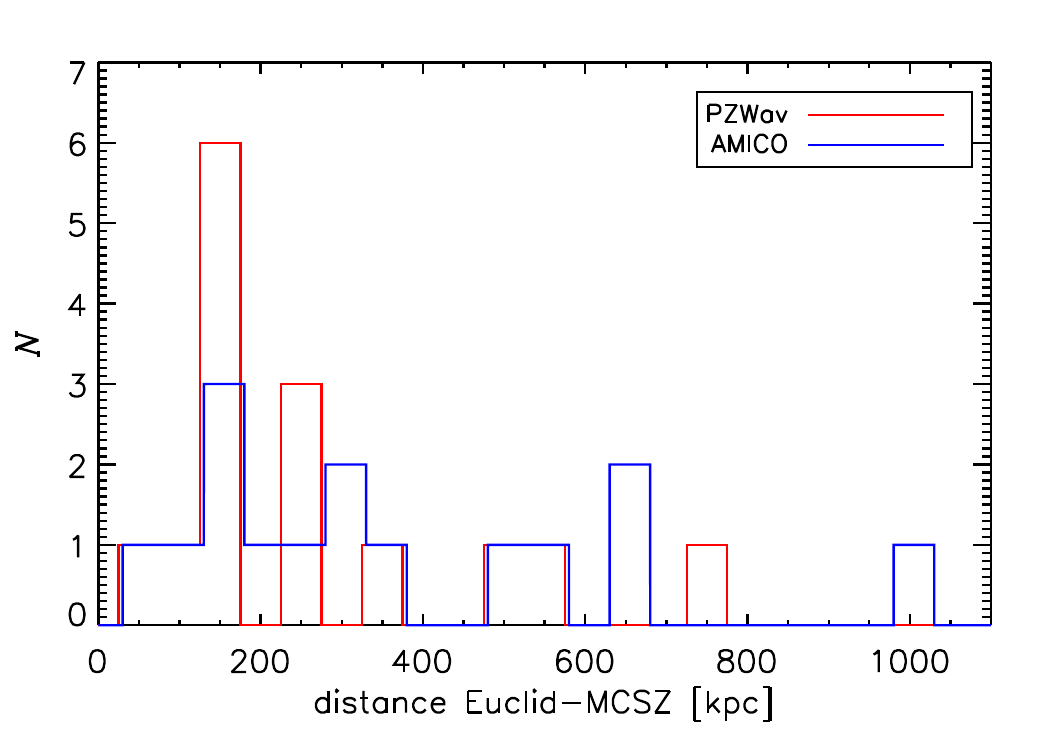}
    \caption{Centring offsets between \Euclid catalogues and external catalogues in the \gls{edfs}. The left and right panels refer to eROSITA and \gls{mcsz}, respectively.}
    \label{fig:distance_external_comparison}
\end{figure*}

We computed the distances between the positions of the \Euclid detections and the positions of the matched clusters in the external catalogues. \cref{fig:distance_external_comparison} shows the histograms in the \gls{edfs} for eROSITA (left) and \gls{mcsz} (right), for positions taken from \PZWav (red) and \AMICO (blue).
For eROSITA, a clear peak is visible in the histogram below $200\, {\rm kpc}$ for \PZWav and \AMICO.

For \gls{mcsz}, the two histograms show a small peak around $200\,{\rm kpc}$. This can be explained by the positional accuracy of the \gls{sz} experiments, which has a typical value of a few times \ang{;0.1;} for the South Pole Telescope \citep[SPT, see][]{Ruhl2004} and the Atacama Cosmology Telescope \citep[ACT, see][]{Fowler2007} clusters, corresponding to about $100\, {\rm kpc}$ at $z=0.5$. The positional accuracy of the \gls{sz} experiments is lower than the positional accuracy of eROSITA (below \ang{;0.1;}, and is expected to further smooth the histogram. This effect is related to the fact that there are only a small number of matches (15) with \gls{mcsz}. We also investigated the dependence of these offsets with the \snr~of the detections for each algorithm and found no significant correlation.

\subsection{Comparison between \Euclid and external cluster observables}
\begin{figure*}
    \centering
    \includegraphics[width=0.33\linewidth]{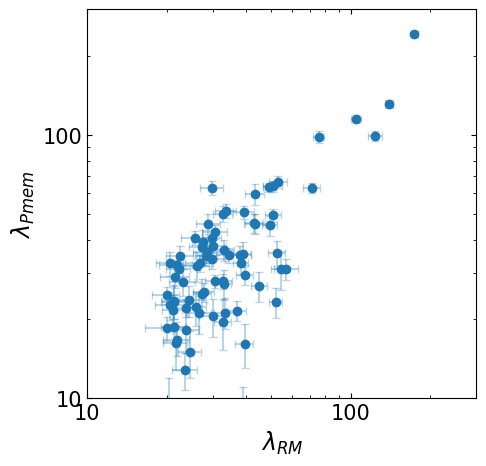}
    \includegraphics[width=0.33\linewidth]{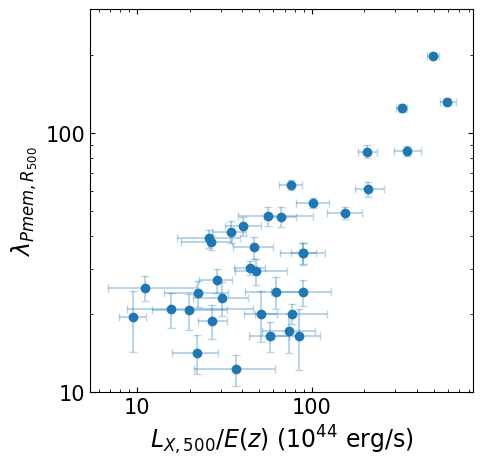}
    \includegraphics[width=0.33\linewidth]{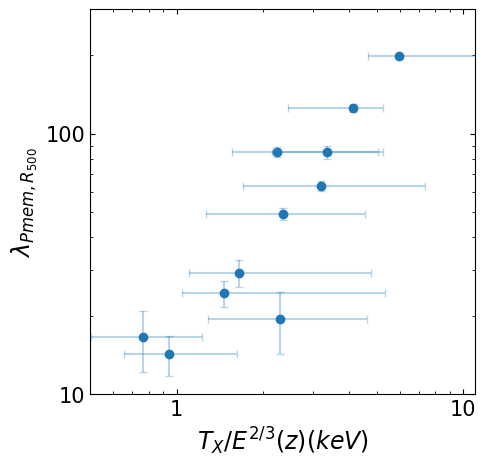}
    \caption{Correlation of the \Euclid richness $\lambda_{Pmem}$ against \redmapper richness (left), eROSITA mass proxies - the X-ray luminosity $L_{X, 500}$ (middle) and X-ray temperature $T_{\rm X,500}$  (right) for the matched systems within the \gls{edfs}.}
    \label{fig:rich_scaling}
\end{figure*}

For the largest cross-matched population of objects in the \gls{edfs}, in addition to the redshift and positional accuracy, we examined the correlation of the \Euclid cluster observables -- principally the richness -- against external mass proxies from multiwavelength data, where available. \cref{fig:rich_scaling} highlights the correlation of the \Euclid richness against three external mass proxies from the \gls{des} Y1 \redmapper and eROSITA data sets, namely the \redmapper richness, eROSITA-based luminosity, and temperature. For the comparison to eROSITA observables, the richness is computed within the same R$_\mathrm{500}$ radius. In all cases, a clear correlation is present, with the tightest scatter observed in the direct comparison of richnesses, with a larger, expected scatter for the optical richness compared to the X-ray luminosity, consistent with results shown in \cite{Kluge2024a}. Finally, the large uncertainties in the X-ray temperatures are due to the limited number of photon counts in the eRASS1 data set.

\begin{figure}
    \centering
    \includegraphics[width=\linewidth]{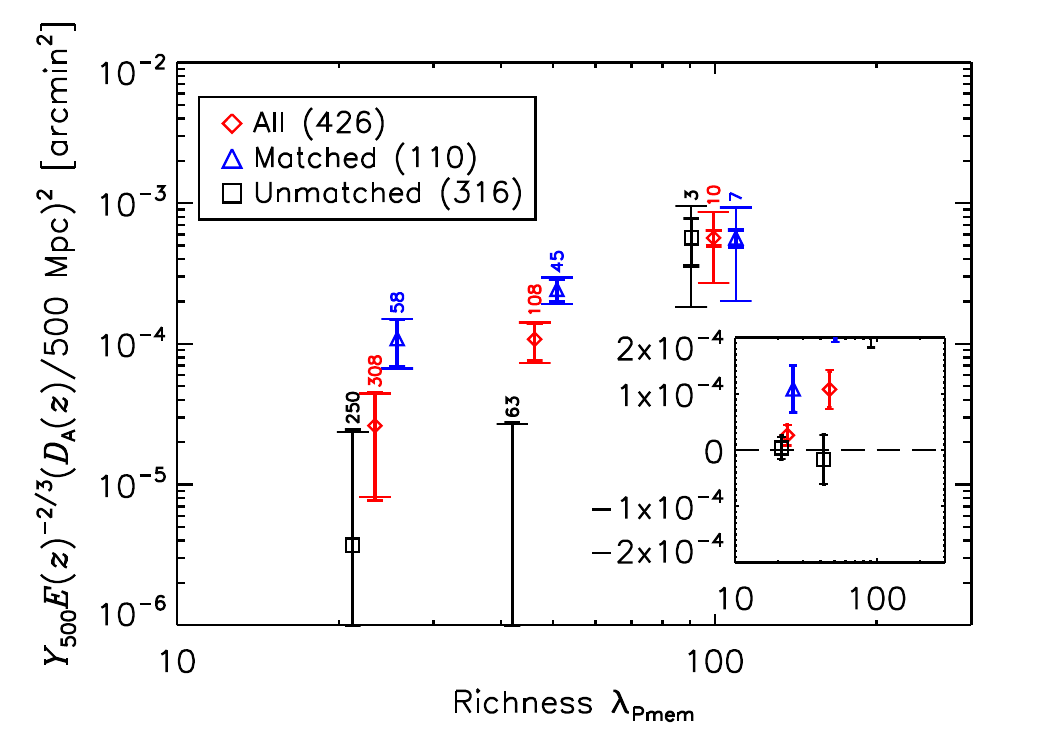}
    \caption{{\em Planck} \gls{sz} flux versus \Euclid richness for the full \Euclid Q1 sample (red diamonds), and for the subsamples matched (blue triangles) and unmatched (black squares) to external catalogues and meta-catalogues. Thick error bars are statistical errors and thin error bars are bootstrap (total) errors. The number on top of each error bar is the number of objects averaged in each bin. The \gls{sz} flux is detected by {\em Planck} in the two highest richness bins for the full sample and in all the three bins for the matched subsample. However, \Planck does not detect the signal in the unmatched subsample, indicating that it contains significantly less hot gas than the matched subsample.}
    \label{fig:planck_sz}
\end{figure}

We also extracted the \gls{sz} flux of the 426 \Euclid detections in the {\em Planck} legacy maps. We averaged the signal in three bins of \Euclid richness ($\lambda_{\rm Pmem}<36.1$, $36.1<\lambda_{\rm Pmem}<93.4$, $\lambda_{\rm Pmem}>93.4$). Results are shown in \cref{fig:planck_sz}. The SZ flux in a sphere of $R_{500}$, $Y_{500}$, is scaled by $E^{-2/3}(z)[D_{\rm A}(z)/500 \, {\rm Mpc}]^2$ to make it proportional to the halo mass $M_{500}$, where $E(z)=H(z)/H_0$ with $H(z)$ the Hubble parameter. Red diamonds show the result for the full sample. 

The \gls{sz} signal is detected in the three richness bins. We also present the results for two subsamples: the matched subsample (blue triangles) containing the 110 detections matched with catalogues and meta-catalogues corresponding to numbers in brackets in~\cref{tab:external_datasets}; and the unmatched subsample (black squares) containing the remaining $426-110=316$ detections. The SZ signal is detected with the same amplitude in the highest richness bin (third) for both subsamples. In the lower richness bins (first and second), the signal is detected for the matched subsample but not detected for the unmatched subsample. In particular, the difference in \gls{sz} flux between the matched and unmatched detections in the first/second bin is $2.3/3.8\sigma$ respectively, indicating that the unmatched subsample contains significantly less hot gas than the matched subsample for richness $\lambda_{\rm Pmem}<93.4$. Investigations with deeper SZ observations (e.g. ACT or SPT-3G) and deep X-ray observations (e.g. XMM or eROSITA) will be necessary to quantify better the gas content of the newly discovered \Euclid detections. 

This difference in the gas content of the matched and unmatched subsample with $\lambda_{\rm Pmem}<93.4$ may be due to the fact that a large fraction of our external catalogues and meta-catalogues are \gls{sz} or X-ray selected, so most probably contain gas-rich systems. The unmatched subsample is thus expected to contain more gas-poor systems. It is also possible that projection effects in the \Euclid detection are boosting richness for intrinsically poor systems, whose mass is consistent with a very low \gls{sz} and X-ray signal. While this effect is expected to be marginal for the richest systems (for which \gls{sz} flux is comparable for matched and unmatched), it could be significant at lower richness. This result should be compared with the results reported in \cite{Popesso2024}, where optically detected (but X-ray undetected) systems identified with eROSITA are predominantly found in filaments, while those detected in X-rays are mainly located in nodes. Recent simulation-based studies supporting this trend are presented in \cite{Cui2024} and \cite{Marini2025}. 

\subsection{Unmatched detections from external data sets}
Because the \gls{q1} cluster sample is intended to serve as a first validation of the cluster workflow during the early phase of the \Euclid mission, it has been constructed on the basis of purity rather than completeness. As a result, we forgo any in-depth analysis on individual clusters from external data sets that are not matched to any of the systems in the Q1 cluster catalogue. Nevertheless, in order to understand the detection limit of the current Q1 sample with respect to external data sets, we construct the histograms of matched and unmatched populations for three external data sets in the \gls{edfs}, because this field contains the largest number of multi-wavelength counterparts; these are the \gls{des} Y1 \redmapper, eROSITA, and \gls{mcsz} samples.  

\begin{figure*}
    \centering
    \includegraphics[width=0.33\linewidth]{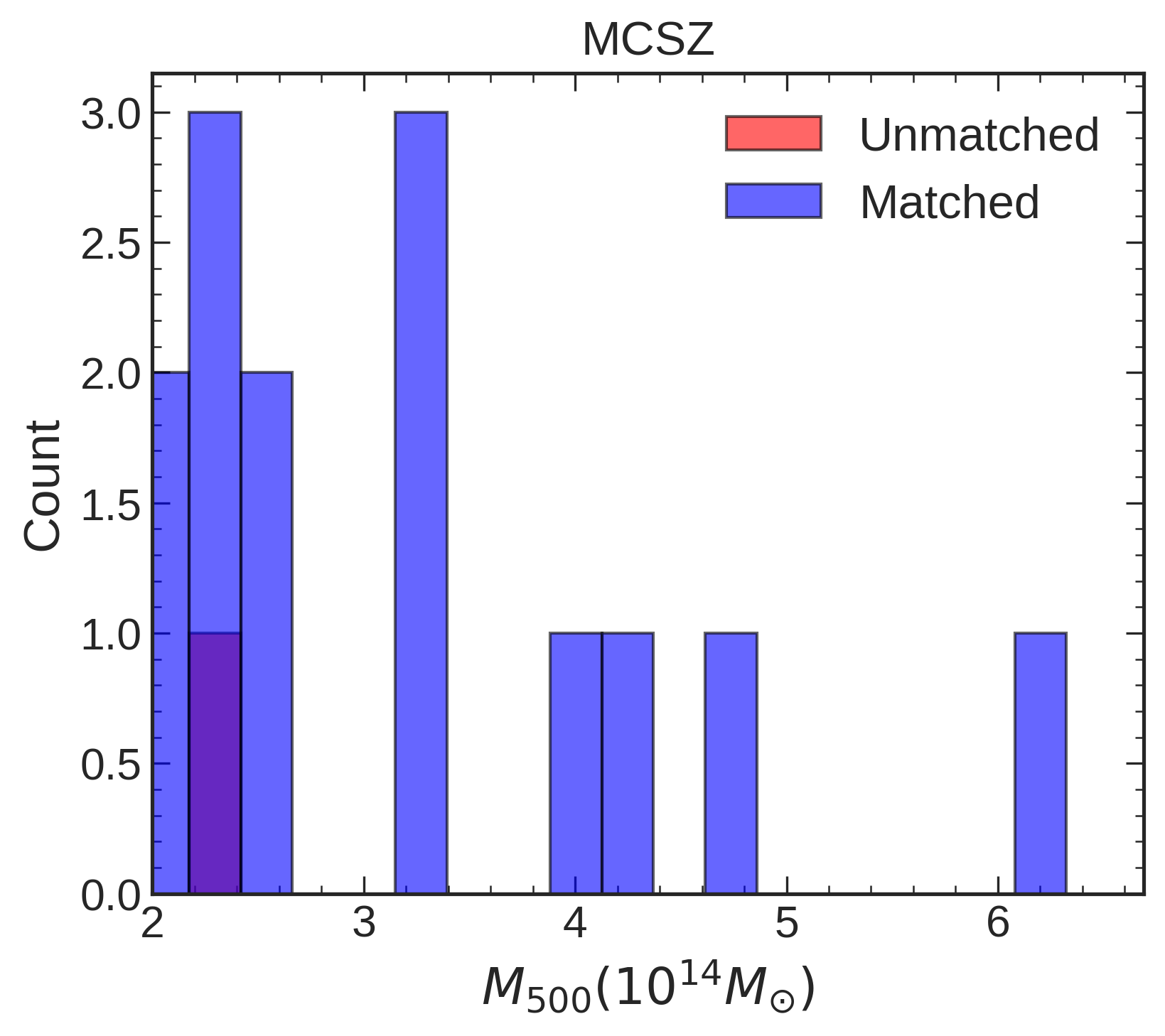}
    \includegraphics[width=0.33\linewidth]{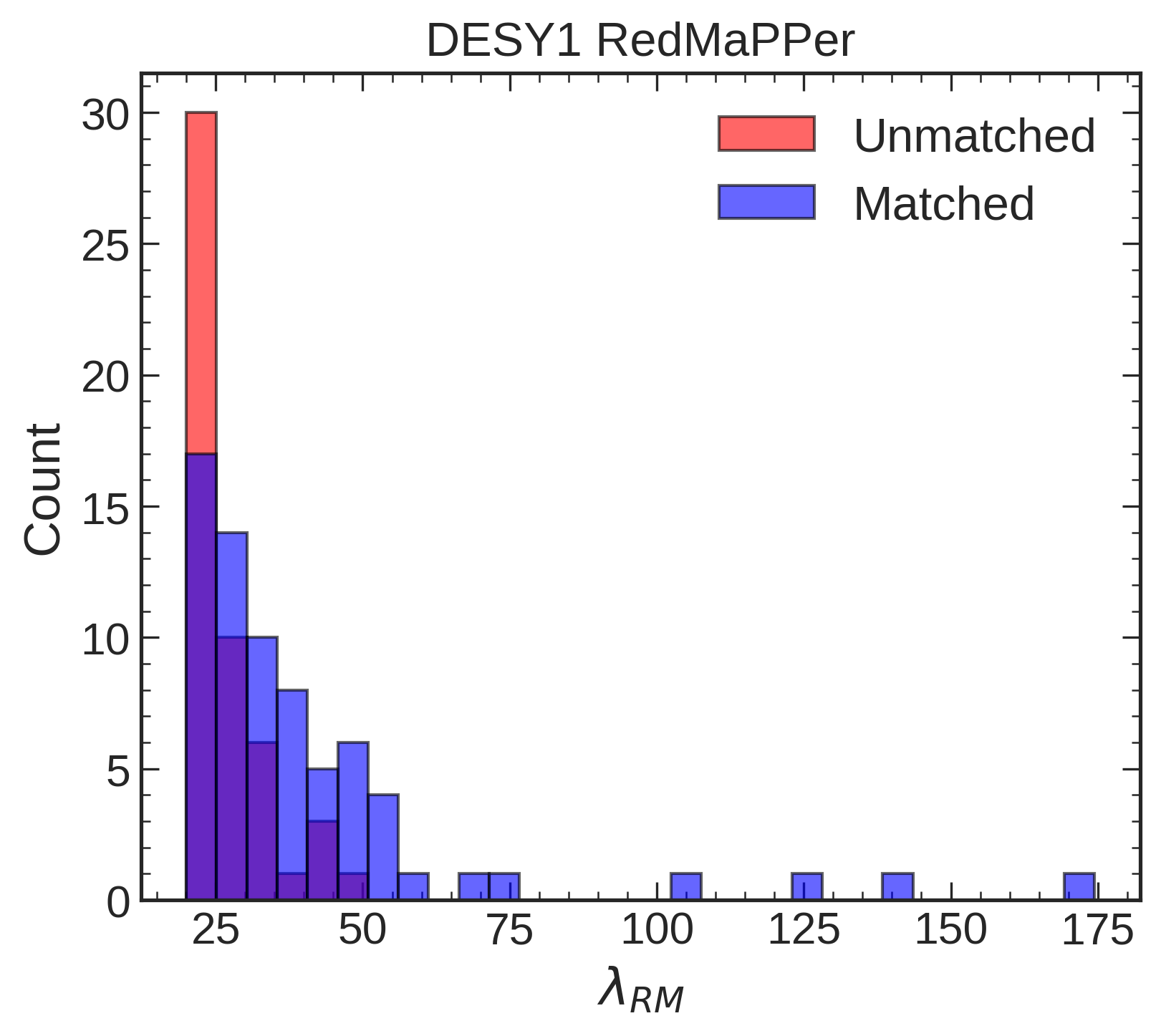}
    \includegraphics[width=0.33\linewidth]{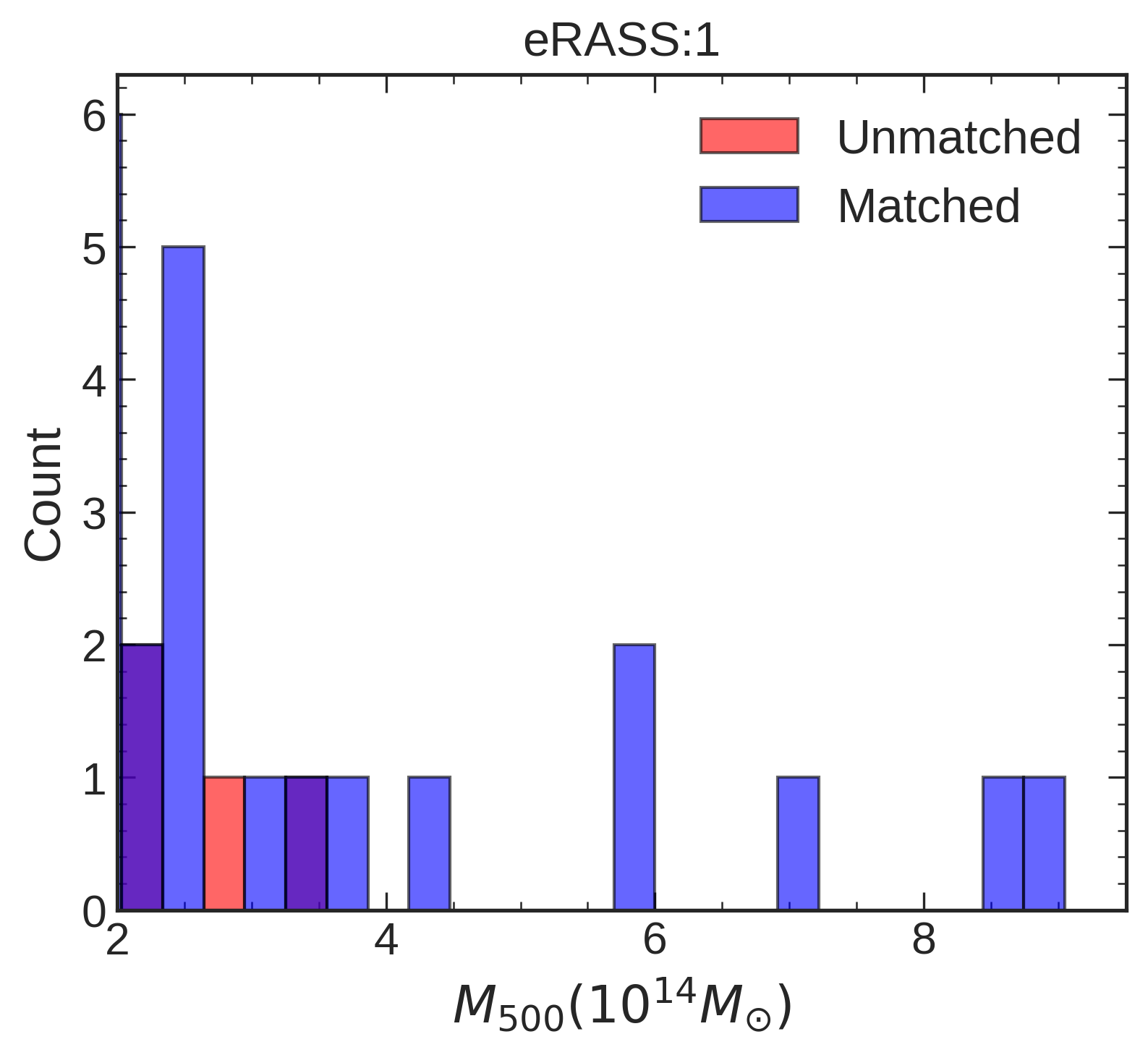}
    \caption{Histograms of matched and unmatched clusters from external multi-wavelength data sets in the \gls{edfs} -- \gls{mcsz} (\emph{left}),  \gls{des} Y1 \redmapper (\emph{middle}), and eROSITA (\emph{right}). The histograms are shown as a function of \gls{sz}-based total mass ($M_{500}$), optical richness ($\lambda$), and X-ray based total mass ($M_{500}$), respectively.}
    \label{fig:erosita_mcsz_comp}
\end{figure*}

\cref{fig:erosita_mcsz_comp} shows the distribution in \gls{sz}-based total mass, \redmapper richness, and eROSITA X-ray based mass estimate, for the matched and unmatched clusters falling within the \gls{edfs}, where the expected coverage is above 80\% (see \cref{sec:galsel}). We see a consistent trend across optical, X-ray and millimetre wavelengths, whereby the unmatched populations systematically occupy the lower richness and lower mass ends of the overall distribution, although within a limited area of approximately 27 square degrees, the statistics are sparse. 

Since the joint \Euclid catalogue is cut at high \snr~ by construction, it does not completely sample the underlying cluster population, particularly at lower richnesses (78\% of the unmatched \redmapper objects have a richness less than 30). This reinforces the hypothesis that incompleteness in the cluster sample at this stage is largely driven by the high \snr~threshold on the individual detection catalogues and the degree of intrinsic scatter shown in \cref{fig:snr_snr}, in addition to known issues of incompleteness in the \Euclid galaxy catalogues at lower redshifts, described in more detail in {\color{blue}Klein et al. (in preparation)}.

\subsection{The potential for new \Euclid cluster detections}\label{subsec:neweuclid}
For the consideration of new, potentially undetected clusters in the \gls{q1} fields, we expanded our search beyond the list of catalogues and meta-catalogues described in \cref{subsec:euclidcounterparts} for the redshift validation and measurement of positional scatter. We performed a cross-match of all the detections within the \gls{ned}\footnote{NED is funded by the National Aeronautics and Space Administration and operated by the California Institute of Technology.}, using a matching radius of two arcminutes, and a redshift constraint of $|\zp-z_{\rm NED}|<3 \sigma_{\zp}$. We also perform a match to the \gls{madcows2}, described in \cite{Thongkham2024} and \gls{wh24} cluster catalogue \citep{Wen2024}, using an identical scheme to the one described above, simply replacing the NED redshifts with those from MaDCoWS2/WH24.

After filtering for overlap between the matches retrieved from NED, \gls{madcows2} and/or \gls{wh24} with the existing meta-catalogues, we find 76, 72, and 91 new matches in the \gls{edfn}, \gls{edff}, and \gls{edfs}, respectively. Detailed information on the supplementary matches is contained within \cref{app:external}. Subtracting the total number of existing matches in the literature, we report 57, 5, and 15 new detections in the \gls{edfn}, \gls{edff}, and \gls{edfs}, totaling 77 potentially new \Euclid clusters. A total of 48 of these systems have a cluster redshift $\zp \geq 1$, 36 of which are in \gls{edfn} and 11 in \gls{edfs}. 

Each of these new detections was visually inspected in both the \Euclid imaging data and \gls{desi} Legacy imaging to confirm the presence of a galaxy overdensity and to check for possible contamination of a detection from artefacts or nearby bright stars.  We also inspected histograms of the photometric redshifts within \ang{;2;} of the cluster centre to look for coherent peaks at the nominal cluster redshift, and used external spectroscopic redshifts to confirm the presence of real clusters. The results of these final checks indicates that 20 out of the total sample of 426 clusters are likely to be spurious. 

\subsection{Beyond $z \geq 1.5$: the case for \Euclid}\label{subsec:beyondz}

While we impose an upper redshift limit on our joint cluster sample for this first analysis due to the limitations in photometric redshift estimation described in \cref{subsec:photozs},
the detection algorithms as well as the richness estimation are nevertheless applied on the nominal \Euclid redshift range, which extends to a redshift of $z = 2$. To this end, we study the number of joint clusters which fall in the redshift range $1.5 \leq z \leq 2 $ to evaluate their reliability as a means to highlight one of the key scientific goals of \Euclid -- the detection and characterisation of the high redshift cluster population, as \cite{Sartoris2016} showed that formally half of the cosmological constraining power of the \Euclid cluster survey comes from clusters in the redshift range $1 \leq z \leq 2$. Due to the increased scatter of galaxy photometric redshifts beyond $z \geq 1.5$ and contamination in the galaxy catalogues, we expect strong incompleteness and impurity in this range. However, we present a subsample of promising candidates that were selected based on a combination of \gls{snr} above the threshold and visual inspection. Because the purity is low at this epoch for the current catalogue, these represent a small subset of high \gls{snr} detections. We anticipate significantly improved purity in this redshift range for DR1.

\begin{figure*}
    \centering
    \includegraphics[width=0.95\linewidth]{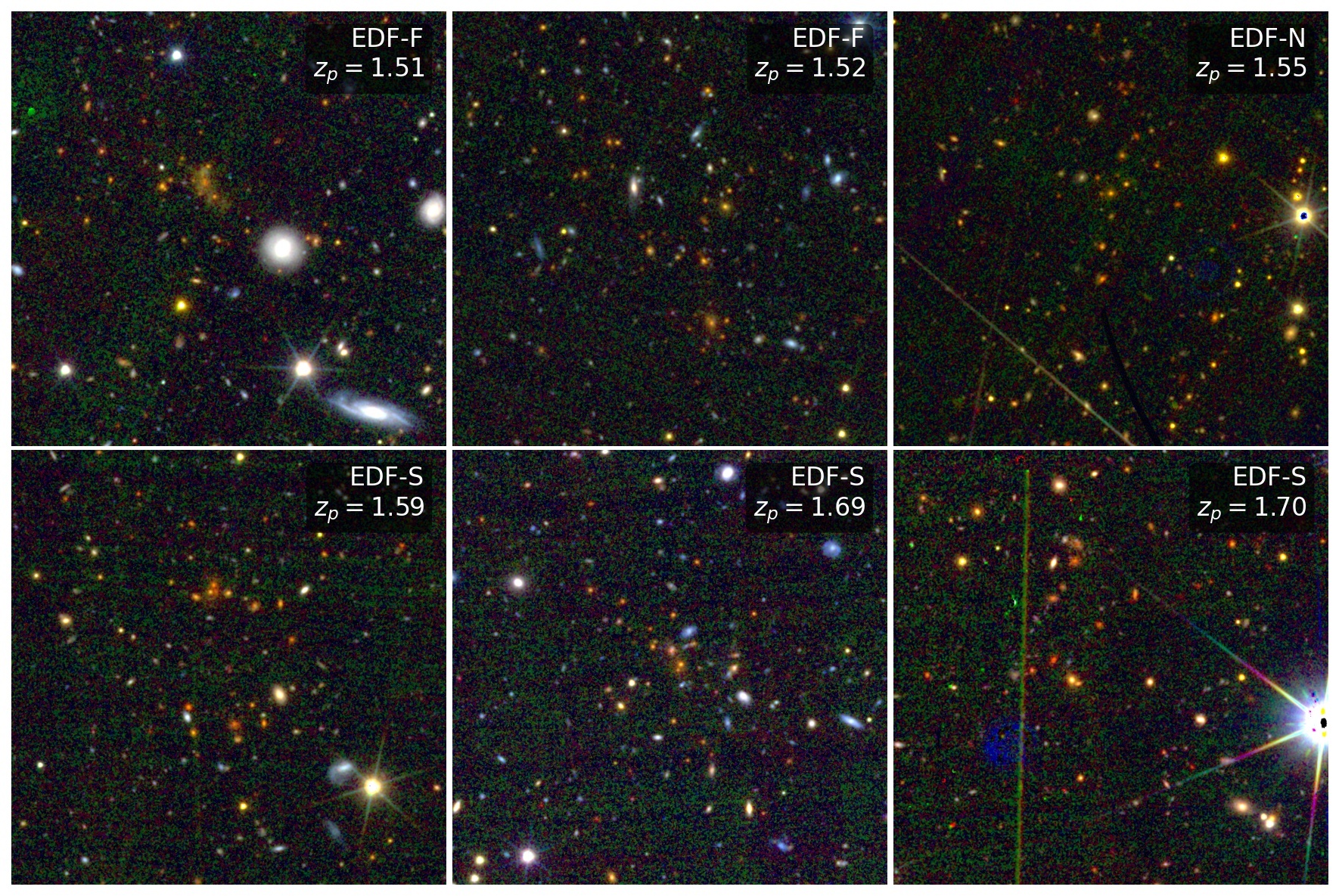}
    \caption{Postage stamps of $z \geq 1.5$ cluster candidates in the three Q1 fields as described in \cref{subsec:beyondz}. The properties, including coordinates, redshifts and richnesses for these candidates are displayed in \cref{table:highzcatalog}.}
    \label{fig:highzgallery}
\end{figure*}

In total, we find 58 systems across the three fields (four in the \gls{edff}, 10 in the \gls{edfs}, and 44 in the \gls{edfn}). The asymmetry in the number of detections between the north and south is driven by the apparent higher rate of contamination in the north in this field, in part due to its higher stellar density. We use the same criteria as described in \cref{subsec:neweuclid} to determine the reliability of these high redshift systems, noting 15 in total, which display clear visual overdensities in \Euclid and external imaging data where applicable. The complete properties, including positions, redshifts and richness estimates for these 15 systems are shown in \cref{table:highzcatalog}. We display postage stamps of six examples of these candidates across the three fields in \cref{fig:highzgallery}.

\section{External spectroscopic validation of clusters}\label{sec:specz}

We used the previously mentioned set of publicly available spectroscopic redshifts described in \cref{subsec:photozs} to estimate the spectroscopic redshifts for cluster detections in both the \gls{edff} and \gls{edfn}. As stated earlier, the spectroscopic coverage of the \gls{edfs} is poor, so we were unable to perform this analysis on that field. Using external spectroscopic data is particularly valuable in the low-redshift regime, which is not covered by \Euclid spectroscopy.

We selected galaxies with external spectroscopic redshift measurements within a cylinder centred on the cluster coordinates provided by the \AMICO and \PZWav detection algorithms. For simplicity, \PZWav detections are used as the cylinder centres, as similar results would be obtained using \AMICO detections. The selection radius was initially set to $1\,\mathrm{Mpc}$ from the centre, with an initial redshift-space window of $0.1(1+\zd)$ centred on $\zd$, where $\zd$ here refers to the photometric redshift measured by \PZWav. 

We focused on the cluster core to identify the peak in redshift space in order to minimise potential  line-of-sight contamination in the outskirts.  The relatively large initial redshift-space window was chosen to mitigate the bias and scatter previously observed between photometric and spectroscopic redshifts. We also applied a photometric redshift filter to disentangle the spectroscopic counterpart corresponding to the detection peak from potential other concentrations, keeping objects in a photometric redshift window $0.04 (1+\zd)$ centred on $\zd$.  

The spectroscopic redshift of the cluster was estimated using the biweight technique which has been shown to be appropriate for a small number of objects \citep{Beers1990}. After identifying the peak in redshift space, the criteria in photometric redshift was relaxed, and the radius was increased to 2\,Mpc in order to improve statistics. Galaxies with a radial velocity offset larger than $v_r =
\SI{3000}{\kilo\meter\per\second}$ in the cluster rest frame were rejected. 
The estimates of the spectroscopic redshift, its error, and the final number of galaxies used within 2 Mpc are provided in \cref{table:catalog} for those clusters with more than three redshifts remaining at the end of the procedure. 

\cref{fig:comp_z} compares the external spectroscopic redshift estimates with the photometric redshift estimates derived from the detections for the subsample of high-\gls{snr} detections in the \gls{edff} and \gls{edfn}, as previously defined. 

Out of 98 (137) cluster candidates with redshift lower than 1.5 in the \gls{edff} (\gls{edfn}), a spectroscopic redshift was measured for 39(51) of them with more than three members. We found 30 cluster candidates (39) with at least five spectroscopic members (shown in blue), while 9 (12) had three or four members (shown in magenta). Spectroscopic redshift estimates are shown to be in good agreement with the photometric redshift estimates. The relationship between the two estimates is fitted using a parametric model, and the region within one standard deviation is shaded in grey in the  \cref{fig:comp_z}. A systematic bias appears in the \gls{edff}, particularly in the redshift range $z\in[0.5,0.9]$, and to a lesser extent below redshift $z = 0.7$ in the \gls{edfn}, reflecting the same trend observed in the galaxy sample (\cref{fig:zpzs_stats}). 
\begin{figure*}
    \includegraphics[width=0.98\linewidth]{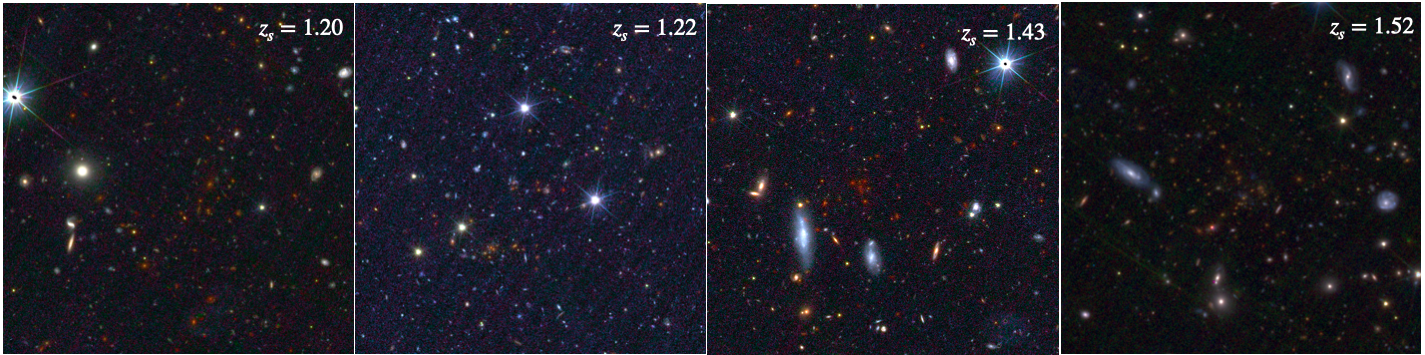}
    \caption{$2 \times 2$ arcminute VIS and NIR band colour cutouts of four \Euclid cluster detections in the \gls{edfn} with an estimated $\zs \geq 1$ based on at least two reliable spectroscopic redshifts per cluster field, in keeping with the methodology described in \cref{sec:specz}. The spectroscopic redshifts shown are derived within an aperture of 2\, Mpc from the cluster position. From left to right, the number of galaxies used for the cluster redshift measurement are two, five, four, and two, respectively.}. 
\end{figure*}

When possible, we performed a supplementary validation of the cluster redshifts using the internal \Euclid \OUSPE framework. However, since the number of concordant galaxies with \Euclid spectra is below five in all cases, we do not deem this sufficient to spectroscopically confirm the cluster detections. Nevertheless, the values obtained with \OUSPE and the \Euclid photometry are largely consistent; the details of this analysis can be found in  \cref{app:ouspe}.

\begin{figure*}
    \centering
    \includegraphics[width=0.42\linewidth]{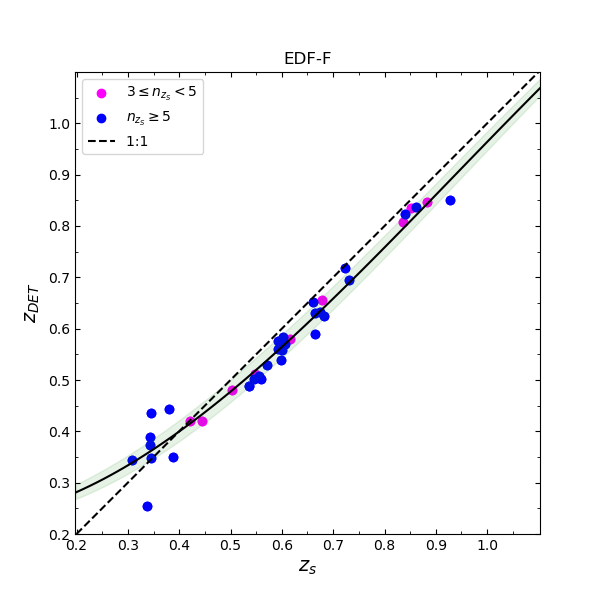}
     \includegraphics[width=0.42\linewidth]{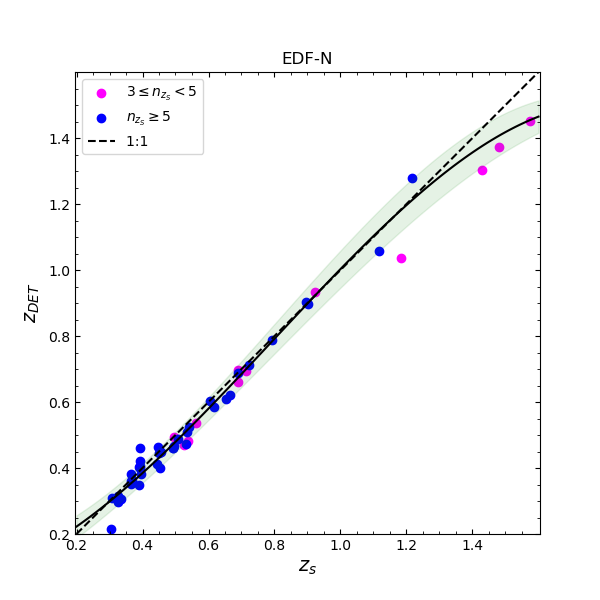}
    \caption{Spectroscopic redshift versus photometric redshift for high \gls{snr} cluster detections covered by spectroscopy in the \gls{edff} (\emph{left}) and the \gls{edfn} (\emph{right}). Cluster candidates with more than five members are shown in blue, and those with three or four members are shown in magenta. An empirical fit of the \( z_{\text{DET}} \)-\( z_s \) relation is shown with a continuous line, along with the \( 1\sigma \) envelope of the residuals (light green).}
    \label{fig:comp_z}
\end{figure*}

\section{Conclusions}\label{sec:conclusions}
We have performed a validation of the \Euclid cluster workflow on the \gls{q1} data release. We emphasise that this is the first time the workflow has been validated on real \Euclid data following its extensive testing on the Flagship mock catalogues presented in {\color{blue}Cabanac et al. (in preparation)}. We applied a galaxy selection and extensive validation of photometric quality on the \Euclid source catalogues for the purposes of optimal cluster detection. Subsequently, we ran the \AMICO and \PZWav cluster finders on the three Q1 fields to produce a joint catalogue consisting of 426 high \snr~detections identified from the two algorithms in a redshift range $0.2 \leq \zp \leq 1.5$. 

We characterise the detections according to their richness, colour-magnitude diagrams, and centring offsets, and we performed visual inspection of each of the sources. We cross-checked the validity of the Q1 catalogue against external photometric and spectroscopic data sets, noting that we have favoured purity heavily over completeness for the purposes of a first validation.  Nevertheless, we explored the limitations of the \Euclid detections, driven in part by incompleteness in the galaxy catalogues and large photometric redshift uncertainties at higher redshift. 

The detections presented in this paper represent an initial validation of the functionality of the \Euclid cluster detection workflow, yielding high \gls{snr} detections detected by both \AMICO and \PZWav. The characterisation of the photometric redshifts, centring, and richnesses further confirms that the detection workflow works as expected. 

We have presented an initial catalogue consisting of 426 reliable cluster detections. Of these, 349 clusters have at least one known counterpart, spanning various external catalogues of multi-wavelength data. The remaining 77 are plausible, new candidates without established counterparts in the literature, verified by \Euclid visual inspection and photometric redshift distributions around the cluster positions. Among them,  8 have a spectroscopic redshift estimated with more than five members within 2 \,Mpc, so seem to be spectroscopically confirmed. 

For all systems reported in this catalogue, we find strong internal agreement between the redshift and centring estimation from the two cluster finders. For the systems matched to external data sets, we find clear, positive correlations of the \Euclid richness with external mass proxies such as the SZ total mass and X-ray luminosity and temperature.   

The main limitations at present arise from residual issues with the input galaxy catalogue that are expected to be resolved by the time of the \gls{dr1} release. Specifically, incompleteness in the galaxy catalogue --  especially for bright, low-redshift galaxies -- and the current large scatter in the photometric redshifts at $z>1.5$ respectively impact the derived richnesses and limit the catalogue to lower redshifts. 

We also foresee improvements in contamination rejection and masking prior to \gls{dr1}, although both of these concerns are already manageable with the \gls{q1} data set. Because we expect that the external band photometry will improve significantly following the observation of more fields to better calibrate OU-PHZ templates, in addition to bias corrections in the photometric pipeline, we can expect that the results will improve considerably for \gls{dr1}, whilst already demonstrating a successful application of multiple aspects of the \Euclid cluster workflow for the detection and characterisation of optically selected galaxy clusters.

   \begin{figure*}
   \centering
    \end{figure*}
%
%
%

\begin{acknowledgements}
SB would like to acknowledge support from the Poincar\'e Fellowship funded by the Observatoire de la Côte d’Azur and the Initiative d'excellence d'Universit\'e C\^ote d'Azur. EM is supported by the PRIN 2022 project `EMC2 - Euclid Mission Cluster Cosmology: unlock the full cosmological utility of the Euclid photometric cluster catalog', funded by the European Union - NextGenerationEU, Mission 4, Component 2, Investment 1.1 -- CUP C53D23001140006, project code: 2022KCS97B. This work made use of the HEALPix software package \citep{healpix}.
\AckQone
\AckEC  
\end{acknowledgements}

%
%

\bibliography{CombinedEuclid}

\begin{appendix}

\section{The joint \Euclid cluster catalogue}

\begin{landscape}
\setlength{\tabcolsep}{5pt}
\begin{table}[p]
\tiny
\caption{The joint \Euclid cluster catalogue. Subset of 35 \PZWav and \AMICO-detected clusters with the highest signal-to-noise ratios.}
\label{table:catalog}
\centering
\begin{tabular}{llrrrrccrrrrr}
\hline
\hline
\noalign{\vskip 2pt}
NAME & 
ID & 
RA$_{\mathrm{PZWav}}$ & Dec$_{\mathrm{PZWav}}$ & RA$_{\mathrm{AMICO}}$ & Dec$_{\mathrm{AMICO}}$ & 
$z_{\mathrm{PZWav}}$ & $z_{\mathrm{AMICO}}$ &
$z_{\mathrm{spec}}$ & $N_{\mathrm{zspec}}$ &
SNR$_{\mathrm{PZWav}}$ & SNR$_{\mathrm{AMICO}}$ & 
$\lambda_{\mathrm{Pmem}}$ \\
 & & 
 (degrees) & 
 (degrees) & 
 (degrees) & 
 (degrees) & 
 & &  & (< 2 Mpc)  & & & \\
\noalign{\vskip 4pt}
\hline
\noalign{\vskip 4pt}

EUCL-Q1-CL J041113.88$-$481928.2 & EUCL-Q1-CL-0 & 62.8089 & $-48.3245$ & 62.8067 & $-48.3245$ & 0.41 & 0.40 & - & - & 40.81 & 89.38 & 241.47$\pm$4.79 \\
EUCL-Q1-CL J041724.90$-$474849.7 & EUCL-Q1-CL-1 & 64.3521 & $-47.8150$ & 64.3554 & $-47.8127$ & 0.55 & 0.56 & - & - & 24.20 & 60.43 & 115.26$\pm$4.13 \\
EUCL-Q1-CL J033747.63$-$275112.8 & EUCL-Q1-CL-2 & 54.4491 & $-27.8497$ & 54.4478 & $-27.8574$ & 0.49 & 0.48 & 0.535$\pm$0.004 & 10 & 23.43 & 51.72 & 81.36$\pm$4.74 \\
EUCL-Q1-CL J032321.22$-$275105.3 & EUCL-Q1-CL-3 & 50.8371 & $-27.8549$ & 50.8397 & $-27.8481$ & 0.42 & 0.41 & - & - & 22.61 & 46.43 & 56.14$\pm$4.73 \\
EUCL-Q1-CL J040151.05$-$502848.7 & EUCL-Q1-CL-4 & 60.4686 & $-50.4780$ & 60.4568 & $-50.4824$ & 0.43 & 0.42 & - & - & 22.38 & 30.49 & 99.03$\pm$4.45 \\
EUCL-Q1-CL J040558.72$-$491549.3 & EUCL-Q1-CL-5 & 61.5018 & $-49.2626$ & 61.4875 & $-49.2648$ & 0.31 & 0.30 & - & - & 22.21 & 52.78 & 131.57$\pm$4.40 \\
EUCL-Q1-CL J040655.77$-$480504.2 & EUCL-Q1-CL-6 & 61.7282 & $-48.0850$ & 61.7365 & $-48.0840$ & 0.69 & 0.70 & - & - & 22.14 & 34.27 & 99.21$\pm$3.72 \\
EUCL-Q1-CL J041343.51$-$480756.5 & EUCL-Q1-CL-7 & 63.4322 & $-48.1305$ & 63.4304 & $-48.1343$ & 1.47 & 1.46 & - & - & 21.74 & 37.54 & 22.95$\pm$2.24 \\
EUCL-Q1-CL J180607.67$+$635615.6 & EUCL-Q1-CL-8 & 271.5285 & $63.9386$ & 271.5355 & $63.9367$ & 1.32 & 1.33 & - & - & 21.53 & 57.26 & 23.13$\pm$2.05 \\
EUCL-Q1-CL J040509.84$-$464902.4 & EUCL-Q1-CL-9 & 61.2955 & $-46.8097$ & 61.2865 & $-46.8250$ & 0.42 & 0.38 & - & - & 20.48 & 37.06 & 99.49$\pm$4.55 \\
EUCL-Q1-CL J035146.84$-$480404.3 & EUCL-Q1-CL-10 & 57.9409 & $-48.0683$ & 57.9494 & $-48.0674$ & 0.56 & 0.56 & - & - & 20.22 & 35.70 & 28.75$\pm$2.73 \\
EUCL-Q1-CL J034959.80$-$481952.6 & EUCL-Q1-CL-11 & 57.5029 & $-48.3288$ & 57.4954 & $-48.3338$ & 1.40 & 1.44 & - & - & 20.14 & 31.49 & 60.20$\pm$1.75 \\
EUCL-Q1-CL J035003.59$-$504145.7 & EUCL-Q1-CL-12 & 57.5123 & $-50.6948$ & 57.5176 & $-50.6973$ & 0.65 & 0.66 & - & - & 20.10 & 23.28 & 33.66$\pm$2.74 \\
EUCL-Q1-CL J033352.73$-$272113.8 & EUCL-Q1-CL-13 & 53.4665 & $-27.3535$ & 53.4728 & $-27.3542$ & 0.57 & 0.55 & 0.605$\pm$0.005 & 8 & 19.98 & 36.64 & 45.67$\pm$3.15 \\
EUCL-Q1-CL J032929.78$-$281930.3 & EUCL-Q1-CL-14 & 52.3730 & $-28.3266$ & 52.3752 & $-28.3236$ & 0.63 & 0.66 & 0.682$\pm$0.007 & 37 & 19.74 & 29.01 & 59.93$\pm$4.68 \\
EUCL-Q1-CL J033412.98$-$282418.7 & EUCL-Q1-CL-15 & 53.5511 & $-28.4065$ & 53.5570 & $-28.4039$ & 0.65 & 0.66 & 0.662$\pm$0.007 & 74 & 18.84 & 25.32 & 97.08$\pm$2.75 \\
EUCL-Q1-CL J033355.96$-$283807.7 & EUCL-Q1-CL-16 & 53.4879 & $-28.6368$ & 53.4785 & $-28.6342$ & 0.59 & 0.56 & 0.665$\pm$0.007 & 38 & 18.70 & 25.15 & 71.24$\pm$3.38 \\
EUCL-Q1-CL J040440.54$-$472444.6 & EUCL-Q1-CL-17 & 61.1691 & $-47.4098$ & 61.1687 & $-47.4150$ & 0.88 & 0.89 & - & - & 18.62 & 33.87 & 50.03$\pm$3.70 \\
EUCL-Q1-CL J033619.94$-$263512.9 & EUCL-Q1-CL-18 & 54.0817 & $-26.5831$ & 54.0844 & $-26.5907$ & 0.50 & 0.50 & 0.546$\pm$0.006 & 11 & 17.66 & 27.53 & 42.73$\pm$2.41 \\
EUCL-Q1-CL J033635.31$-$292553.3 & EUCL-Q1-CL-19 & 54.1464 & $-29.4327$ & 54.1478 & $-29.4303$ & 0.87 & 0.90 & - & - & 17.47 & 30.25 & 42.64$\pm$3.09 \\
EUCL-Q1-CL J041838.86$-$455252.9 & EUCL-Q1-CL-20 & 64.6645 & $-45.8807$ & 64.6593 & $-45.8821$ & 0.69 & 0.62 & - & - & 17.19 & 30.79 & 76.28$\pm$4.23 \\
EUCL-Q1-CL J033851.49$-$284807.8 & EUCL-Q1-CL-21 & 54.7174 & $-28.7999$ & 54.7117 & $-28.8045$ & 1.04 & 1.04 & - & - & 16.97 & 24.47 & 44.61$\pm$2.97 \\
EUCL-Q1-CL J042117.89$-$484550.8 & EUCL-Q1-CL-22 & 65.3239 & $-48.7647$ & 65.3252 & $-48.7635$ & 1.33 & 1.39 & - & - & 16.82 & 23.87 & 33.74$\pm$1.83 \\
EUCL-Q1-CL J033526.63$-$291824.9 & EUCL-Q1-CL-23 & 53.8620 & $-29.3015$ & 53.8599 & $-29.3124$ & 0.51 & 0.51 & 0.555$\pm$0.010 & 9 & 16.51 & 34.45 & 39.71$\pm$3.00 \\
EUCL-Q1-CL J175712.26$+$680259.0 & EUCL-Q1-CL-24 & 269.3108 & $68.0460$ & 269.2914 & $68.0535$ & 0.38 & 0.40 & - & - & 16.46 & 31.82 & 43.42$\pm$4.31 \\
EUCL-Q1-CL J033538.84$-$270305.5 & EUCL-Q1-CL-25 & 53.9143 & $-27.0511$ & 53.9094 & $-27.0520$ & 0.35 & 0.35 & 0.344$\pm$0.002 & 11 & 16.43 & 34.13 & 50.39$\pm$4.01 \\
EUCL-Q1-CL J041109.36$-$490953.7 & EUCL-Q1-CL-26 & 62.7910 & $-49.1650$ & 62.7870 & $-49.1649$ & 0.42 & 0.35 & - & - & 16.42 & 31.19 & 31.01$\pm$5.12 \\
EUCL-Q1-CL J034251.87$-$491114.0 & EUCL-Q1-CL-27 & 55.7173 & $-49.1902$ & 55.7150 & $-49.1842$ & 1.40 & 1.49 & - & - & 16.38 & 31.36 & 30.49$\pm$1.25 \\
EUCL-Q1-CL J040406.81$-$481302.2 & EUCL-Q1-CL-28 & 61.0243 & $-48.2198$ & 61.0324 & $-48.2148$ & 0.44 & 0.40 & - & - & 16.37 & 38.93 & 98.53$\pm$5.27 \\
EUCL-Q1-CL J033055.76$-$294746.3 & EUCL-Q1-CL-29 & 52.7296 & $-29.7977$ & 52.7351 & $-29.7948$ & 0.84 & 0.84 & - & - & 16.36 & 21.79 & 38.51$\pm$2.72 \\
EUCL-Q1-CL J034001.03$-$284959.7 & EUCL-Q1-CL-30 & 55.0111 & $-28.8358$ & 54.9975 & $-28.8307$ & 0.25 & 0.32 & 0.337$\pm$0.004 & 34 & 16.23 & 23.99 & 17.76$\pm$2.84 \\
EUCL-Q1-CL J041839.71$-$483832.0 & EUCL-Q1-CL-31 & 64.6662 & $-48.6298$ & 64.6647 & $-48.6547$ & 1.41 & 1.50 & - & - & 16.19 & 18.74 & 34.17$\pm$2.46 \\
EUCL-Q1-CL J174731.98$+$663438.5 & EUCL-Q1-CL-32 & 266.8740 & $66.5803$ & 266.8924 & $66.5744$ & 0.42 & 0.42 & 0.392$\pm$0.007 & 7 & 15.96 & 55.86 & 14.38$\pm$2.76 \\
EUCL-Q1-CL J181458.48$+$645725.1 & EUCL-Q1-CL-33 & 273.7412 & $64.9563$ & 273.7461 & $64.9577$ & 0.52 & 0.52 & - & - & 15.86 & 28.17 & 36.98$\pm$3.22 \\
EUCL-Q1-CL J040332.81$-$482819.1 & EUCL-Q1-CL-34 & 60.8779 & $-48.4695$ & 60.8955 & $-48.4745$ & 0.37 & 0.36 & - & - & 15.85 & 36.24 & 63.50$\pm$2.45 \\

\hline

\noalign{\vskip 2pt}
\end{tabular}
\tablefoot{z$_{\mathrm{spec}}$ is the estimate of the spectroscopic cluster redshift from external spectroscopy data derived with the methodology detailed in \cref{sec:specz}.}  
\end{table}
\end{landscape}

\begin{landscape}
\setlength{\tabcolsep}{5pt}
\begin{table}[p]
\caption{Most reliable \PZWav and \AMICO-detected clusters at redshifts $z \geq 1.5$}
\label{table:highzcatalog}
\centering
\begin{tabular}{llrrrrccrrr}
\hline
\hline
\noalign{\vskip 2pt}
NAME &
ID &  
RA$_{\mathrm{PZWav}}$ & Dec$_{\mathrm{PZWav}}$  & RA$_{\mathrm{AMICO}}$ & Dec$_{\mathrm{AMICO}}$ & 
$z_{\mathrm{PZWav}}$ & $z_{\mathrm{AMICO}}$ &
SNR$_{\mathrm{PZWav}}$ & SNR$_{\mathrm{AMICO}}$ & 
$\lambda_{\mathrm{Pmem}}$ \\
 &  & (degrees) & (degrees) & (degrees) & (degrees) & & & & & \\
\noalign{\vskip 4pt}
\hline
\noalign{\vskip 4pt}
EUCL-Q1-CL J180351.79$+$674957.7 & EUCL-Q1-HzCL-0 & 270.9647 & $67.8332$ & 270.9669 & $67.8322$ & 1.56 & 1.56 & 22.91 & 34.18 & 44.15$\pm$1.60 \\
EUCL-Q1-CL J180720.52$+$653940.4 & EUCL-Q1-HzCL-1 & 271.8386 & $65.6585$ & 271.8324 & $65.6640$ & 1.97 & 1.98 & 21.01 & 34.07 & 10.39$\pm$2.24 \\
EUCL-Q1-CL J175448.16$+$654559.8 & EUCL-Q1-HzCL-2 & 268.7050 & $65.7638$ & 268.6963 & $65.7694$ & 1.55 & 1.55 & 14.17 & 23.50 & 17.40$\pm$1.02 \\
EUCL-Q1-CL J034654.61$-$480825.0 & EUCL-Q1-HzCL-3 & 56.7254 & $-48.1447$ & 56.7297 & $-48.1359$ & 1.70 & 1.74 & 13.32 & 17.41 & 34.83$\pm$1.25 \\
EUCL-Q1-CL J181953.75$+$651241.1 & EUCL-Q1-HzCL-4 & 274.9701 & $65.2106$ & 274.9779 & $65.2122$ & 1.77 & 1.80 & 11.93 & 19.57 & 17.33$\pm$1.37 \\
EUCL-Q1-CL J181349.12$+$674651.3 & EUCL-Q1-HzCL-5 & 273.4570 & $67.7727$ & 273.4523 & $67.7891$ & 1.56 & 1.56 & 10.80 & 18.27 & 33.25$\pm$1.19 \\
EUCL-Q1-CL J180347.27$+$640019.4 & EUCL-Q1-HzCL-6 & 270.9477 & $64.0089$ & 270.9462 & $64.0019$ & 1.95 & 1.96 & 10.24 & 25.00 & 7.74$\pm$1.52 \\
EUCL-Q1-CL J034533.23$-$500806.7 & EUCL-Q1-HzCL-7 & 56.3879 & $-50.1344$ & 56.3889 & $-50.1360$ & 1.69 & 1.77 & 9.78 & 20.01 & 13.92$\pm$1.24 \\
EUCL-Q1-CL J033056.82$-$284246.3 & EUCL-Q1-HzCL-8 & 52.7357 & $-28.7109$ & 52.7378 & $-28.7148$ & 1.47 & 1.52 & 9.59 & 17.63 & 17.68$\pm$1.51 \\
EUCL-Q1-CL J041546.25$-$495532.5 & EUCL-Q1-HzCL-9 & 63.9408 & $-49.9269$ & 63.9447 & $-49.9244$ & 1.59 & 1.57 & 9.45 & 16.89 & 12.15$\pm$1.46 \\
EUCL-Q1-CL J033730.18$-$283827.6 & EUCL-Q1-HzCL-10 & 54.3729 & $-28.6438$ & 54.3786 & $-28.6382$ & 1.51 & 1.49 & 9.17 & 16.39 & 12.05$\pm$1.88 \\
EUCL-Q1-CL J040832.33$-$473418.8 & EUCL-Q1-HzCL-11 & 62.1376 & $-47.5722$ & 62.1318 & $-47.5716$ & 1.79 & 1.81 & 8.89 & 17.58 & 8.51$\pm$1.51 \\
EUCL-Q1-CL J033722.21$-$283404.8 & EUCL-Q1-HzCL-12 & 54.3415 & $-28.5675$ & 54.3435 & $-28.5685$ & 1.70 & 1.70 & 8.82 & 22.31 & 8.97$\pm$1.48 \\
EUCL-Q1-CL J175433.97$+$674611.6 & EUCL-Q1-HzCL-13 & 268.6438 & $67.7704$ & 268.6393 & $67.7694$ & 1.55 & 1.54 & 8.73 & 21.09 & 12.90$\pm$0.36 \\
EUCL-Q1-CL J034923.66$-$481352.5 & EUCL-Q1-HzCL-14 & 57.3443 & $-48.2336$ & 57.3529 & $-48.2289$ & 1.85 & 1.89 & 8.59 & 16.40 & 9.94$\pm$0.70 \\
\hline
\noalign{\vskip 2pt}
\end{tabular}
\end{table}
\end{landscape}



\section{Detailed breakdown of Q1 cluster matches to external data sets}\label{app:external}

In order to perform the external validation of the Q1 cluster sample, the Q1 clusters were matched to various existing data sets using the methodology outlined in \cref{sec:external}. The tables detailed in this Appendix illustrate the breakdown of matches from the meta-catalogues in this analysis, and contains the results from the extended search detailed in \cref{subsec:neweuclid}, from which the assessment of potentially new \Euclid clusters has been determined. 

\begin{table*}
\caption{Cluster match breakdown in \gls{edfs}}
\label{tab:results_edfs}
\centering
\begin{tabular}{lcc}
  Cluster sample & Matches with \Euclid Q1 & Unique matches with \Euclid Q1 \\
  \hline
  MCXC-II & 0 & 0  \\
  MCSZ & 15 & 5  \\
  ComPRASS & 3 & 0 \\
  eROSITA & 39 & 6 \\
  MCCD & 2 & 0 \\
  ${\rm LC}^2$ & 0 & 0 \\
  Abell & 3 & 0 \\
  DES Y1 RM & 71 & 41 \\
  \hline
  In two catalogues & & 23 \\
  In three catalogues & & 6 \\
  In four catalogues & & 3 \\
  In five catalogues & & 1 \\
  In six catalogues &  & 0 \\
  Unmatched & & 106 \\
  \hline
  Total & & 191 \\
\end{tabular}
\tablefoot{No additional match from NED, 40 additional matches from MaDCoWS2, and 51 additional matches from WH24.
}
\end{table*}

\begin{table*}
\caption{Cluster match breakdown in \gls{edff}}
\label{tab:results_edff}
\centering
\begin{tabular}{lcc}
  Cluster sample & Matches with \Euclid Q1 & Unique matches with \Euclid Q1 \\
   \hline
   MCXC-II & 3 & 1  \\
   MCSZ & 3 & 1  \\
   ComPRASS & 0 & 0 \\
   eROSITA & 10 & 3 \\
   MCCD & 0 & 0 \\
   ${\rm LC}^2$ & 0 & 0 \\
   Abell & 1 & 0 \\
   DES Y1 RM & 13 & 9 \\
   \hline
   In two catalogues & & 5 \\
   In three catalogues & & 2 \\
   In four catalogues & & 0 \\
   Unmatched & & 77 \\
   \hline
   Total & & 98 \\
\end{tabular}
\tablefoot{20 additional matches from NED, 23 additional matches from MaDCoWS2, and 29 additional matches from WH24.
}
\end{table*}

\begin{table*}
\caption{Cluster match breakdown in \gls{edfn}}
\label{tab:results_edfn}
\centering
\begin{tabular}{lcc}
  Cluster sample & Matches with \Euclid Q1 & Unique matches with \Euclid Q1 \\
   \hline
   MCXC-II & 4 & 3  \\
   MCSZ & 0 & 0  \\
   ComPRASS & 0 & 0 \\
   eROSITA & 0 & 0 \\
   MCCD & 0 & 0 \\
   ${\rm LC}^2$ & 0 & 0 \\
   Abell & 1 & 0 \\
   DES Y1 RM & 0 & 0 \\
   \hline
   In two catalogues & & 1 \\
   In three catalogues & & 0 \\
   Unmatched & & 133 \\
   \hline
   Total & & 137 \\
\end{tabular}
\tablefoot{29 additional matches from NED, No additional match from MaDCoWS2, and 7 additional matches from WH24.
}
\end{table*}

\section{Spectroscopic validation with \OUSPE}\label{app:ouspe}
The detection algorithms estimate cluster redshifts based on the mean photometric redshift of member galaxies. To confirm and further refine these estimates, we examined the \Euclid spectroscopy of galaxies in the cluster fields, as provided in the Q1 data release. We limited our inspections of the \Euclid spectroscopy to those candidate clusters with $\zd > 0.9$, the lower limit at which the H$\alpha$ lines can be detected given the spectral sensitivity range of the \Euclid red grism \citep{EuclidSkyOverview}. This leaves us with 100 cluster candidates (17 in the \gls{edff}, 43 in the \gls{edfs}, and 40 in the \gls{edfn}).

The \Euclid \OUSPE package provides five $z$ estimates for each galaxy; in the following, we only use the highest ranked estimate -- \texttt{spec.spec\_rank=0} in the \gls{sas} -- when selecting objects for verification of their spectroscopic redshifts. We restrict our analysis to galaxies with \OUSPE redshift estimates with signal-to-noise ratio (\texttt{spe\_line\_snr\_gf} in the \gls{sas})
$>3.5$, probability (\texttt{spe\_z\_prob} in the \gls{sas}), $>0.99$, with line flux $<10^{-14}\,\mathrm{erg\,cm^{-2}\,s^{-1}}$ to avoid spurious features, and based on more than 300 pixels (\texttt{spe\_npix>300}). 
Of the galaxies passing these spectroscopic quality criteria, we only considered those within 4\,arcmin from the candidate cluster centre (corresponding to $\simeq 2\,\mathrm{Mpc}$ in the range $z=0.9-1.5$),
and with spectroscopic redshifts within a range of
$\pm 0.05 \hide{\times} \, (1+\zd)$, corresponding to about three times the estimated pre-launch cluster $\zd$ uncertainty \citep{Adam-EP3}. 

In total, we selected 527 galaxies (87 in \gls{edff}, 197 in \gls{edfs}, and 243 in \gls{edfn}), and after visual inspection of their spectra we retained 170 redshift estimates as reliable (35 in \gls{edff}, 65 in \gls{edfs}, and 70 in \gls{edfn}), that is, one third of the total. This means we deal with, on average, less than two reliable redshifts per cluster field.

We then look for concordant redshifts in each cluster field,
where by `concordant' we mean that their redshifts must be separated by $\leq 0.0133 \, (1+\zd)$, which corresponds to a velocity separation of \SI{4000}{\kilo\meter\per\second}
in the cluster rest frame. This value corresponds to $\sim \pm 3$ times the velocity dispersion of a Virgo-like cluster.  We were able to identify $\geq 2$ concordant-$z$ members in 20 cluster candidates, 4 in the \gls{edff}, 
7 in the \gls{edfs}, and 9 in the \gls{edfn}. They are listed in \cref{t:specconf}. 
We estimated the spectroscopic cluster redshifts as the average of the redshifts of the concordant galaxies. 
For one cluster in \gls{edfn}, we identify two groups of galaxies at different mean $z$, suggesting that this detections might be partly the result of overlapping structures along the line of sight.

\begin{table}
\centering
\caption{Cluster candidates with concordant spec-$z$ galaxies}
\label{t:specconf}
\begin{tabular}{lrlcc}
\toprule
Candidate & N$_c$ & $\overline{z}$  & \multicolumn{2}{c}{$\zd$} \\
   id.    &              & (spectroscopic) & (PZWav) & (AMICO) \\
\midrule
EUCL-Q1-CL-21   & 4 & $1.0746 \pm 0.0039$ & 1.04 & 1.04 \\
EUCL-Q1-CL-200  & 3 & $1.0931 \pm 0.0054$ & 1.12 & 1.13 \\
EUCL-Q1-CL-211  & 3 & $0.9199 \pm 0.0028$ & 0.98 & 0.96 \\
                & 4 & $0.9932 \pm 0.0016$ & & \\
EUCL-Q1-CL-415  & 2 & $0.9644 \pm 0.0048$ & 1.04 & 1.04 \\
EUCL-Q1-CL-128  & 2 & $0.9395 \pm 0.0018$ & 0.92 & 0.89 \\
EUCL-Q1-CL-231  & 2 & $0.9782 \pm 0.0001$ & 0.95 & 0.92 \\
EUCL-Q1-CL-248  & 2 & $1.0162 \pm 0.0035$ & 0.95 & 0.98 \\
EUCL-Q1-CL-277  & 3 & $1.2610 \pm 0.0003$ & 1.26 & 1.26 \\
EUCL-Q1-CL-285  & 3 & $1.3223 \pm 0.0048$ & 1.21 & 1.26 \\
EUCL-Q1-CL-333  & 3 & $0.9464 \pm 0.0020$ & 1.04 & 1.02 \\ 
EUCL-Q1-CL-403  & 3 & $1.0079 \pm 0.0031$ & 0.94 & 0.93 \\
EUCL-Q1-CL-288  & 2 & $0.9777 \pm 0.0011 $ & 0.93 & 0.99 \\
EUCL-Q1-CL-154  & 4 & $1.0439 \pm 0.0006 $ & 1.00 & 0.96 \\
EUCL-Q1-CL-286  & 3 & $1.1001 \pm 0.0007 $ & 1.04 & 1.05 \\
EUCL-Q1-CL-45   & 2 & $1.1966 \pm 0.0011 $ & 1.18 & 1.16 \\
EUCL-Q1-CL-346  & 2 & $1.2113 \pm 0.0009 $ & 1.32 & 1.32 \\
EUCL-Q1-CL-351  & 3 & $1.3512 \pm 0.0012 $ & 1.37 & 1.38 \\
EUCL-Q1-CL-87   & 2 & $1.3667 \pm 0.0013 $ & 1.38 & 1.37 \\
EUCL-Q1-CL-247  & 2 & $1.5068 \pm 0.0014$ & 1.45 & 1.46 \\
EUCL-Q1-CL-424  & 3 & $1.5162 \pm 0.0010 $ & 1.45 & 1.43 \\

\hline
\end{tabular}
\tablefoot{$N_\mathrm{c}$ is the number of galaxies with concordant spectroscopic redshifts as determined from \Euclid spectra. Candidate EUCL-Q1-CL-211 has two groups of three and four galaxies with concordant $z$'s along the line of sight.} 
\end{table}

\end{appendix}

\end{document}